\documentclass{cdclass}
\usepackage[table,xcdraw]{xcolor}
\usepackage{enumitem} 
\usepackage{bbm,amssymb,amsmath,amsfonts,gensymb}
\usepackage{graphicx}
\usepackage{etoolbox} 

\usepackage{color}
\usepackage{multirow}
\usepackage{pdflscape} 
\usepackage{longtable}  
\usepackage{tabu}

\begin{document}

\title{Pedestrian Crowd Management Experiments: A Data Guidance Paper}

\author{
  Ann Katrin Boomers\authorlabel{1} \and 
  Maik Boltes\authorlabel{1} \and 
  Juliane Adrian\authorlabel{1} \and 
  Mira Beermann\authorlabel{2} \and 
  Mohcine Chraibi\authorlabel{1} \and 
  Sina Feldmann\authorlabel{1} \and 
  Frank Fiedrich\authorlabel{4} \and 
  Niklas Frings\authorlabel{4} \and 
  Arne Graf\authorlabel{1} \and 
  Alica Kandler\authorlabel{1} \and 
  Deniz Kilic\authorlabel{1} \and 
  Krisztina Konya\authorlabel{2} \and 
  Mira Küpper\authorlabel{3} \and 
  Andreas Lotter\authorlabel{4} \and 
  Helena Lügering\authorlabel{1} \and 
  Francesca Müller\authorlabel{4} \and  
  Sarah Paetzke\authorlabel{1} \and 
  Anna-Katharina Raytarowski\authorlabel{1} \and 
  Olga Sablik\authorlabel{3} \and 
  Tobias Schrödter\authorlabel{1} \and 
  Armin Seyfried\authorlabel{1,3} \and 
  Anna Sieben\authorlabel{2,5} \and 
  Ezel Üsten\authorlabel{1} 
}
\authorrunning{A.K. Boomers et al.}
\institute{
  \authorlabel{1} Institute for Advanced Simulation, Forschungszentrum Jülich, Jülich, Germany
  \authoremail{1}{a.boomers@fz-juelich.de}, 
  \href{mailto:m.boltes@fz-juelich.de}{m.boltes@fz-juelich.de}
  \and
  \authorlabel{2} Chair of Social Theory and Social Psychology, Ruhr-Universität Bochum, Bochum, Germany
   \and
  \authorlabel{3} Faculty of Architecture and Civil Engineering, University of Wuppertal, Wuppertal, Germany
  \and
  \authorlabel{4} Faculty of Mechanical Engineering and Safety Engineering, University of Wuppertal,\\ 
  \settowidth{\emailtab}{\mbox{\authorlabel{1}}}\hspace*{\emailtab} Wuppertal, Germany
 \and
 \authorlabel{5} School of Humanities and Social Sciences, University of St. Gallen, St. Gallen, Switzerland
}

\date{year}{date1}{date2}{date3} 
\ldoi{10.17815/CD.20XX.X} 
\volume{V}  
\online{AX} 

\maketitle

\begin{abstract}
Understanding pedestrian dynamics and the interaction of pedestrians with their environment is crucial to the safe and comfortable design of pedestrian facilities.
Experiments offer the opportunity to explore the influence of individual factors.
In the context of the project CroMa (Crowd Management in transport infrastructures), experiments were conducted with about 1000 participants to test various physical and social psychological hypotheses focusing on people's behaviour at railway stations and crowd management measures. 
The following experiments were performed: i) Train Platform Experiment, ii) Crowd Management Experiment, iii) Single-File Experiment, iv) Personal Space Experiment, v) Boarding and Alighting Experiment, vi) Bottleneck Experiment and vii) Tiny Box Experiment. 
This paper describes the basic planning and implementation steps, outlines all experiments with parameters, geometries, applied sensor technologies and pre- and post-processing steps. All data can be found in the pedestrian dynamics data archive.
\end{abstract}


\section{Introduction}
\label{sec:intro}
This paper gives general information about the experiments performed within the project CroMa (Crowd Management in transport infrastructures) \cite{CroMaProject2022}. In addition, experiments from other projects such as CrowdDNA \cite{CrowdDNAProject2022} were carried out in the context of this experiment series as well as experiments that cannot be assigned to a third party funded project. The paper includes information about the overall organization, the experimental site, the procedure and timeline, the participants, the data collection technique and gives an overview of all experiments. Further detailed information of single experiments, especially data analysis, will be found in focused papers on these experiments. All data gathered by the sensors used will be made freely accessible in the pedestrian dynamics data archive \cite{ForschungszentrumJulich2022h} with the publication of the first scientific results at the latest. General data mentioned in this paper like the overall composition of the test persons based on handed out questionnaires and a measurement course can also be found in the data archive \cite{ForschungszentrumJulich2022}.

\noindent The CroMa project itself is focused on developing and enhancing different strategies, such as building regulations, crowd management, and innovative action strategies to increase the efficiency of pedestrian facilities in railway and underground stations. These strategies aim to increase the robustness and efficiency of railway stations during peak load and to avoid crushes in the event of critical crowd densities. Research within the framework of CroMa includes the investigation of pedestrian flow in traffic facilities and the study of pedestrian behaviour within dense crowds. These research fields have also been assessed by means of the large-scale experiments described in this paper in which several external (structural) and internal (characteristics regarding the test person sample) parameters have been varied on a controlled basis.


\section{Preparation of Experiments}
The CroMa-experiments were conducted from October 8, until October 11, 2021 in the Mitsubishi Electric Hall (MEH) in Düsseldorf, Germany. The MEH is a multipurpose event hall with an interior hall size of 3500\,m\textsuperscript{2} and an additional main and side foyer.
The planning and preparation were divided into two interlocked parts. One being the overall organization and provision of rooms and material and the other being the scientific planning of the individual experiments. Several preparatory meetings were held to discuss issues related to the variety of tested scenarios and statistical significance relative to available time and personnel. A temporal and spatial setup was developed to account for the level of information given to the participants about the aims of the study as well as their learning effect over the course of the day. The experimental plans were simplified, concretized and the times for conducting the experiments, announcements, walking routes, filling in the questionnaires and taking small breaks were calculated to test the feasibility of the setup. 
\noindent Originally, the experiments were planned for March 2020 and had to be postponed due to the growing SARS-CoV-2 (Covid-19) pandemic. When the experiments were conducted in October 2021, the setup was revised regarding compliance with safety measures and expanded to include a hygiene and safety concept.

\subsection{General Framework}
\label{subsec:GeneralFramework}
\begin{figure}[h!]
	\centering
 	\includegraphics[width=0.8\textwidth]{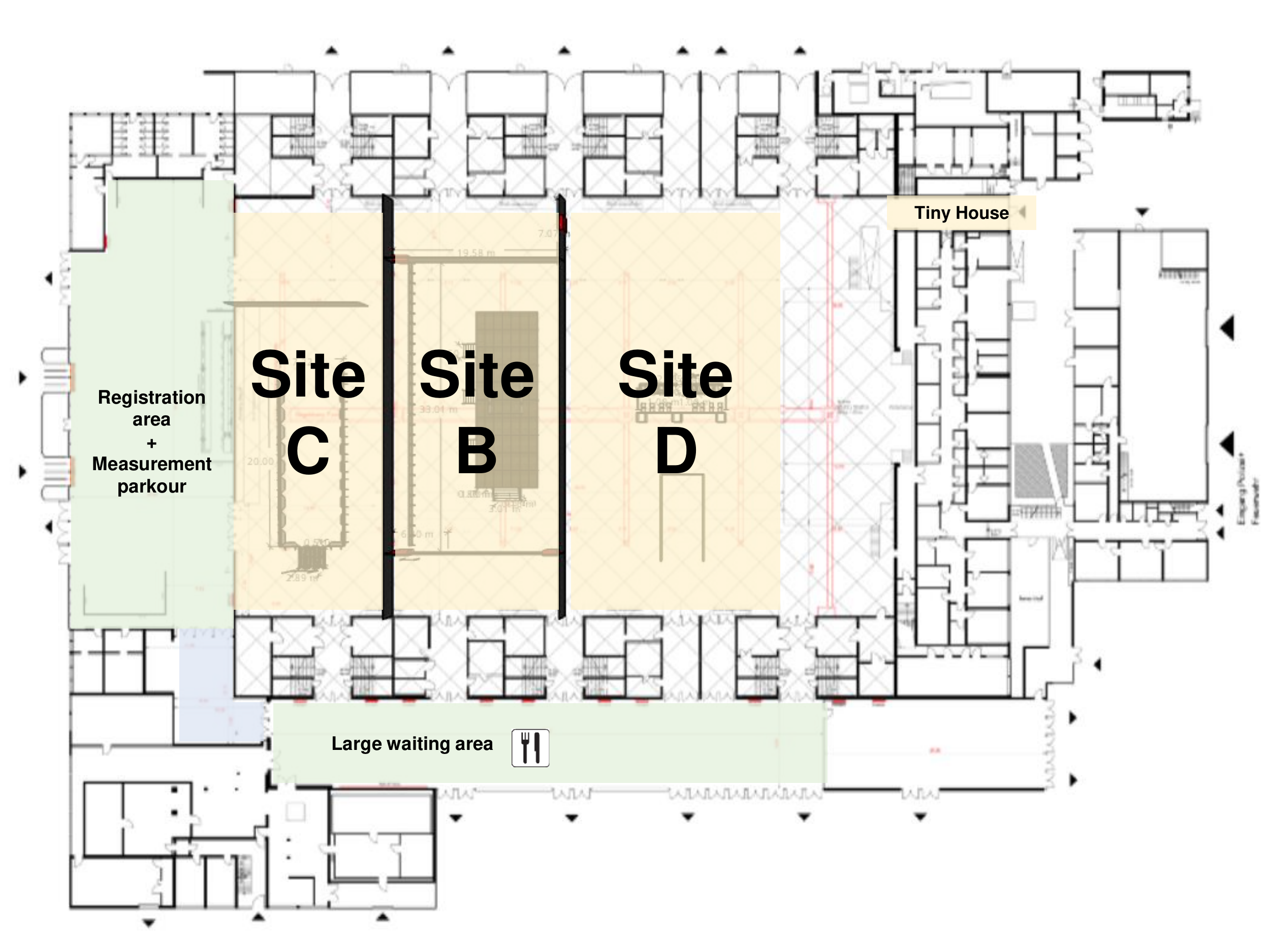}
 	 \caption{Room plan of experiment hall. Yellow patches indicate experimental sites in the interior hall. Green patches indicate open areas. Blue patch indicates corridor between open areas that was used as an `icebreaker' experiment c.f. \sref{subsec:corona}.}
  	\label{fig:sites} 
\end{figure}

\begin{figure}[h!]
	\centering
 	\subfigure[]{\includegraphics[height=0.3\textheight]{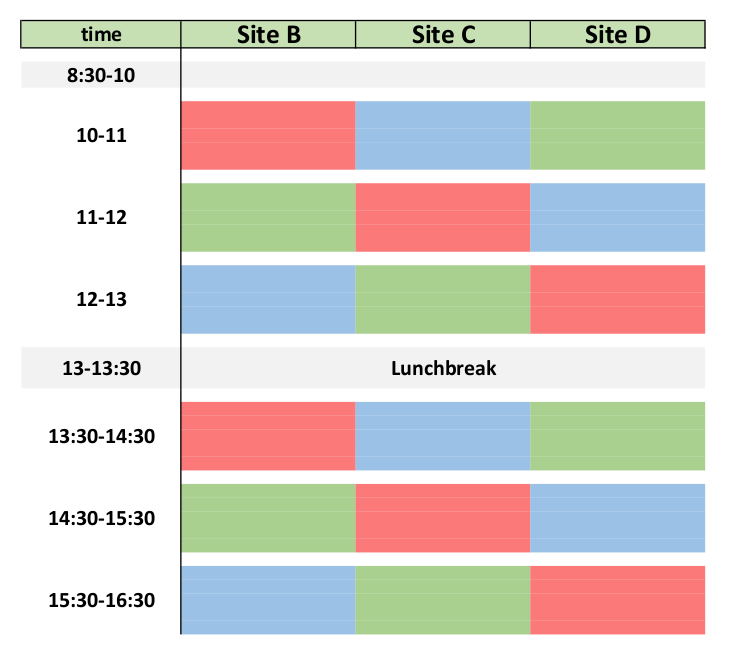}}
 	\subfigure[]{\includegraphics[height=0.3\textheight]{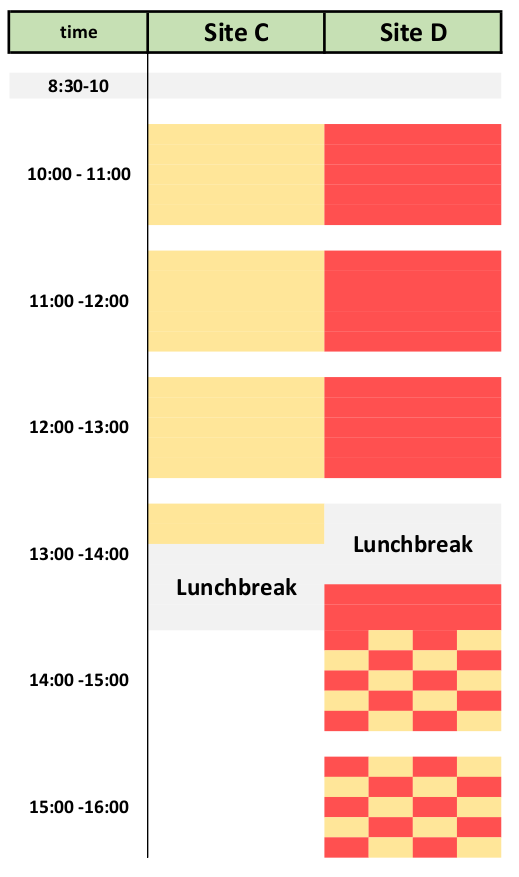}}
 	 \caption{a) Rough time schedule of days 1-3 showing which group (red, green, blue) attended which experimental site at each time slot. b) rough time schedule of day 4 colored by the group affiliation.}
  	\label{fig:timeschedule} 
\end{figure}

\noindent The experiments were performed in a  circuit training model. This means that three experimental setups were performed at the same time at three different sites, and participants were guided from one site to the other in designated groups. The three groups were marked with wristband colors: red, green or blue. The experimental sites were labeled alphabetically `B', `C' and `D' (Fig.\,\ref{fig:sites}). The experimental sites were separated by black curtains that shielded the view but were not sound proof. To limit the view on to the experiment, the waiting areas within the experimental sites were shielded by curtains as well. Each day (day 1-3) consisted of six experimental time slots lasting 1\,hour each, therefore participants attended each experimental site twice a day, but never participated twice in the same experimental setup, as those changed from one time slot to the next. A rough time schedule is shown in Fig.\,\ref{fig:timeschedule}.\\

\noindent The following experiments were performed in site B:
\begin{itemize}
\itemsep0em 
	\item Train Platform Experiment (day 1-3; \sref{subsec:TrainPlatformExperiments})
\end{itemize}
\noindent The following experiments were performed in site C:
\begin{itemize}
\itemsep0em 
	\item Crowd Management Experiment (day 1-3; \sref{subsec:CrowdManagementExperiments})
	\item Single-File Experiment (day 4; \sref{subsec:OvalExperiments})
	\item Personal Space Experiment (day 4; \sref{subsec:PersonalSpaceExperiments})
\end{itemize}
\noindent The following experiments were performed in site D:
\begin{itemize}
\itemsep0em 
	\item Boarding and Alighting Experiment (day 1-3; \sref{subsec:BoardingAndAlightingExperiments})
	\item Tiny Box Experiment (day 1-3; \sref{subsec:TinyHouseExperiments})
	\item Bottleneck Experiment (day 4; \sref{subsec:BottleneckExperiments})
\end{itemize}

\noindent Day 4 was different in that participants were divided into only two groups: Group yellow consisted of 80 people and group red of 120 people. Group red took part in the experiments at site D in all six time slots. Group yellow took part in the experiments at site C for the time slots\,1 to 3 and also came to site D for time slots\,4 to 6. 

\subsection{Recruiting Process}
Participants were recruited by spreading information via various channels including print\-ed and social media as well as e-mail lists of former experiments. The information included a short summary of the project, dates and payment. A QR code and link to a registration website was provided. The website included further information on conditions of participation, days available as well as a registration form. The conditions of participation (originally in German) included:
\begin{itemize}
\itemsep0em 
	\item minimum age of 18 years and recommended age of younger than 75 years
	\item body height of 1.5\,m to 2.0\,m
	\item not being affected by limited mobility or claustrophobia
	\item wearing dark clothes without lettering and not wearing large bags/backpacks
	\item agreeing to being filmed and the material to be published in a data repository 
\end{itemize}

\noindent After submitting the registration form, the potential participants received an e-mail that their registration had been received. People were assigned to days based on their statements of availability and evenly divided among the days if they were available for multiple days. People were only able to register for one of the three first days and additionally for the fourth day. \\
After allocation to the days, participants received an e-mail with allocation information. Two weeks prior to the experiments, a reminder was sent including current information on the hygiene and safety concept. The hygiene concept to protect against infection by Covid-19 was a necessary requirement of the authorities and institutions involved. One week prior to the experiments, participants received an e-mail with a reminder of their personal assigned dates and important things to remember and bring along with them (e.g. ID card, comfortable shoes, wear dark clothes). 

\subsection{Registration and Measurement Course}
\begin{figure}[h!]
	\centering
 	\subfigure[]{\includegraphics[width=0.48\textwidth]{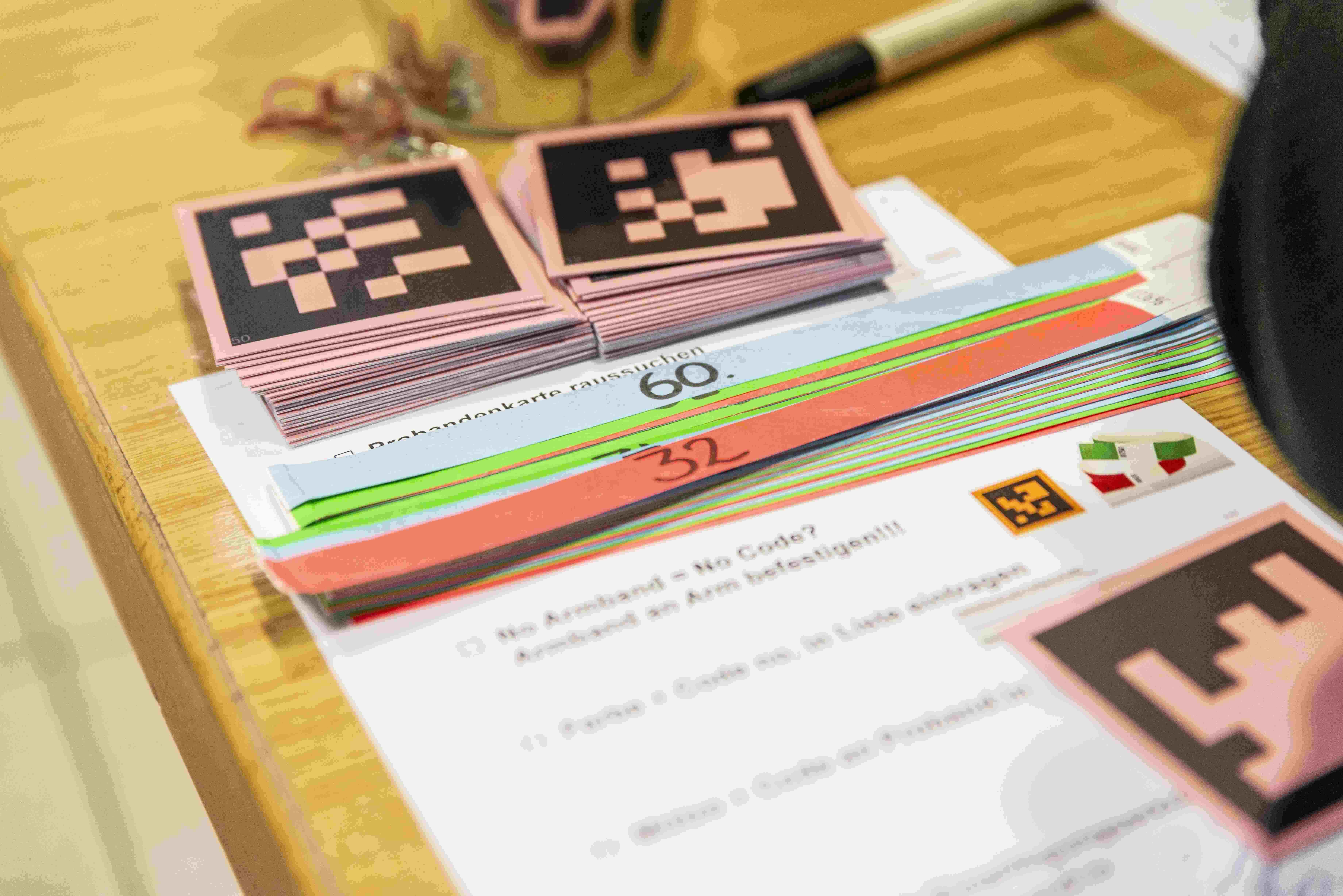}}
 	\subfigure[]{\includegraphics[width=0.48\textwidth]{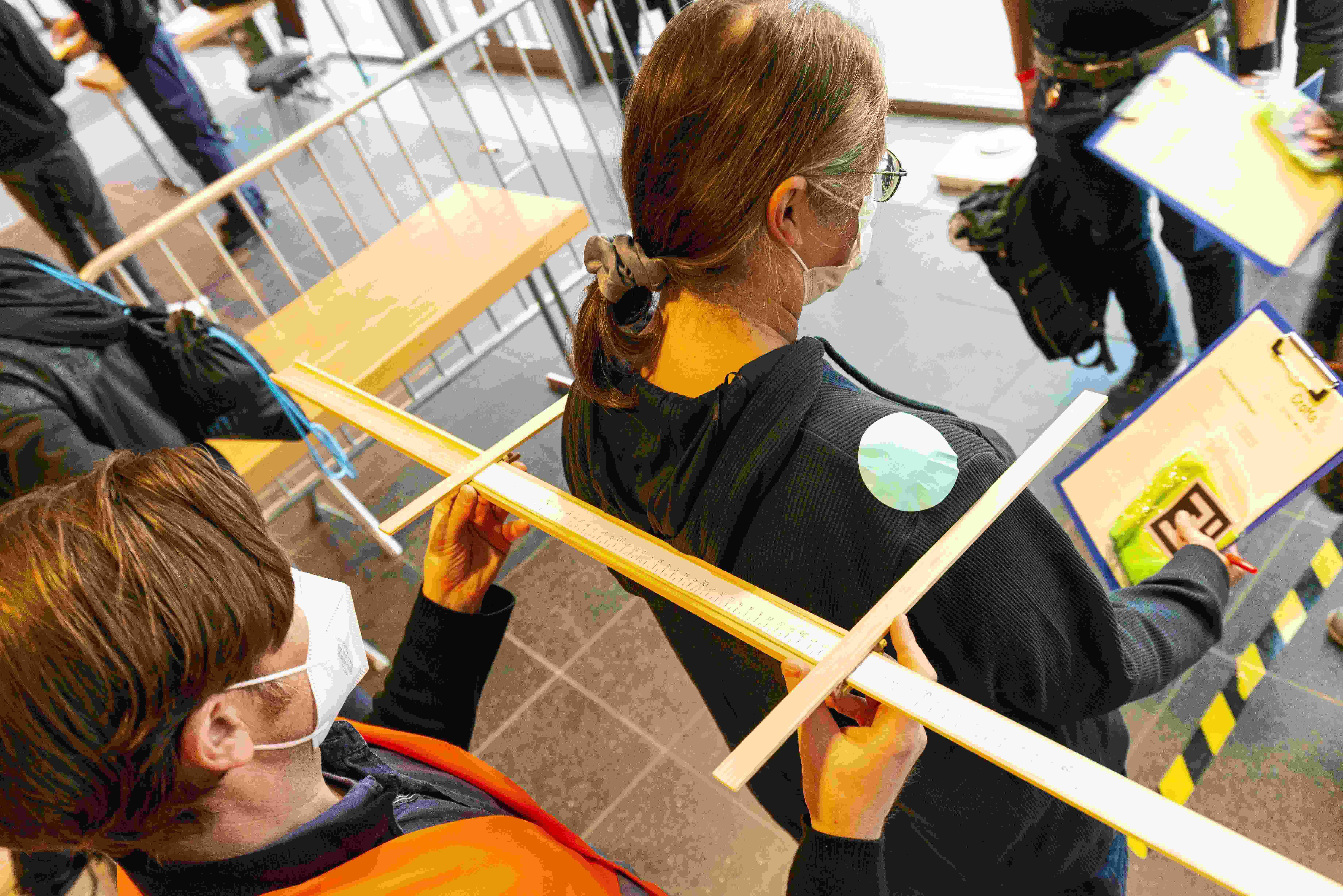}}
 	\subfigure[]{\includegraphics[width=0.3\textwidth]{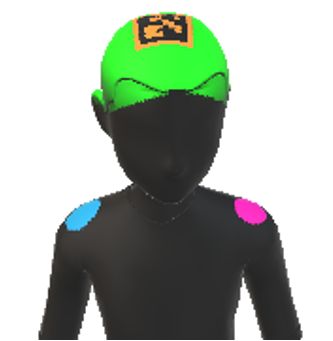}}
 	\subfigure[]{\includegraphics[width=0.48\textwidth]{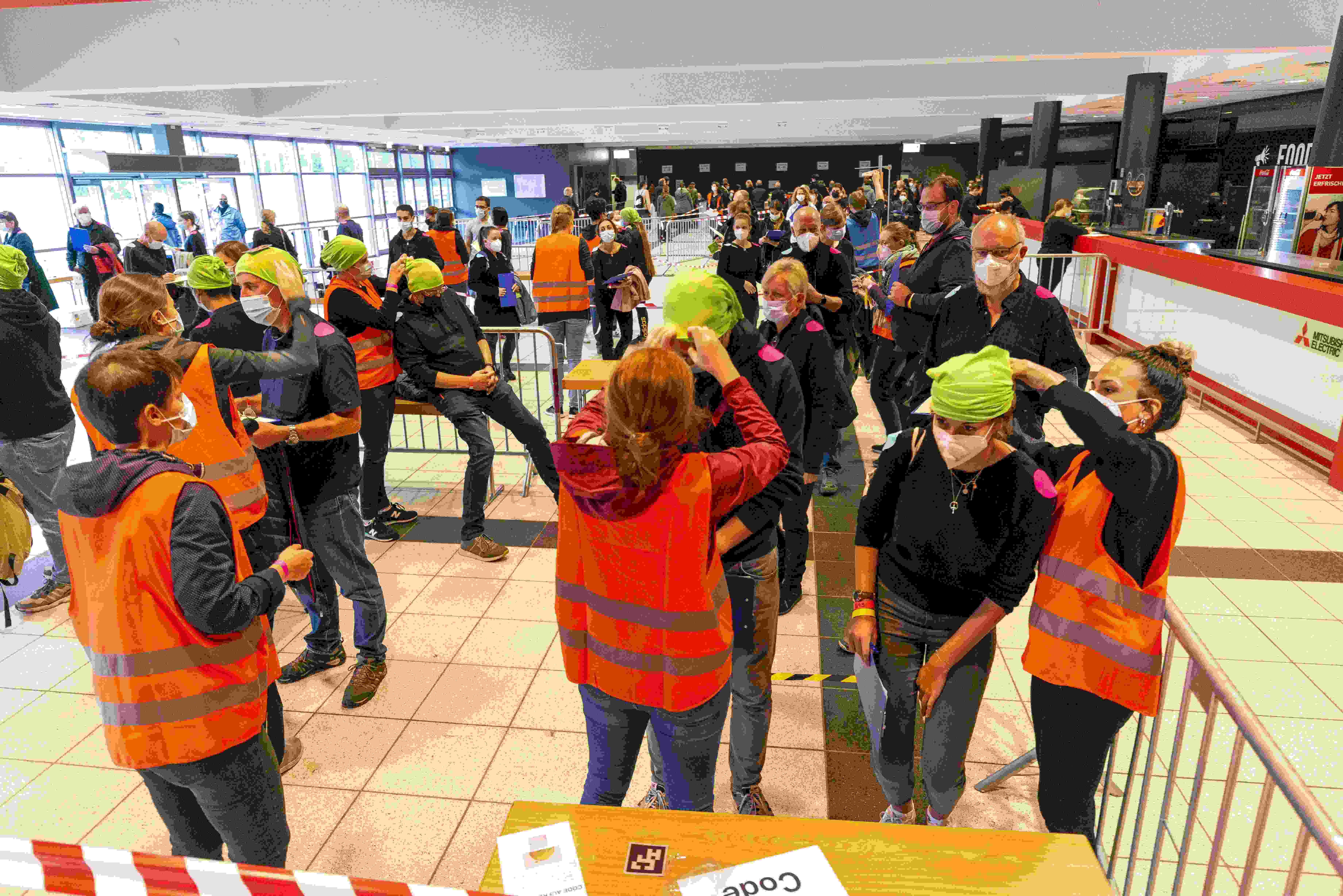}}
 	 \caption{Figure showing pictures taken in the measurement course: a) shows a picture of personal ID codes with matching colored wristbands labeled with the corresponding ID number that were handed out at the registration desks, b) picture showing helper measuring shoulder width at a station of the measurement course, c) schematic of accessories each participant was equipped with, namely a green hat, a personal ID code on top of the hat and markers on the shoulders (pink at left, blue at right shoulder), d) picture showing helpers checking correct fit of the hats and codes. }
  	\label{fig:registration} 
\end{figure}

At the time of arrival participants had to undergo a rapid Covid-19 test outside of the building where the experiments took place. The testing was performed by qualified hired personnel. The Covid rapid test stations were open starting from 8 a.m. People could only enter the building with a negative Covid-19 certificate. Registration started at 8:30 a.m.
After checking the Covid certificate and an identity document, people could enter the Mitsubishi Electric Hall and proceed to the registration desk in the main foyer. During registration the identity documents were checked again, participants signed forms that they consented to the conditions of participation, and they were handed a green hat, personal ID code (Aruco Code dict\_6X6\_1000 \cite{Garrido-Jurado2016}) and corresponding number on a wristband (Fig.\,\ref{fig:registration}\,a) as well as a clipboard with questionnaires and declaration of informed consent to be filled out. \\

\noindent The wristbands had three different colors (red, blue, green) and were handed out alternately. That way participants were divided into three experimental groups on arrival. Participants who arrived in social groups were therefore split among our experimental groups, although it cannot be completely ruled out that people who know each other were in the same experimental group. The wristbands were labeled with numbers (Fig.\,\ref{fig:registration}\,a) referring to the number associated with the personal ID code. The number was used for all questionnaires throughout the course of the day, to allocate sensor information and trajectories to a participant without revealing personal information. 

\noindent After registration participants entered a course that led them to a sequence of stations. At these stations, information was collected and subjects were provided with markers and utensils: 
\begin{itemize}
\itemsep0em 
	\item measuring height 
	\item measuring shoulder width (\fref{fig:registration}\,b)
	\item measuring weight
	\item checking color of top and handing out black long sleeve shirt if necessary
	\item applying shoulder markers to top (\fref{fig:registration}\,c)
	\item putting on green hat  and attaching personal code (\fref{fig:registration}\,c,\,d)
	\item checking correct fit of hat with code and telling people to leave the hat on for the entire day
	\item time for questions
	\item final check if declaration of informed consent was signed and questionnaires were filled out correctly 
	\item targeted addressing of suitable people to ask if they were willing to wear additional sensors (3D motion capturing suit, heart rate sensor)
\end{itemize}

\noindent After completing the measurement course participants could check their bags at a cloakroom and proceed to a large waiting area. 

\subsection{Test Person Sample}
\begin{figure}[h!]
	\centering
 	\includegraphics[width=\textwidth]{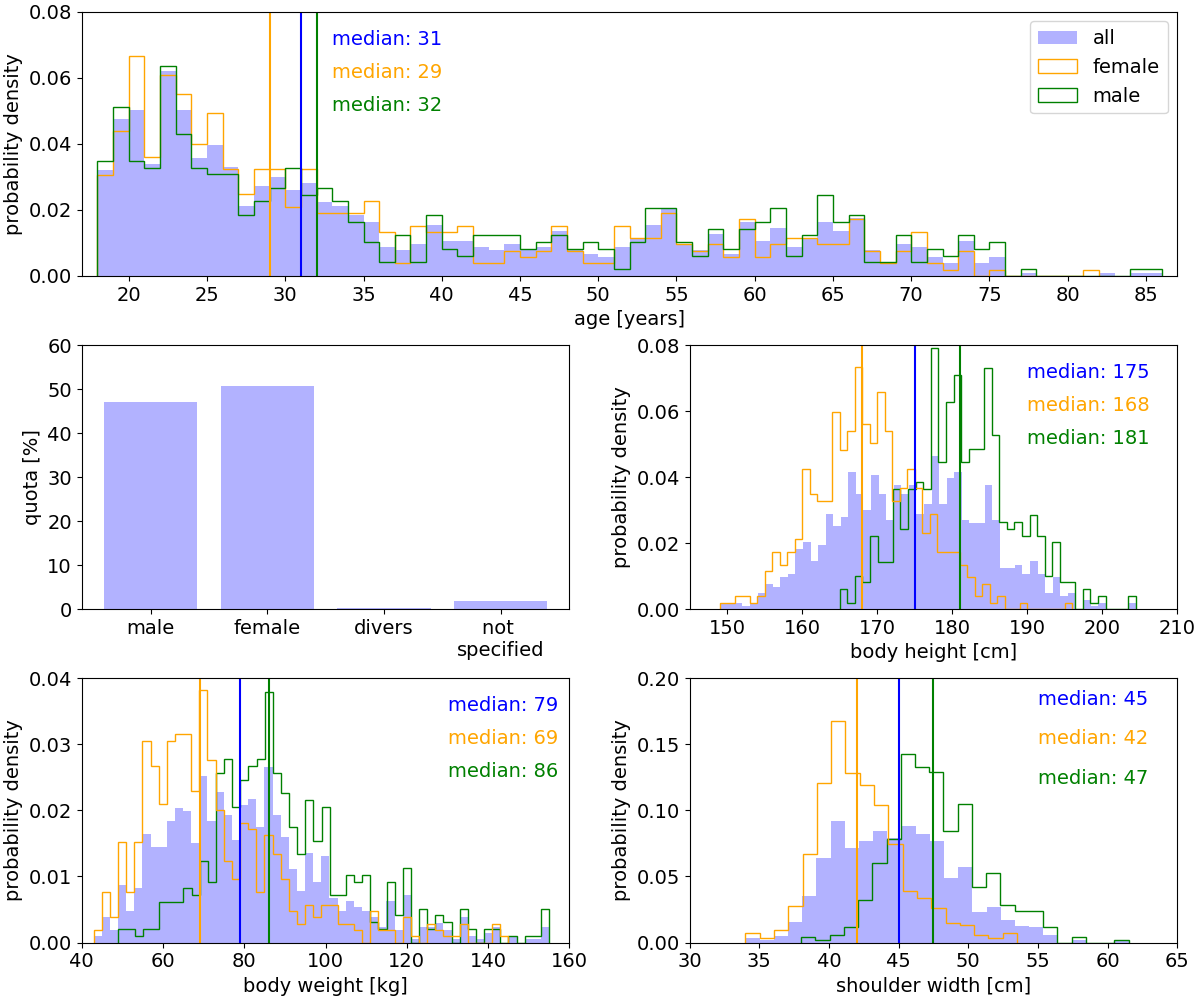}
 	 \caption{Figure showing histograms of different demographic factors of the participants for all four experimental days: age (top), gender (middle left), body height (middle right), body weight (bottom left) and shoulder width (bottom right). Data displayed in blue includes all data, data displayed in yellow includes data of all female participants and data displayed in green data of all male participants. Respective medians are shown in the same color as the data.}
  	\label{fig:testperson_sample} 
\end{figure}

To perform the experiments we accepted 1500 people for the duration of the four days. Of these, 1038 people attended the experiments. The sample of test persons included people from the age of 18 to 85 years (median=31,  $\sigma = \pm$17), with 47\,\% being male, 51\,\% female and 2\,\% not specified. Some of the distributions of the demographic data collected via questionnaire are shown in Figure\,\ref{fig:testperson_sample}. Data that had to be measured such as body height, body weight and shoulder width were collected by employees during the measurement course (\fref{fig:registration}\,b). On average, the participants were 1.75\,m ($\sigma = \pm$0.1) tall, weighed 79\,kg ($\sigma = \pm$21) and had a shoulder width of 45\,cm ($\sigma = \pm$4). Female participants were shorter, more lightweight and slimmer at the shoulders on average. Further personal data were collected via questionnaires. Differences in distributions for the different days and experimental groups can be found in the Appendix (\ref{app:testperson_stats}). More data can be found in the archive \cite{ForschungszentrumJulich2022}.


\subsection{Notes Related to Covid-19 Pandemic}
\label{subsec:corona}
\begin{figure}[h!]
	\centering
 	\includegraphics[width=0.8\textwidth]{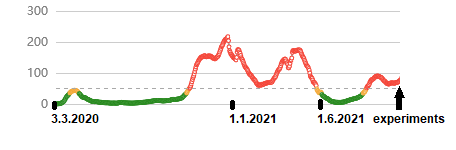}
 	 \caption{Course of 7 days incidence of Covid-19 pandemic for Germany. (\cite{RobertKochInstitut2021}, modified)}
  	\label{fig:corona_course} 
\end{figure}

At the time recruiting started as well as at the time of the experiments, Germany was at the beginning of a third Covid-19 wave (see \fref{fig:corona_course}). Of the enrolled people, 90\,\% declared that they had been fully vaccinated. \\
Due to the pandemic, a number of precautions were taken: 

\begin{itemize}
\itemsep0em 
	\item a hygiene and safety concept was developed by the team and approved by the crisis committee of Forschungszentrum Jülich and the competent regulatory authority of the city of Düsseldorf
	\item participants had to be recovered, vaccinated or tested (referred to as ``3G''  in Germany) 
	\item everyone was tested at the time of arrival and people were only allowed to enter the building with a negative test result
	\item participants had to wear surgical masks at all times (except when eating or drinking at a seat in the waiting area)
\end{itemize} 

\noindent In the first year of the Covid-19 pandemic, Germany's regulations prohibited people from getting together in large groups and required them to keep a distance of 1.5\,m between people not belonging to the same household. During the summer of 2021 these restrictions were dropped for vaccinated people. In public it could be observed that people kept wearing masks and kept their distance to other people to a large extent on a voluntary basis. \\

\noindent Because our experiments were designed for situations in which high densities may occur, we attempted to mitigate the behavioural changes described above. In order to get people accustomed to larger groups and higher densities again, we performed an `icebreaker' experiment. 
Participants were not informed about the icebreaker experiment, which was performed as part of the walkway to the first experiment. After registration people waited in a large waiting area (\fref{fig:sites}\,green shading). When the registration was finished up to 100 people were asked to walk to the first experiment at a time (based on the color of their wristband). A  person in charge, responsible for an experimental area, walked them into a corridor with two doors (\fref{fig:sites}\,blue shading). The person in charge asked people to wait until the last participant had arrived in the corridor. The rear door was closed when everyone had arrived (the density in the room was about 1\,P/m\textsuperscript{2}). Then the person in charge waited for another few minutes before releasing participants into another open space. The icebreaker was performed once every morning for each group.\\

\noindent To assess the extent to which participants were influenced in their actions by the thought of the pandemic, everyone was given a questionnaire about perceptions of various risks (focusing, in particular, on perceptions of risks of Covid-19 infection) and about the potential influence of the Covid pandemic on the experiments, at the end of each experimental day. The questions were answered on a 7-point scale from ``strongly disagree'' (1) to ``strongly agree'' (7). The questionnaire started with the general items about whether participants felt uncomfortable in the crowds during the experiments. Participants were then able to rate how much the following seven factors influenced their discomfort: Crowding, concern about contracting Covid, concern about contracting an other illness, unclear instructions, physical exertion, stress caused by the experimenters, being with many people. Two more questions directly addressed the mental engagement with Covid. In addition, participants could indicate in which setting (e.g., in the registration course) and in which type of experiment (e.g. bottleneck) they were most concerned about Covid with a yes/no answering format. In the two final questions, subjects estimated whether they would have behaved differently before the pandemic and indicated whether they had already been in a crowd before the experiments since the onset of the pandemic. N\,=\,1000 participants filled out the questionnaire on the four experimental days. Descriptive statistics are reported for the questionnaire data (\tref{tab:covidStatistics}). 

\noindent The results of the questionnaire suggest that the Covid pandemic did not have a major impact on the answers of the participants. The concern about infection was reported to be low (mean value M between 1.98 and 2.45). Subjects self-reported that their actions were not significantly different than before the pandemic (M\,=\,2.69). A self-selection effect certainly plays an important role for these results: Presumably only people who were not very concerned about Covid signed up for the experiments. This can also be seen in the last question. The statement that they have already been in a crowd elsewhere was often agreed with (M\,=\,4.02). Furthermore, these results reflect the extensive safety measures that apparently reduced the fear of contagion. In general, a low expression of discomfort was reported in the experiments (M\,=\,2.61). The factors that caused the most discomfort were crowding (M\,=\,3.37) and physical exertion (M\,=\,3.49). Subjects thought most strongly of Covid in the morning registration and measurement course (38\,\% answered with yes). We explain this by the fact that the registration came immediately after a mandatory Covid test. This meant that the topic of the pandemic was very present at that moment. On days 1-3 participants thought about Covid most often in experiment site\,C (26.2\,\%), followed by site\,D (16.1\,\%) and B (10.4\,\%). On day 4 participants thought about Covid quite a lot in the experiment site\,D (37.9\,\%) and very little in experiment site\,C (6.6\,\%).

\section{Configuration of Experiments}

\subsection{Train Platform Experiments}
\label{subsec:TrainPlatformExperiments}
In this experimental site two different experiments were performed. The first one investigated the waiting behaviour of people on a simulated train platform under varying physical or social psychological factors, the second one investigated types of social influence in ambiguous situations on train station platforms. 
\begin{figure}[h!]
	\centering
	\subfigure[]{\includegraphics[width=\textwidth]{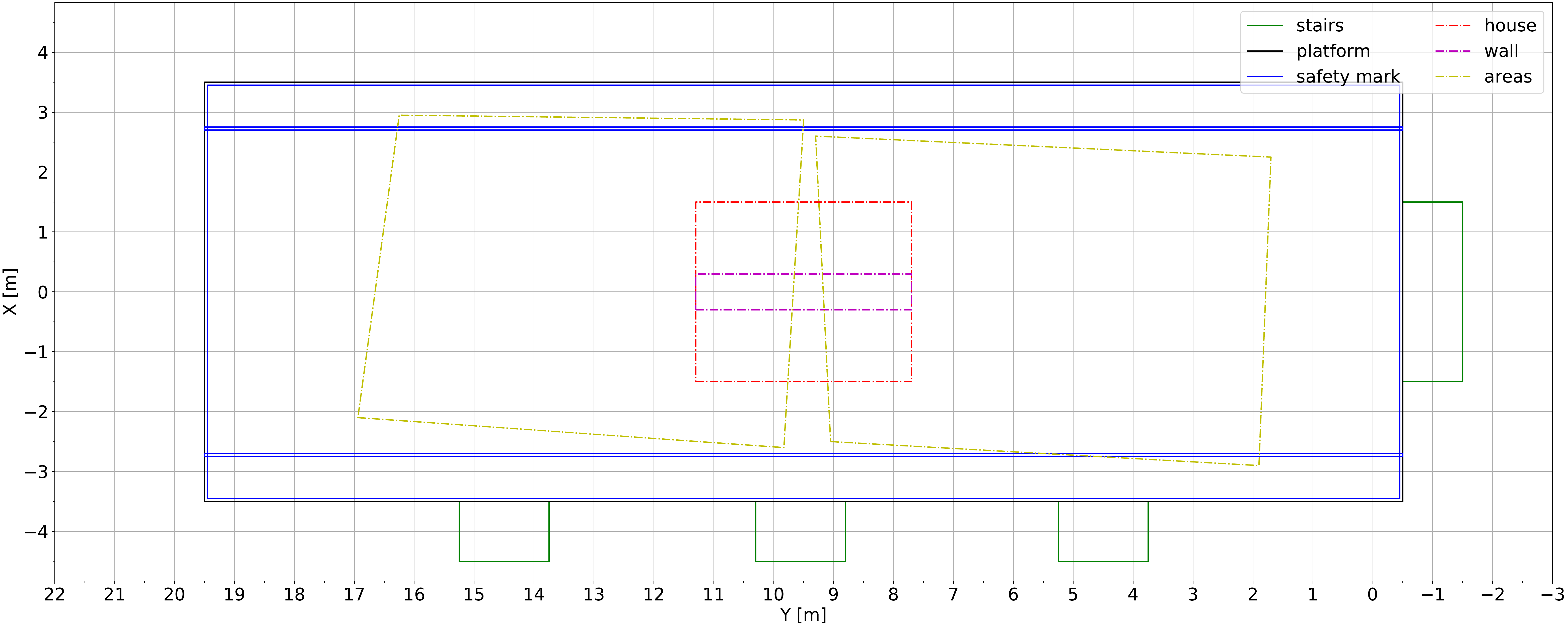}}
 	\subfigure[]{\includegraphics[width=0.48\textwidth]{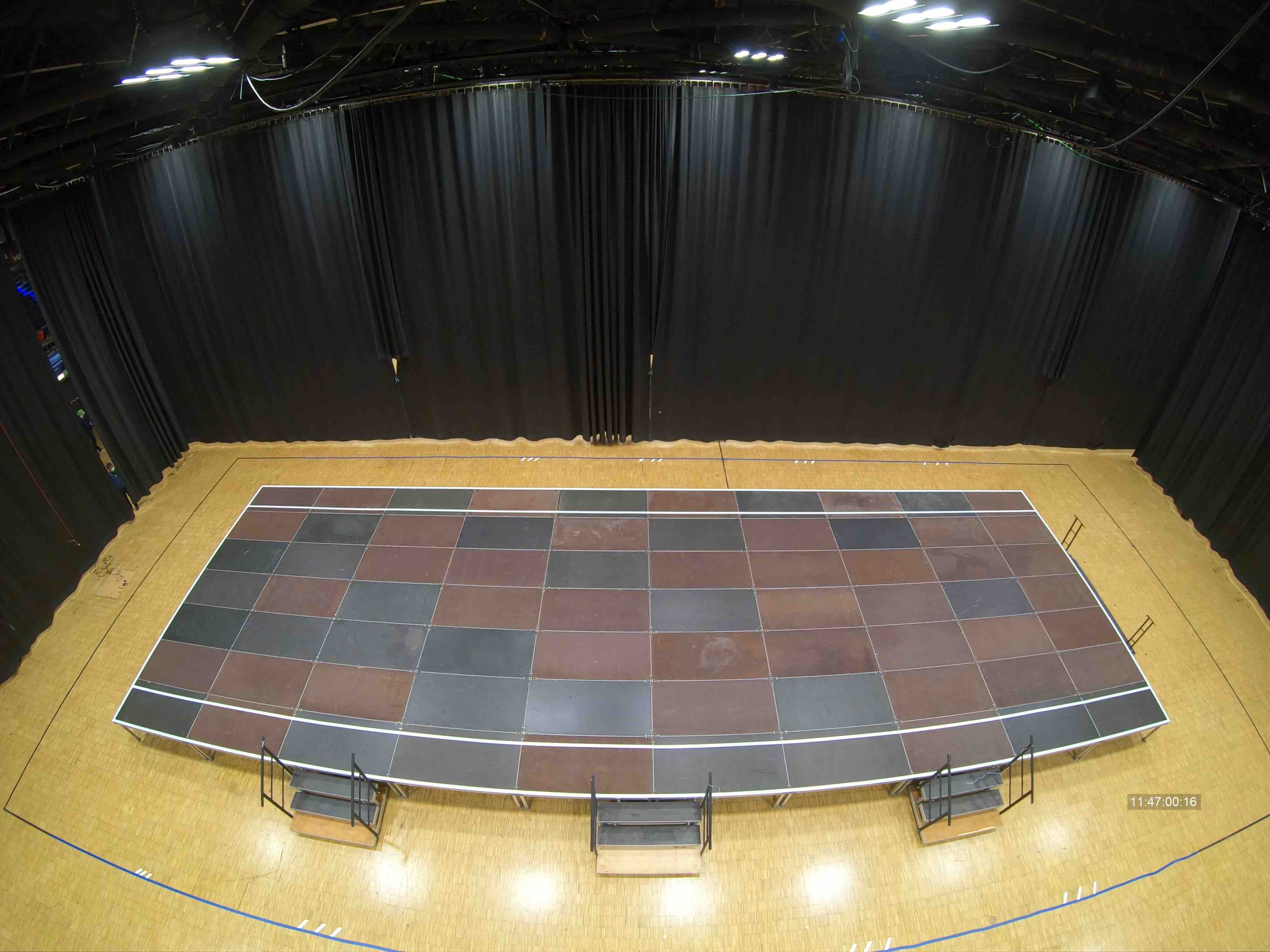}}
 	\subfigure[]{\includegraphics[width=0.48\textwidth]{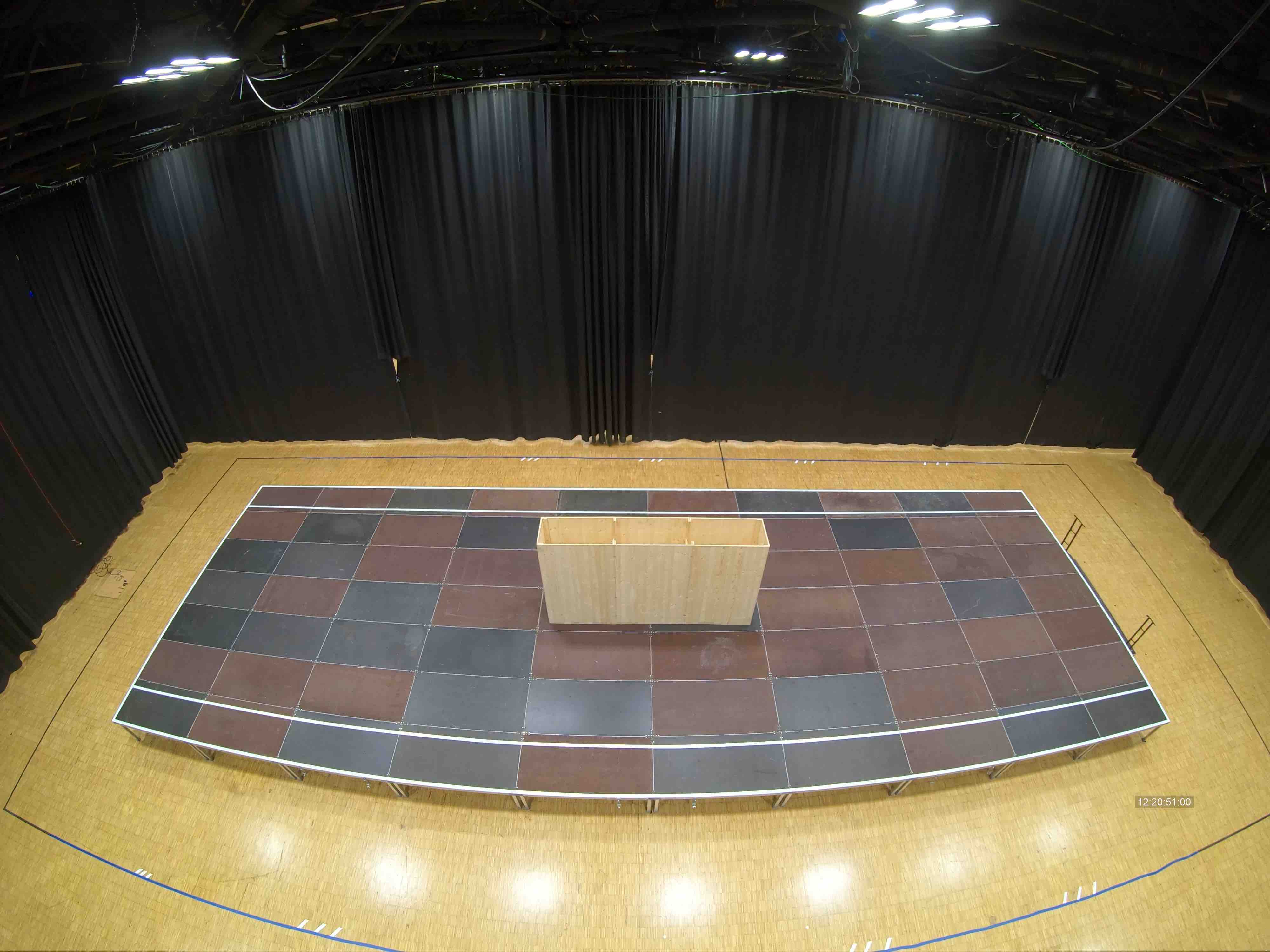}}
 	\subfigure[]{\includegraphics[width=0.48\textwidth]{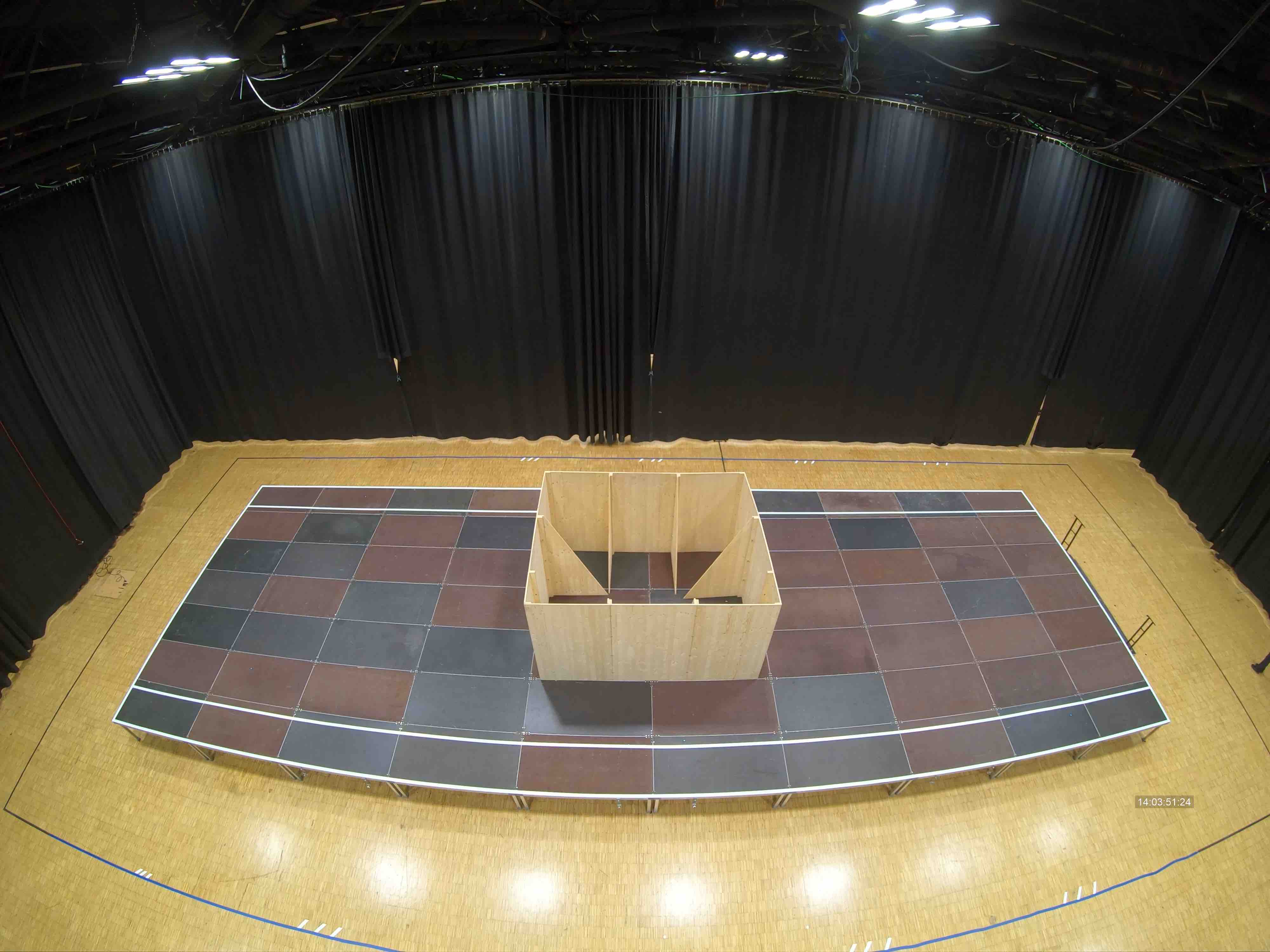}}
 	\subfigure[]{\includegraphics[width=0.48\textwidth]{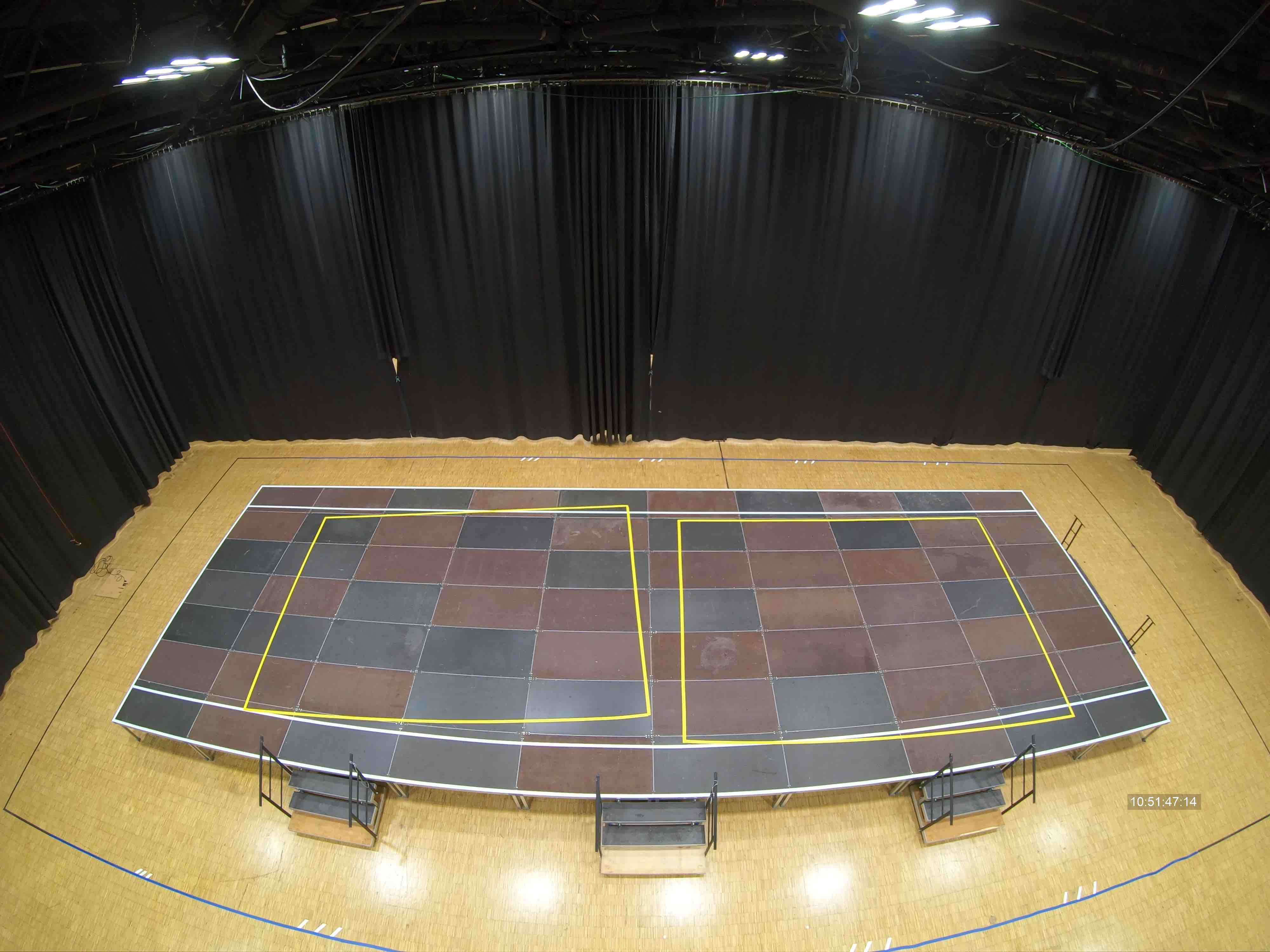}}
 	 \caption{a) Schematic showing coordinate system of simulated train platform with all obstacles as well as screenshot of setup of experimental runs b) without obstacles with stairs present, c) with a wall, d) with a house, e) with marked areas on the floor.}
  	\label{fig:setup_siteBd1-3} 
\end{figure}

\noindent The outer dimensions of the platform were 7\,m\,x\,20\,m\,x\,0.8\,m. The ascent and descent was realized by stairs secured with railings and organized in a way that resembled one primary access and three stairs to ``board the train" that only a few people could stand on at the same time. The stairs at the narrow side were 3\,m wide while the stairs at the long side of the platform were each 1.5\,m wide. The smaller stairs were movable and the positions for save attachment to the platform were only visible for helpers moving the stairs. 
The platform's edge was marked with adhesive tape (width 0.05\,m). White adhesive tape marks the safety distance (0.8\,m) from the edge of the platform, which is a standard for platforms in Germany.  A loudspeaker box with recordings of railroad sounds was placed under the platform during all of the experiments to reduce the influence of sound from the neighbouring experimental sites. \\

\noindent Cameras were mounted to record the experiment. They  are listed in Table\,\ref{tab:cameras_siteBd1-3}. 
Experimental runs in which 3D motion capturing data were recorded are listed in Table\,\ref{tab:mocap}. The mood of the participants (cf.\,\sref{subsec:moodbutton}) was recorded for all runs. 
Trajectories were generated as described in Section\,\ref{subsec:CameraAndTrajectories}. The coordinate origin's location was centred in relation to and 0.5\,m from the short platform border (Fig.\,\ref{fig:setup_siteBd1-3}a), with the y-axis pointing to the left, parallel to the longer edge and aligned with the midpoint of the large stairs.
The data of the Train Platform Experiments are provided online \cite{ForschungszentrumJulich2022a}.

\paragraph{Waiting Experiments} 

\noindent In this experiment instructions were given in the waiting area of the experimental site directly in front of the entrance to the experiment. The waiting area was separated from the experiment by a black curtain and thus the participants could not see the setup during the waiting phase and the instructions. In runs in which questionnaires were completed, this was done after the respective runs in a second waiting area at the opposite side. With the instructions, the participants were informed that the train they intend to board would arrive on the left hand side of the platform. The following parameters were varied (a detailed list of performed runs and combinations of parameters can be found in Appendix \ref{app:Trainplatform}): 
\begin{itemize}
\itemsep0em 
	\item number of participants: 40, 80-100, (140-180)
	\item obstacle on platform: none (blank), wall, house (\fref{fig:setup_siteBd1-3}\,b-d)
	\item inflow: every 2-3\,seconds, in groups of ten
	\item waiting time on platform: 2, 4\,minutes
\end{itemize}
The measurements of the wall were 0.6\,m\,x\,3.6\,m\,x\,2\,m and of the house 3\,m\,x\,3.6\,m\,x\,2\,m. Both were aligned symmetrically to the borders of the platform. 
In runs with 40 people, participants waited either 2 or 4\,minutes. The waiting time was counted from the moment the last participant entered the platform. Additionally, the inflow sequence to the platform was varied, for the experimental runs without obstacles. The participants entered individually or in groups of ten. The groups of ten entered the platform with an interval of 35\,seconds.
For the platform without an obstacle and the setup with the house, additional runs with an larger  number of participants were performed. In those runs the group of participants assigned to the corresponding experimental slot entered the platform first. After those participants had positioned themselves on the platform, participants from another group were brought in. Therewith the total number of passengers on the platform was unknown to all participants.

\paragraph{Experiments on Decision Making} 

\noindent For the experiment, the participants of each run were instructed directly in front of the platform. They entered the platform using the stairs on the long side. On the platform, two areas were marked, one slightly larger than the other. Participants were instructed to wait in the larger area but not instructed on how they should know which area was the larger one (because it was not easily visible). The curtains to the waiting area of the experimental site prevented the participants who did not take part in that specific run from observing the active participants of the run while carrying out the task. Questionnaires were completed for all runs both before entering and after leaving the area. 

The following parameters were varied (a detailed list of performed runs and combinations of parameters can be found in Appendix \ref{app:Trainplatform}): 
\begin{itemize}
\itemsep0em 
	\item number of participants: 10 (small groups), 23-41 (other groups)
	\item special design: 2 marked areas (\fref{fig:setup_siteBd1-3}\,e)
	\item inflow: all at once per group
	\item waiting time on platform: up to 5 minutes
	\item degree of familiarity of the participants with each other: no connection at all, short acquaintance before starting the experiment, being in the same group for hours before the experiment
	\item special announcements: talking allowed, talking prohibited
\end{itemize}
\noindent  The marked areas were placed on the platform asymmetrically (\fref{fig:setup_siteBd1-3} a) or e)), and had a size of 35 m\textsuperscript{2}  (left) and 36.7 m\textsuperscript{2}  (right).
 \subsection{Crowd Management Experiments}
\label{subsec:CrowdManagementExperiments}
\begin{figure}[h!]
	\centering
	\includegraphics[width=\textwidth]{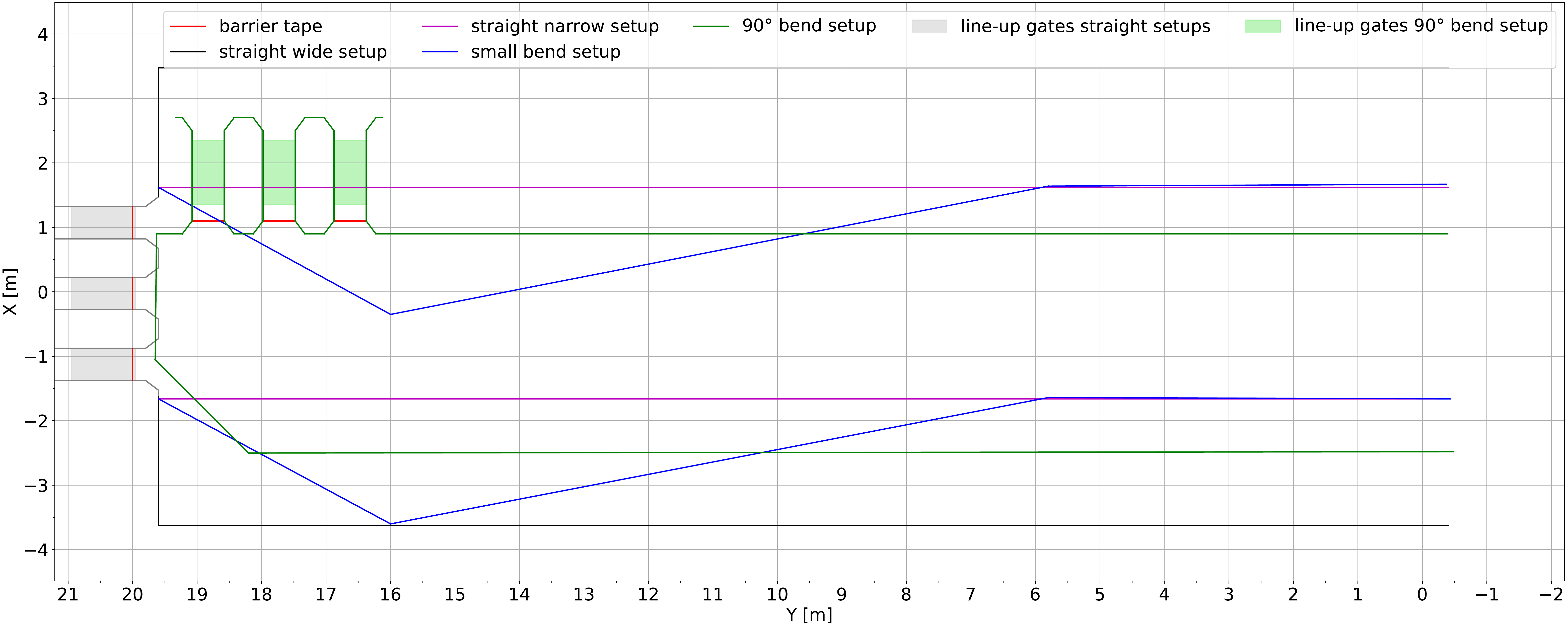}
	 \caption{Schematic showing coordinate system and setup of all Crowd Management Experiments performed on day 1-3 in one.}
	 \label{fig:setup_siteCd1-3_scetch} 
\end{figure}

\begin{figure}[ht!]
	\centering
 	\subfigure[]{\includegraphics[width=0.48\textwidth]{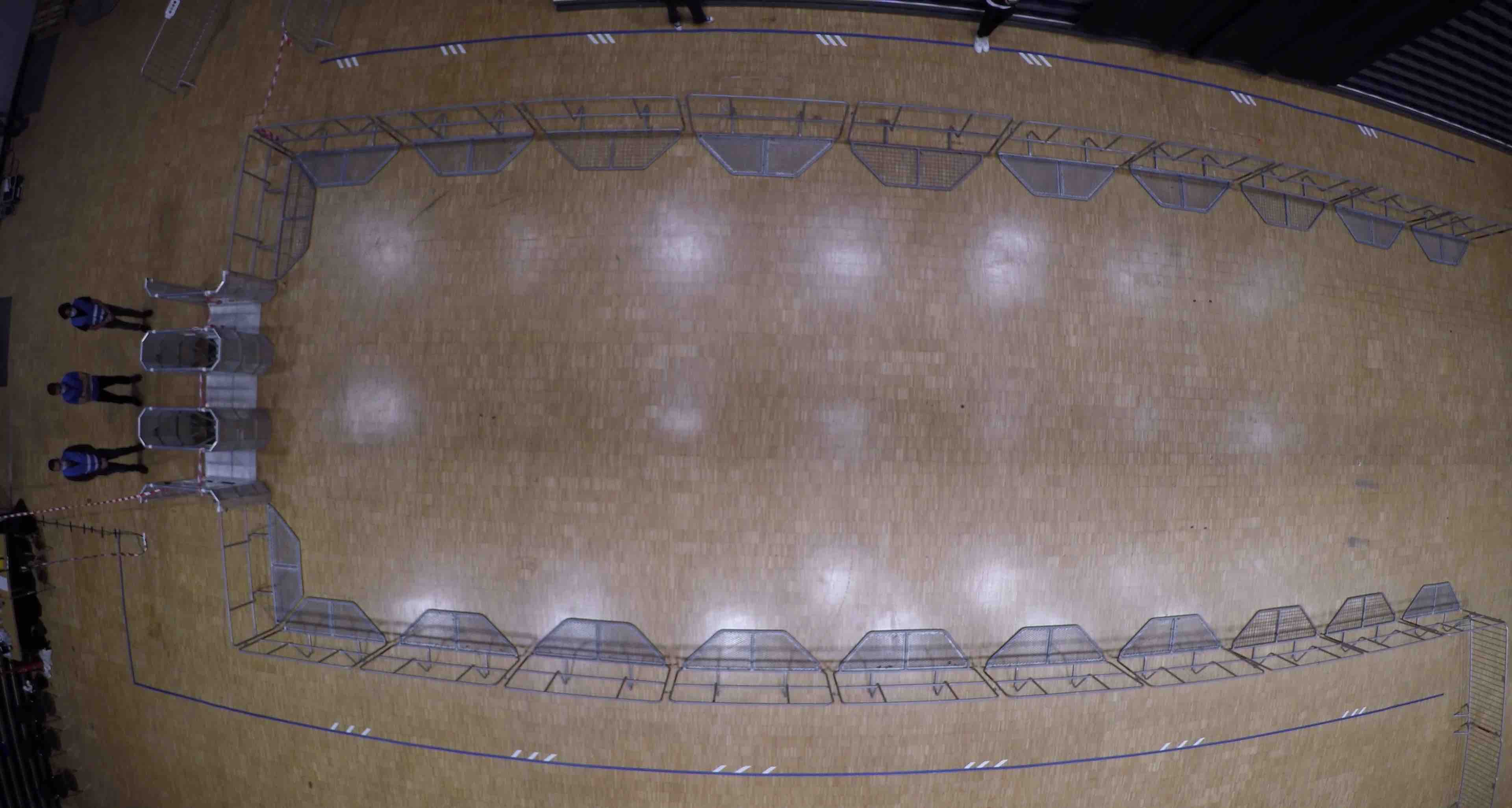}}
 	\subfigure[]{\includegraphics[width=0.48\textwidth]{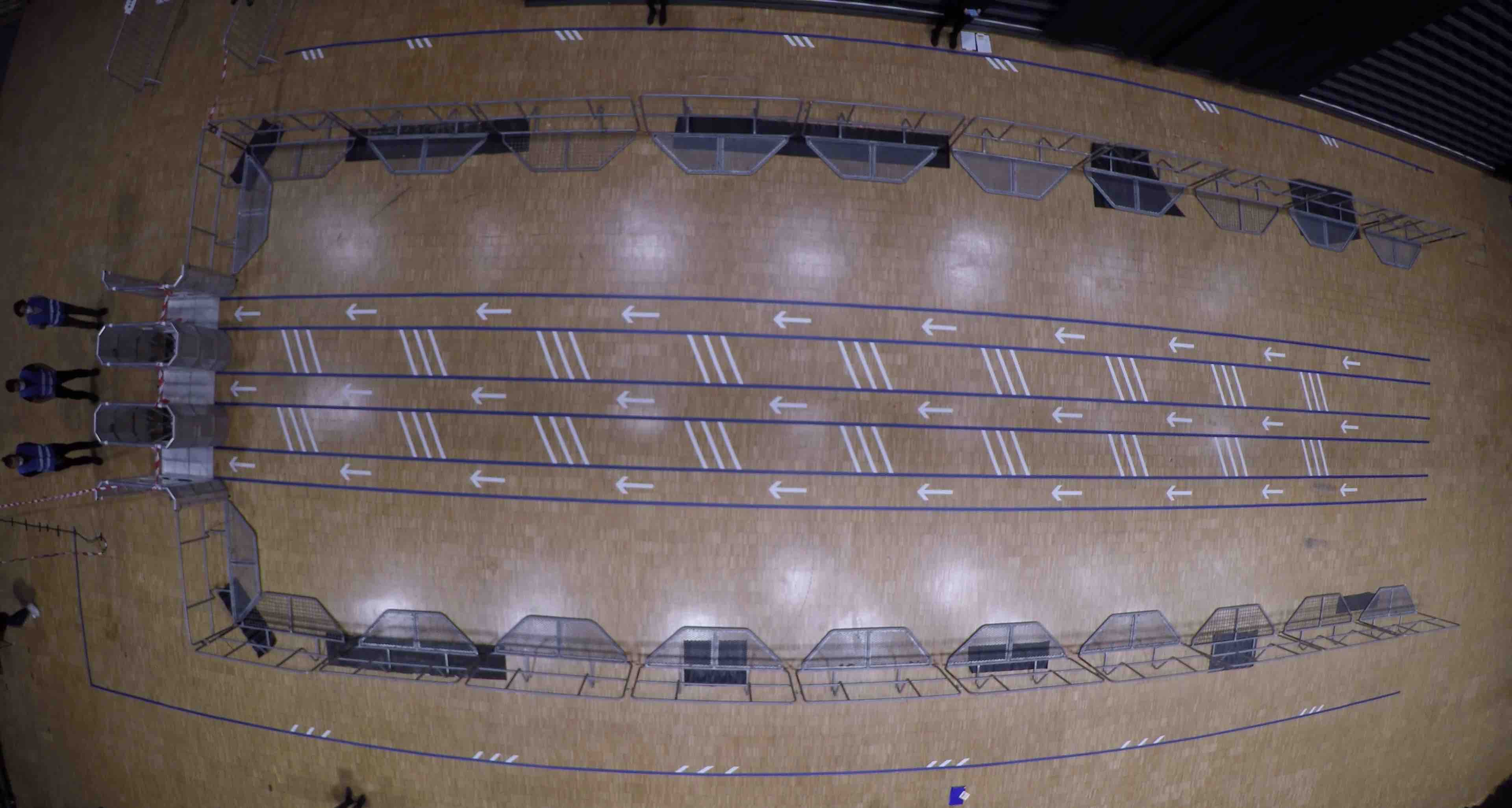}}
 	\subfigure[]{\includegraphics[width=0.48\textwidth]{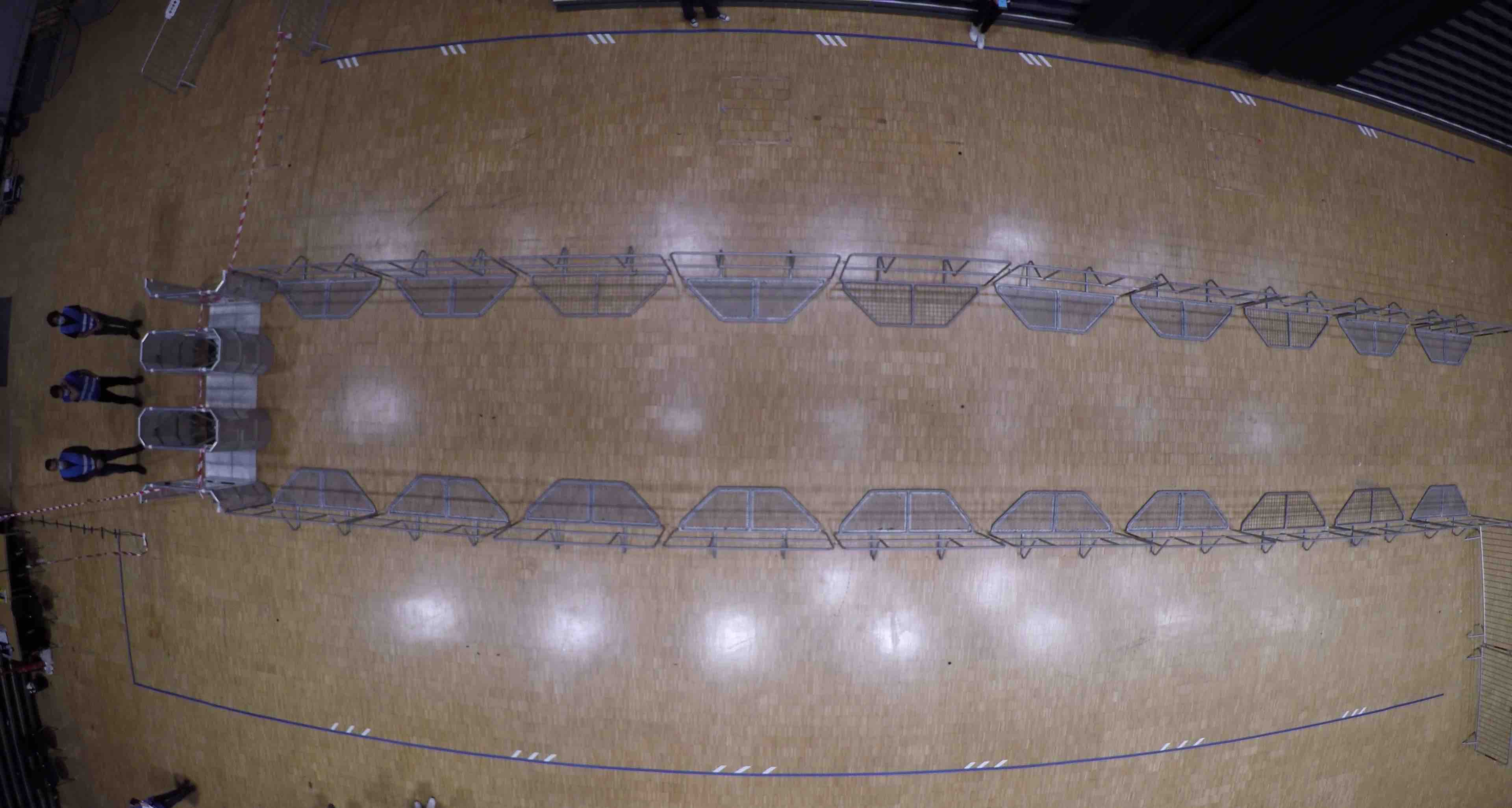}}
 	\subfigure[]{\includegraphics[width=0.48\textwidth]{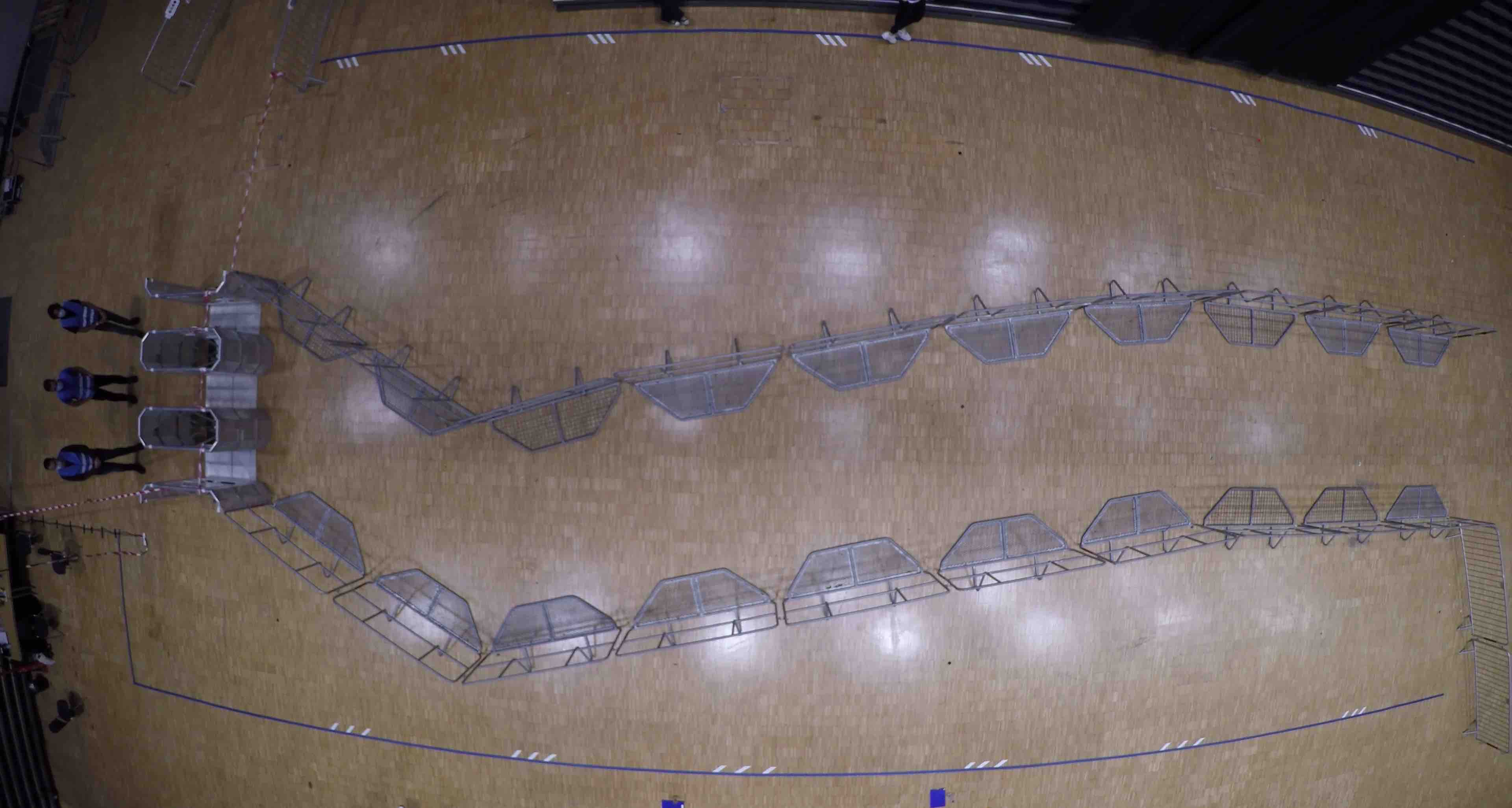}}
 	\subfigure[]{\includegraphics[width=0.48\textwidth]{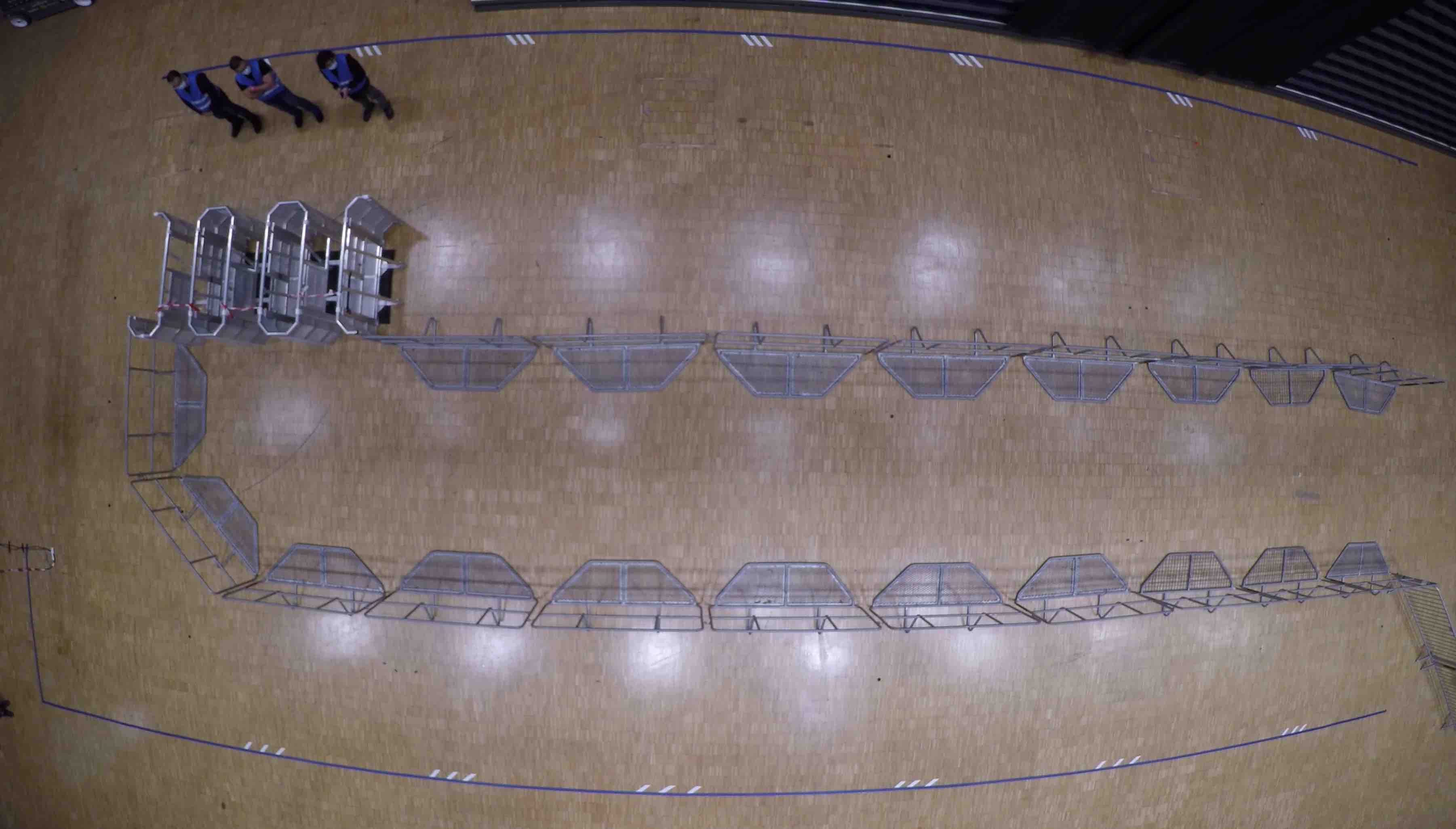}}
 	\subfigure[]{\includegraphics[width=0.48\textwidth]{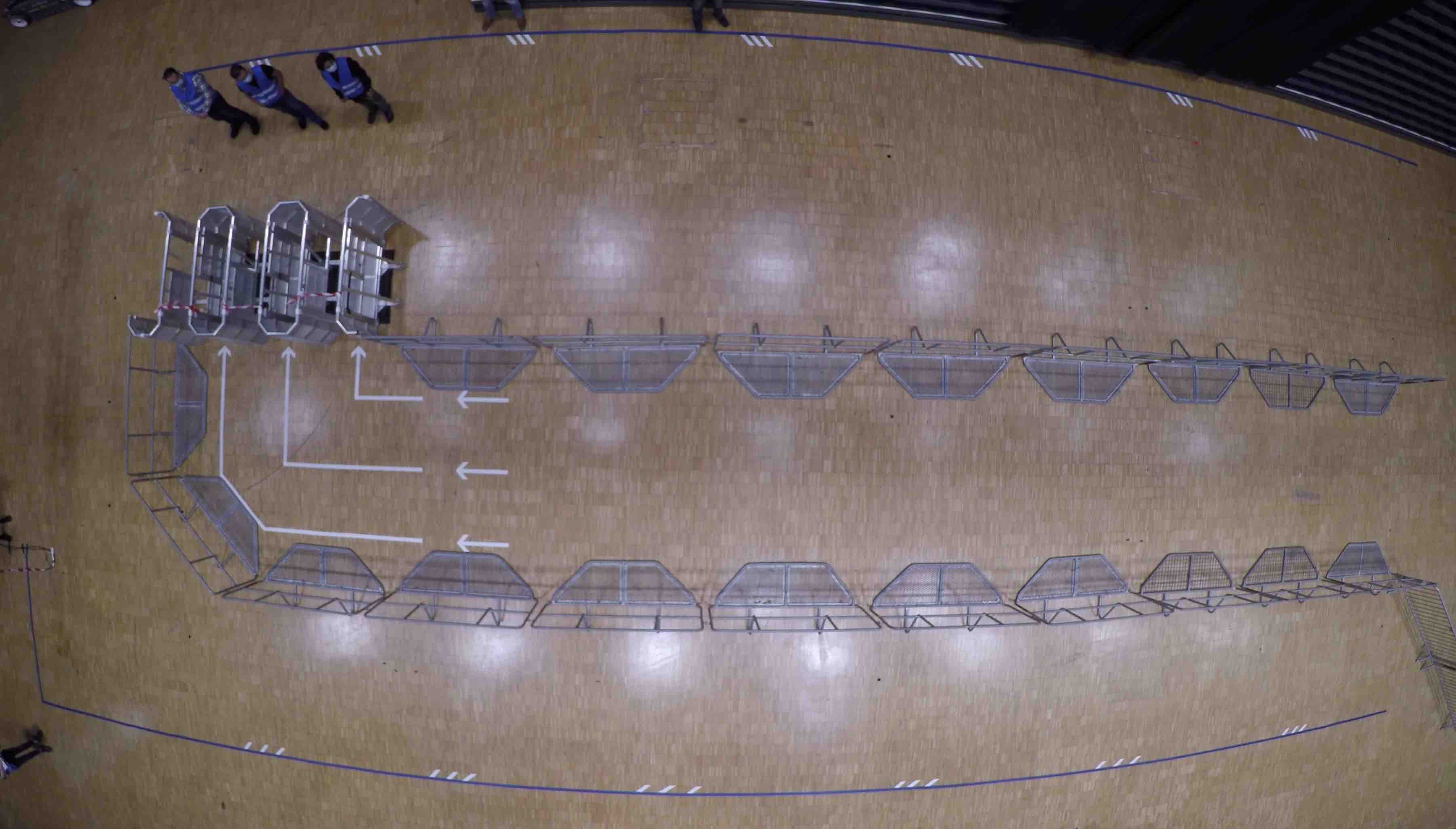}}
 	 \caption{a) - f) Snapshots of overview cameras showing a) setup straight wide, b) setup straight wide with lines, c) setup straight narrow, d) setup small bend, e) setup 90\degree\, bend and f) setup 90\degree\,  bend with lines.}
  	\label{fig:setup_siteCd1-3_snapshots} 
\end{figure}

This series of experiments investigated the extent to which physical parameters such as the number of line-up gates or the width and the shape of the barrier layouts influence the formation and the density of a queue. For this purpose, an admission situation under the assumption of ``admission to the concert of your favorite artist''  was simulated using barriers and line-up gates typical to those used at large events. Furthermore, non-physical parameters were considered. 
The following parameters were varied (a detailed list of performed runs and combinations of parameters can be found in Appendix \ref{app:CrowdManagement}): 
\begin{itemize}
\itemsep0em 
	\item setup structure grid (narrow): straight, small bend, 90\degree -right bend with and without lines (floor markings) (\fref{fig:setup_siteCd1-3_scetch}, \ref{fig:setup_siteCd1-3_snapshots}\,c-)
	\item setup structure grid (wide): none, lines, signs (above the line-up gates construction) (\fref{fig:setup_siteCd1-3_scetch}, \ref{fig:setup_siteCd1-3_snapshots}\,a-b)
	\item no. of open line-up gates: 1, 3
	\item motivation: low (enough time and guaranteed seat ticket), high (the general admission ticket)
	\item norm specification: none, 70\%, 85\%, 95\%, 100\%
	\item special announcements: no interruption (nI), with interruption (wI) (along with HRV sensors (\sref{subsec:HeartRateVariability}))
	\item reference runs to record free walking speed (solo\_ref)
\end{itemize}


\noindent Instructions according to the motivation and the run were given directly in front of the entrance to the experiment before participants could see the experimental setup. At the time of the instructions, participants were separated from the experiment by a black curtain. If a norm specification was given, every participant was given a slip of paper with a note on it saying either ``In the following experiment, behave as you always would'' or ``Imagine that you are a very selfish person. Push yourself to the front during the experiment''. The percentage refers to the amount of paper notes prescribing normal behaviour. After each run, participants were directed back to the area where the instructions were given and questionnaires distributed. HRV Sensors were collected (if given) after each group had finished their respective runs.
Within the framework of the experiments, a comparison of the estimated and the physically measured density of people was carried out. Some assessors had been previously trained to the Level of Service and had been given further knowledge of density estimation, while others were untrained. Time-dependent densities ranging from low to medium to high were estimated \cite{FeuerwehrDuesseldorf2020} and documented. In addition, positive and negative factors influencing the density estimation of the observers were surveyed using questionnaires. 

\noindent The outer dimensions of the experimental area were 7\,m\,x\,20\,m. The line-up gates construction was 2.5\,m\,x\,3.3\,m x\,1.18\,m (length\,x width\,x height) with a passage width of 0.5\,m each. The line-up gate construction at the 90\degree bend setup was 30 cm wider. Police barriers with dimensions of 2.0\,m\,x\,0.94\,m\,x\,1.1\,m were used to set up the structure grid. \\

\noindent Cameras were mounted to record the experiment and are listed in Table\,\ref{tab:cameras_siteCd1-3}. 
Experimental runs in which 3D motion capturing data were recorded are listed in Table\,\ref{tab:mocap} and runs in which HRV data were recorded in Table\,\ref{tab:hrv}. 
The mood of the participants (cf.\sref{subsec:moodbutton}) was recorded for all runs. 
Trajectories were generated as described in Section\,\ref{subsec:CameraAndTrajectories}. The coordinate origin was located where participants enter the experimental area, with the y-axis pointing in walking direction and aligned with the midpoint of the middle entry gate (\fref{fig:setup_siteCd1-3_scetch}). The data of the Crowd Management Experiments are provided online \cite{ForschungszentrumJulich2022c}.
\subsection{Single-File Experiments}
\label{subsec:OvalExperiments}

This series of experiments investigated how walking speed and density affect physiological arousal. For this purpose, subjects were equipped with electrodermal activity (EDA) and heart rate variability (HRV) sensors (cf. Sec.\,\ref{subsec:Electrodermal-Activity}, \ref{subsec:HeartRateVariability}). Additionally the effect of gender on walking speed and density was investigated. The experiments were performed in the setup of classical single-file experiments, where people walked in ovals behind each other. Overtaking was prohibited.
The following parameters were varied (a detailed list of performed runs and combinations of parameters can be found in Appendix \ref{app:Oval}): 
\begin{figure}[b]
	\centering
 	\subfigure[]{\includegraphics[width=0.46\textwidth]{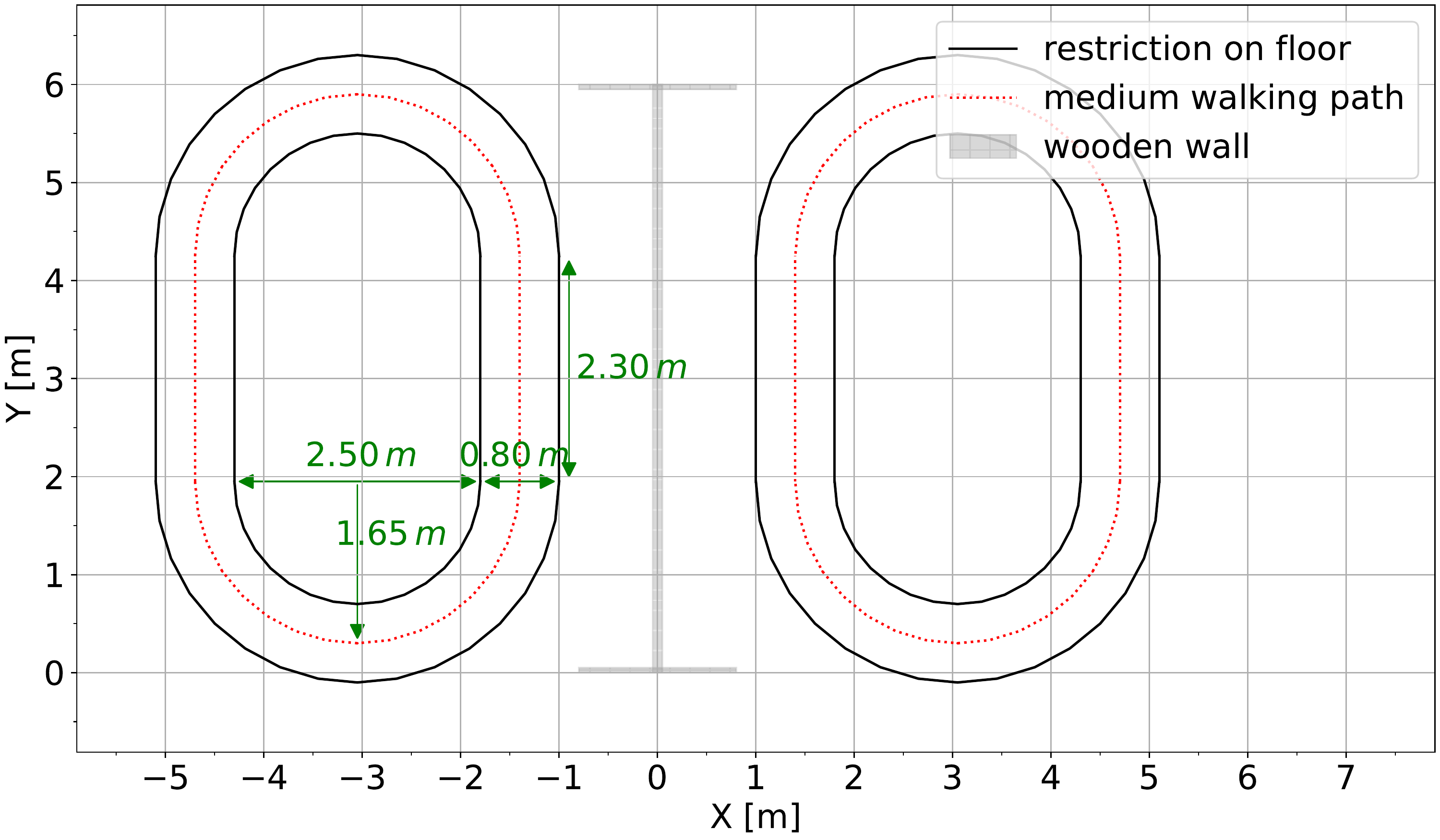}}
 	\subfigure[]{\includegraphics[width=0.48\textwidth]{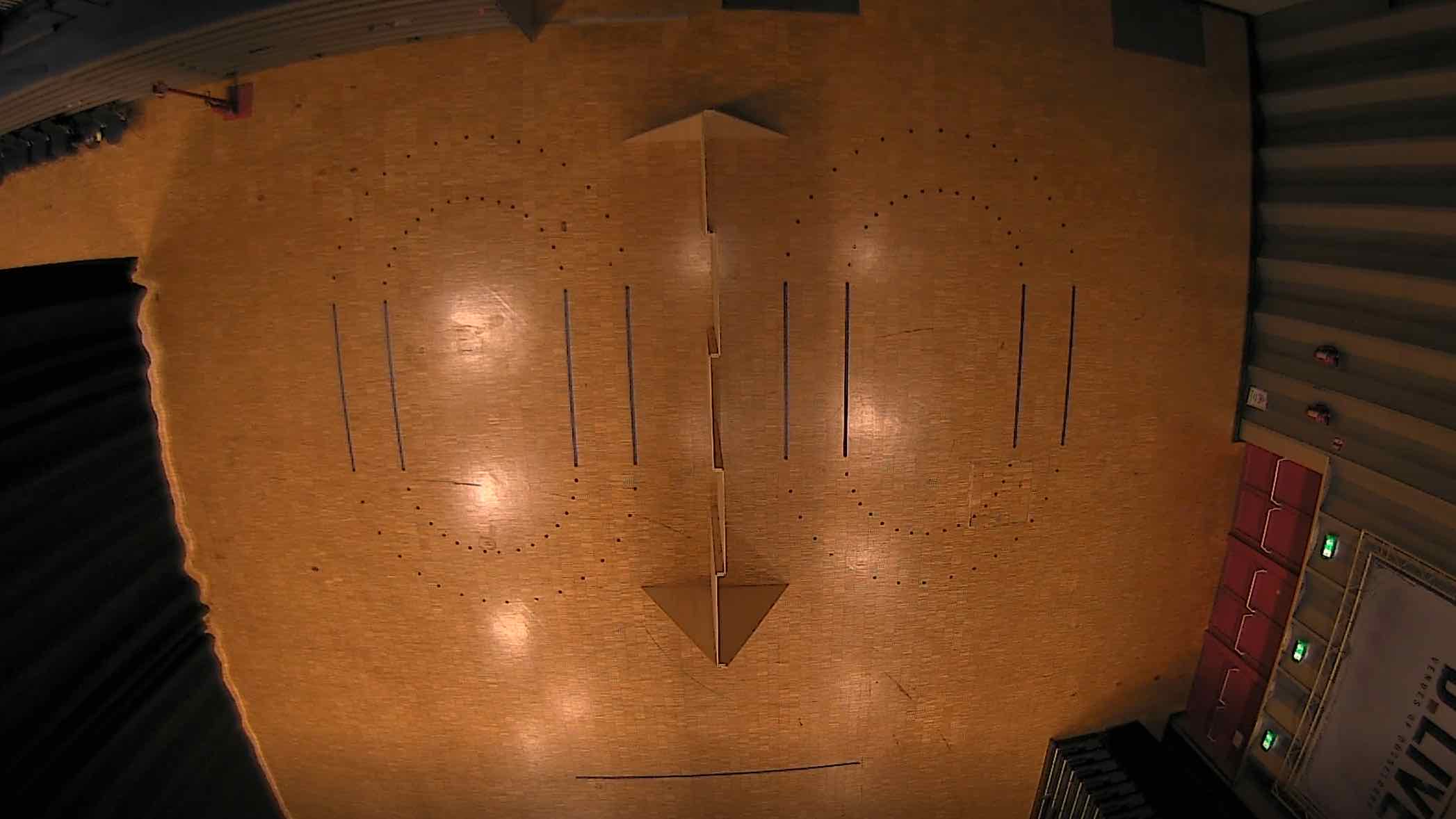}}
 	 \caption{a) Schematic showing the dimensions of the oval. b) Screenshot from Camera showing the setup of two ovals separated by a wooden wall.}
  	\label{fig:setup_siteCd4} 
\end{figure}
\begin{itemize}
\itemsep0em 
	\item number of participants
	\item gender: male, female, mixed
	\item running order by gender: random, alternating
\end{itemize}

\noindent Instructions to start or to stop were given by a person standing between the two experimental setups without using technical amplification. Participants walked in the oval for at least 2\,minutes or until they had walked one round at very high densities. The width of the ovals walking paths was 0.8\,m with a circumference of 14.97\,m as measured from the middle of the indicated walking width (Fig.\,\ref{fig:setup_siteCd4}a). The course was indicated by colored markers on the floor. Two oval experiments were performed at the same time. They were separated by a wooden wall (Fig.\,\ref{fig:setup_siteCd4}b). \\

\noindent Cameras were mounted to record the experiment and are listed in Table\,\ref{tab:cameras_siteCd4_oval}. Experimental runs in which EDA and HRV sensors were recorded are listed in Tables\,\ref{tab:eda} and \ref{tab:hrv}. Trajectories were generated as described in Section\,\ref{subsec:CameraAndTrajectories}. The coordinate origin was located at the lower side of the two ovals' in the axis of the screen wall (Fig.\,\ref{fig:setup_siteCd4}a). The data of the Single-File Experiments are provided online \cite{ForschungszentrumJulich2022d}.
 \subsection{Personal-Space Experiments}
\label{subsec:PersonalSpaceExperiments}
This experimental series investigated physiological arousal when personal space is violated at low densities. Seven participants were positioned within an area marked on the floor and then passed by ten other participants (individually or several simultaneously) from all directions without being touched. All participants assigned to be standing in the designated spots were equipped with electrodermal activity (EDA) and heart rate variability (HRV) sensors (cf. Sec.\,\ref{subsec:Electrodermal-Activity}, \ref{subsec:HeartRateVariability}).

\begin{figure}[h!]
	\centering
	\subfigure[]{\includegraphics[width=0.46\textwidth]{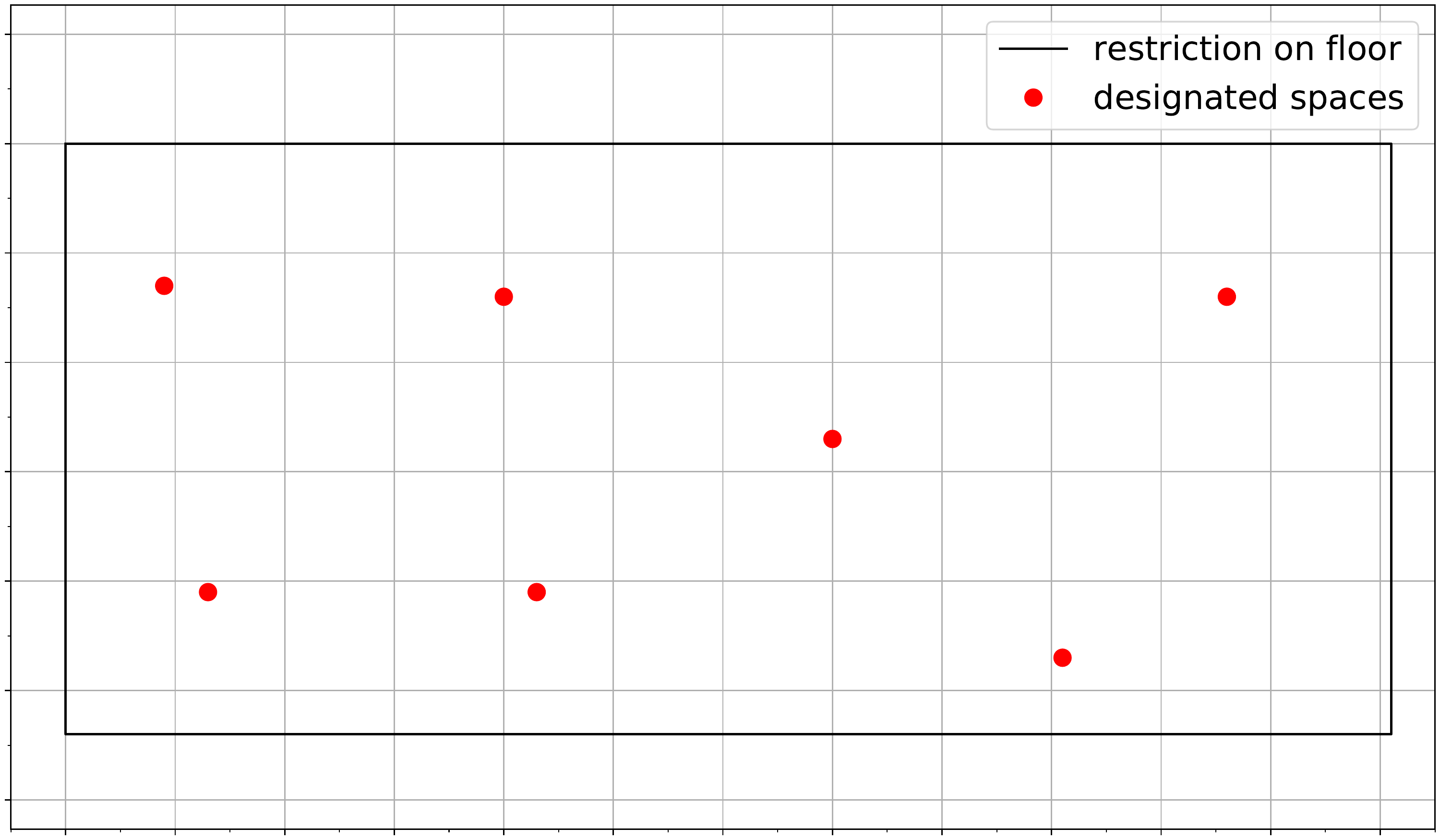}}
 	\subfigure[]{\includegraphics[width=0.48\textwidth]{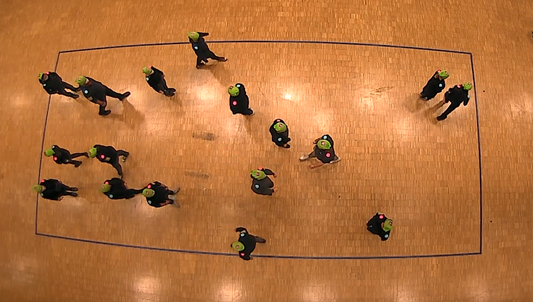}}
 	 \caption{a) Schematic showing the dimensions of the experimental area on a 1\,m\,x\,1\,m grid with positions of standing participants. b) Snapshot of experiment with standing and passing participants.}
  	\label{fig:setup_personalSpace} 
\end{figure}

\noindent Instructions were given without using technical amplification as the groups were small. Eight runs of four minutes each were performed in total (a list of performed runs can be found in Appendix \ref{app:PSpace}). 

\noindent The dimensions of the experimental area were 12.1\,m\,x\,5.3\,m. The experiment was performed next to the Oval Experiments outlined in the previous subsection. There was no visual shielding between the two experiments and it was possible for participants to pass unhindered between the two experimental sites. Participants with EDA and HRV sensors were placed at positions indicated as red dots in Figure\,\ref{fig:setup_personalSpace}a. Passing participants were participants who were not currently on runs in the neighbouring Oval Experiment. Questionnaires were completed at the end of the whole experiment set including the runs of the Oval Experiment.\\

\noindent Cameras were mounted to record the experiment and are listed in Table\,\ref{tab:cameras_siteCd4_pspace}. Experimental runs in which EDA and HRV sensors were recorded are listed in Tables\,\ref{tab:eda} and \ref{tab:hrv}. No trajectories were exported for this experiment. The data of the Personal Space Experiments is provided online \cite{ForschungszentrumJulich2022e}.

 \subsection{Boarding and Alighting Experiments}
\label{subsec:BoardingAndAlightingExperiments}
This series of experiments investigated how different parameters influence the boarding and alighting process of a train car. For this purpose, the boarding area of a local train was mimicked. Sliding doors could be opened from outside the experimental area via ropes without interfering with participants. Different parameters were varied (a detailed list of performed runs and combinations of parameters can be found in Appendix \ref{app:BoardingAlighting}): 
\begin{itemize}
\itemsep0em 
	\item number of persons boarding/alighting/staying in train car
	\item luggage: none, backpack, suitcase, baby stroller, mix
	\item cooperation: none, work together for fast boarding and alighting
	\item motivation: normal, hurry, pushing, distracted by using phone
	\item line-up: none, throng, corridor, 45\degree
	\item groups: none, pairs, groups of 3, groups of 5, mix
	\item norm: none, 70\%, 80\%, 100\%
\end{itemize}
\begin{figure}[t!]
	\centering
	\subfigure[]{\includegraphics[width=\textwidth]{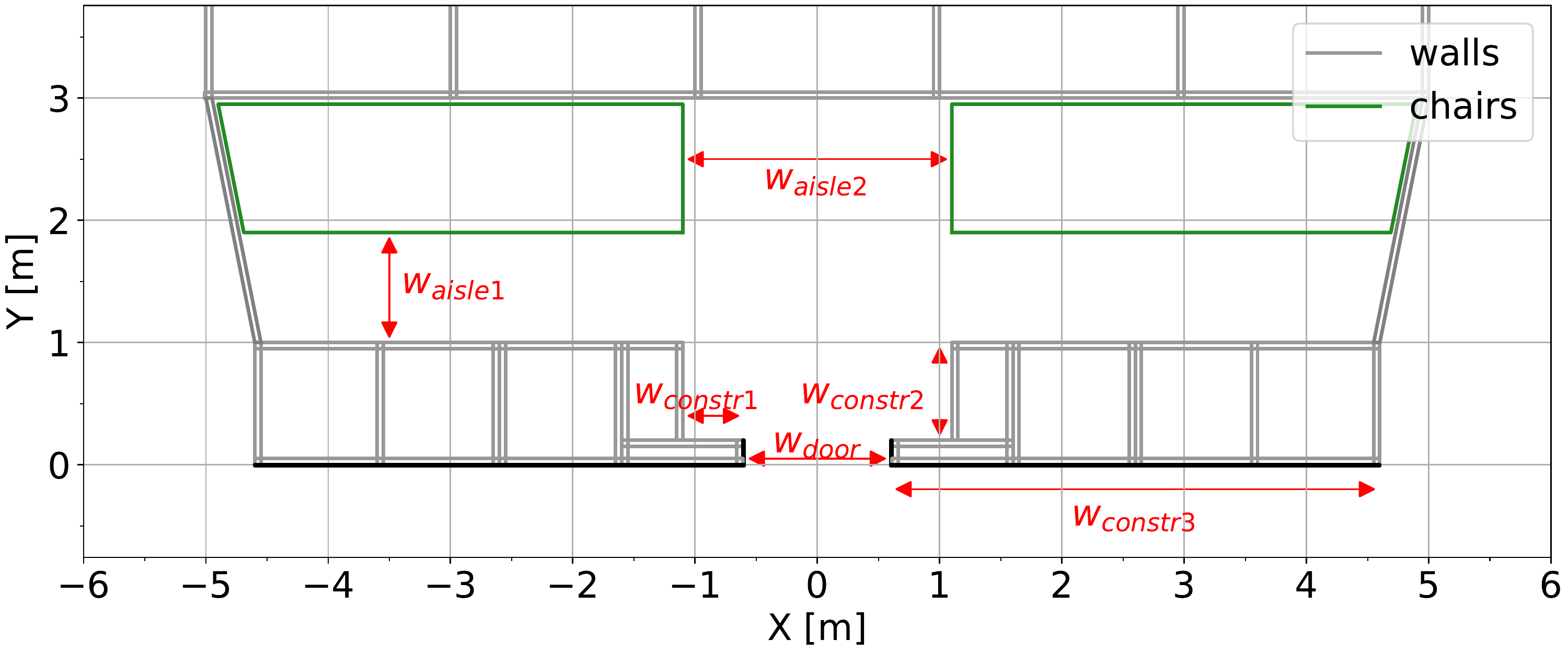}}
 	\subfigure[]{\includegraphics[width=0.48\textwidth]{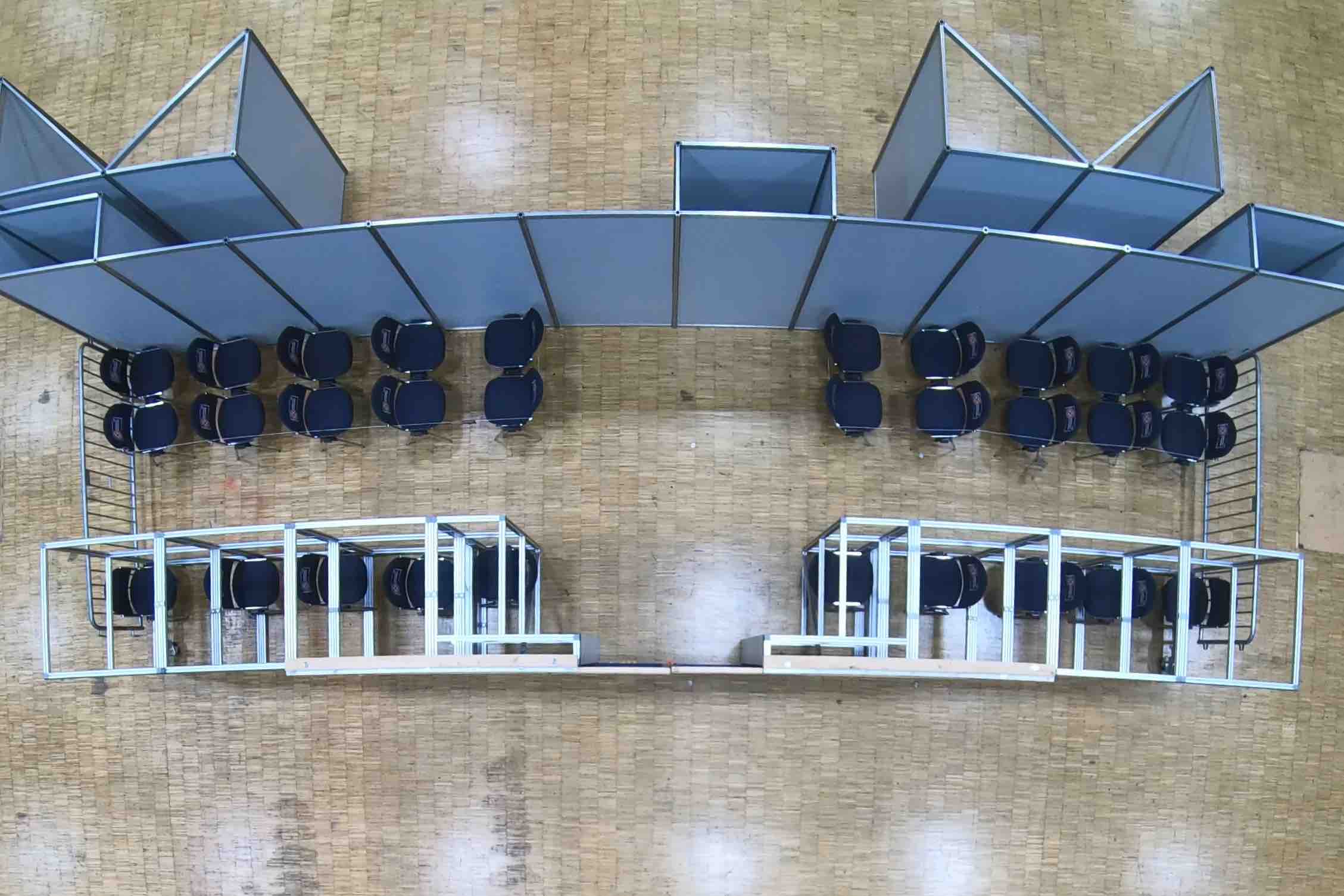}}
 	\subfigure[]{\includegraphics[width=0.48\textwidth]{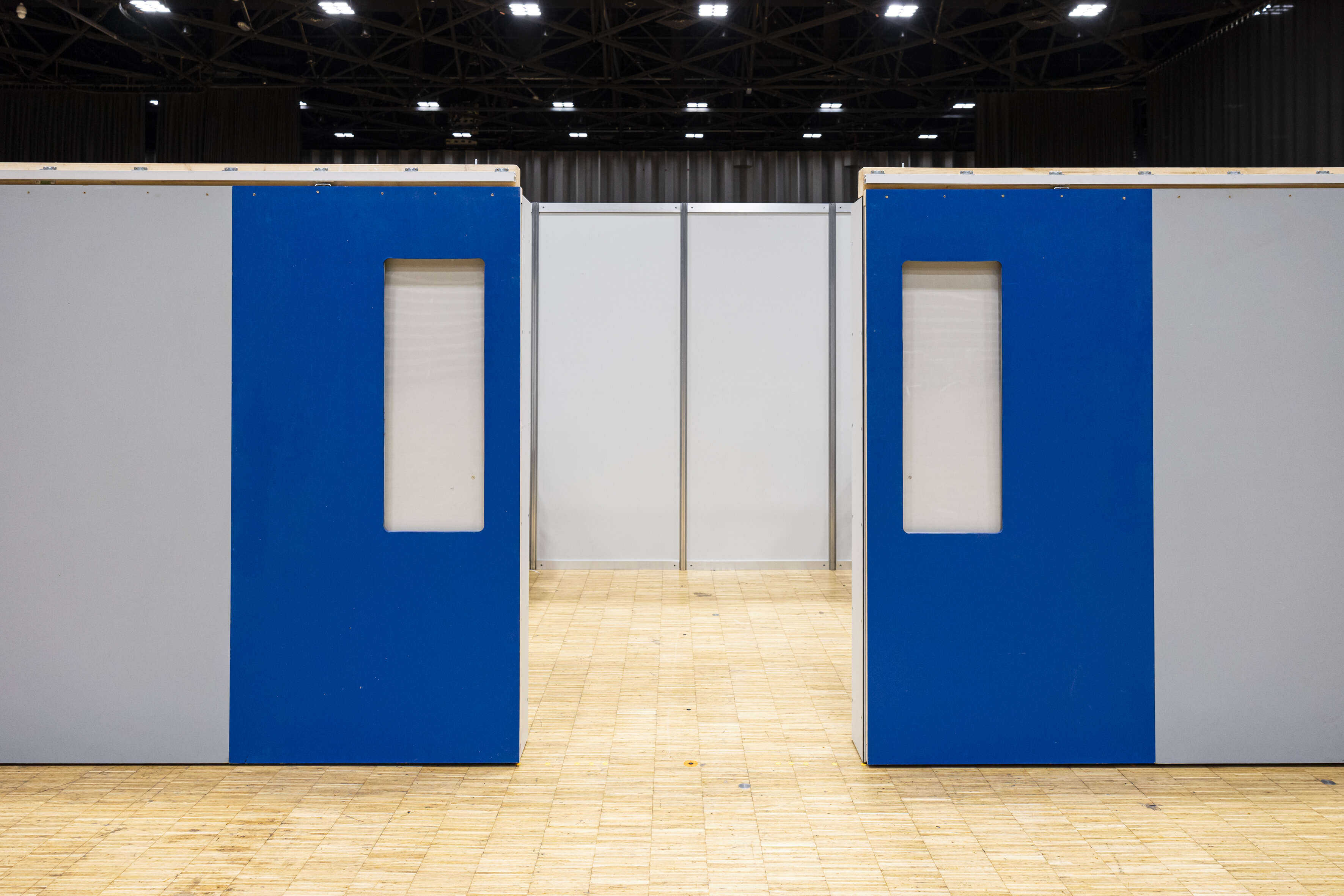}}
 	 \caption{a) Schematic showing coordinate system of simulated train car as well as screenshots of the setup in b) top view and c) front view.}
  	\label{fig:setup_siteDd1-3} 
\end{figure}
\noindent Instructions indicating `the arrival' and `the departure' of the train as well as the opening and the closing of the doors were given without using technical amplification from a person standing behind the waiting/boarding pedestrians. Special announcements were made via slips of paper or by directly addressing individuals by the investigators to make individual, targeted announcements when necessary to achieve the study objective. These kinds of announcements as well as handing out luggage were done in the waiting area to the left of the experimental area. Persons assigned to a group got sticky dots of the same color and had the instruction to stay together during the boarding process. For the variation of norm, the percentage in the list above refers to the amount of paper notes prescribing `normal', considerate behaviour whereas the remaining persons got the information that pushing is allowed. Where questionnaires were completed, this was done after the respective runs in the waiting area.\\

\noindent The outer dimensions of the experimental area were 20\,m\,x\,20\,m. The inner dimensions of the train car were approximately 9.2\,m\,x\,3\,m (the exact dimensions can be extracted from Fig.\,\ref{fig:setup_siteDd1-3}a) and aim to mimic a typical local train in Germany with the measurements $w_{door} = 1.2\,m$, $w_{const1} = 0.5\,m$, $w_{const2} = 0.8\,m$,  $w_{const3} = 4.0\,m$, $w_{aisle1} = 0.9\,m$, $w_{aisle2} = 2.2\,m$ as indicated in Figure\,\ref{fig:setup_siteDd1-3}a. \\

\noindent Cameras were mounted to record the experiment and are listed in Table\,\ref{tab:cameras_siteDd1-3}. 
Experimental runs in which 3D motion capturing data were recorded are listed in Table\,\ref{tab:mocap}. 
The mood of the participants (c.f.\sref{subsec:moodbutton}) was recorded for all runs. Trajectories were generated as described in Section\,\ref{subsec:CameraAndTrajectories}. The coordinate origin was located on the axis of the front of the bottleneck in the middle between the two bottleneck sides. The data of the Boarding and Alighting Experiments are provided online \cite{ForschungszentrumJulich2022f}.
 \subsection{Tiny Box Experiments}
\label{subsec:TinyHouseExperiments}
This experimental series investigated the relationship between density and physiological arousal while waiting. For this purpose up to eight participants waited in `tiny boxes' under different conditions. All participants were equipped with electrodermal activity and heart rate variability Sensors (cf. Sec.\,\ref{subsec:Electrodermal-Activity}, \ref{subsec:HeartRateVariability}). The following parameters were varied (a detailed list of performed runs and combinations of parameters can be found in Appendix \ref{app:TinyHouse}): 
\begin{itemize}
\itemsep0em 
	\item number of people in box
	\item communication: speaking allowed, speaking prohibited
\end{itemize}

\begin{figure}[h!]
	\centering
 	\includegraphics[width=0.8\textwidth]{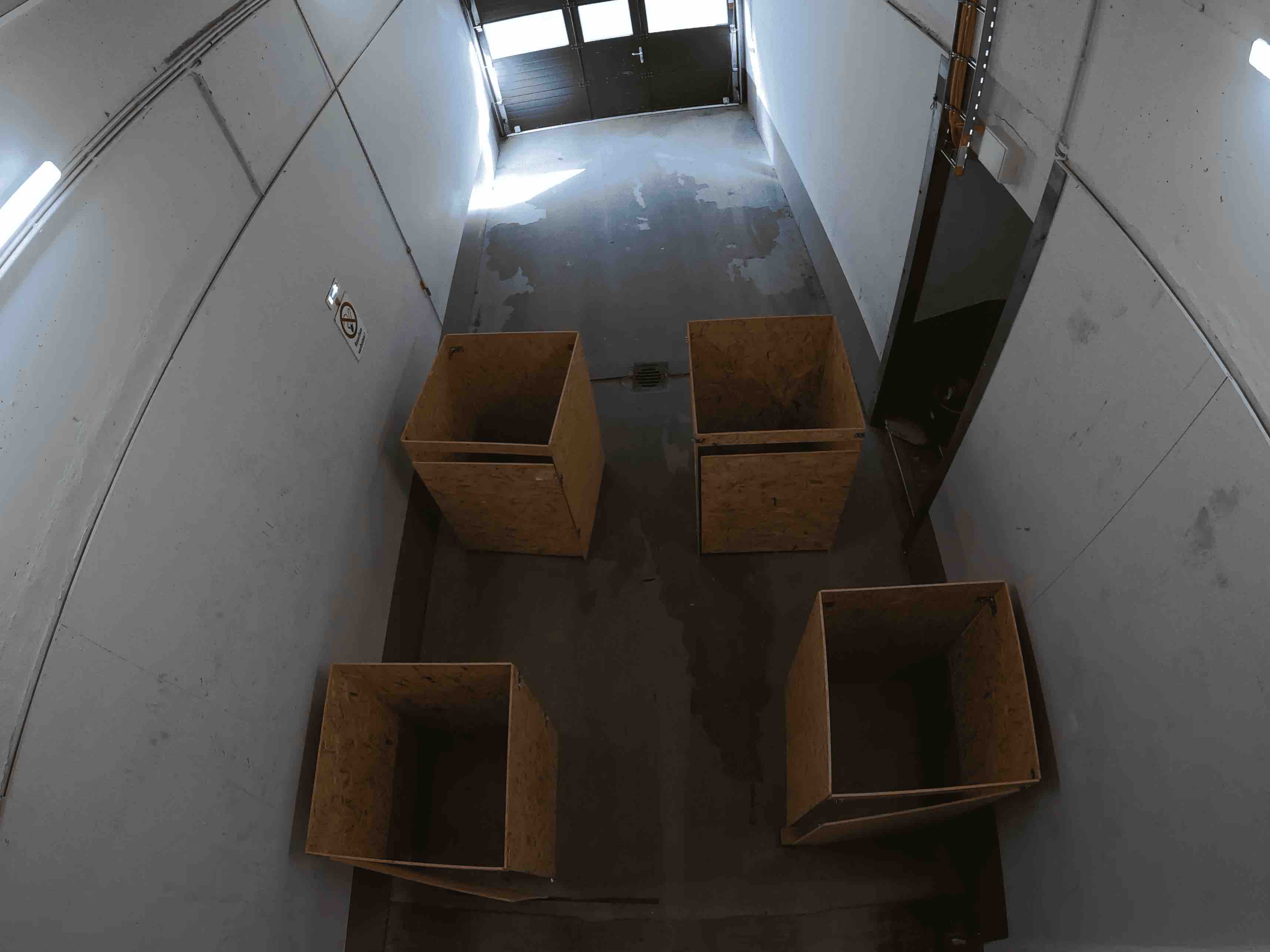}
 	 \caption{Snapshot of the tiny box experiment without participants. }
  	\label{fig:setup_siteTinyHouse} 
\end{figure}

\noindent The tiny boxes were four wooden boxes of 1\,m\,$\times$\,1\,m and a height of 1.5\,m (Fig.\,\ref{fig:setup_siteTinyHouse}). Participants could enter and exit the boxes through a one-sided swinging door. The experiments were performed in a delivery channel close to experimental site\,D (cf. Fig.\ref{fig:sites}) to shield participants as well as possible from acoustic and visual influence of the experiments performed at the same time. Participants were chosen based on age and gender from the respective groups taking part in the experiments at site\,D.  Instructions were given without technical amplification as the groups were small. Where questionnaires were completed, this was done after the respective runs in an area in front of the delivery channel. \\

\noindent Cameras were mounted to record the experiment and are listed in Table\,\ref{tab:cameras_siteTinyhouse}. Experimental runs in which EDA and HRV sensors were recorded are listed in Tables\,\ref{tab:eda} and \ref{tab:hrv}. No trajectories were exported for this experiment. The data of the Tiny Box Experiments are provided online \cite{ForschungszentrumJulich2022b}.
 \subsection{Bottleneck Experiments}
\label{subsec:BottleneckExperiments}
This series of experiments investigated different physical and social-psychological aspects in a bottleneck scenario. The following experimental parameters were varied (a detailed list of performed runs and combinations of parameters can be found in Appendix \ref{app:Bottleneck}): 
\begin{itemize}
\itemsep0em 
	\item bottleneck width: 0.6\,m, 0.7\,m, 0.8\,m, 1.0\,m, 1.2\,m, 1.6\,m
	\item bottleneck length: 0.2\,m, 2.0\,m
	\item motivation: normal, hurry, full commitment
	\item number of participants
	\item initial line-up: directly at the bottleneck, 2\,m semi-circle, special positions at 4\,m circle
	\item special announcements: none, active pushing / slowing, abort signal, interruption (with information passing on)
\end{itemize}

\begin{figure}[h!]
	\centering
 	\subfigure[]{\includegraphics[width=0.48\textwidth]{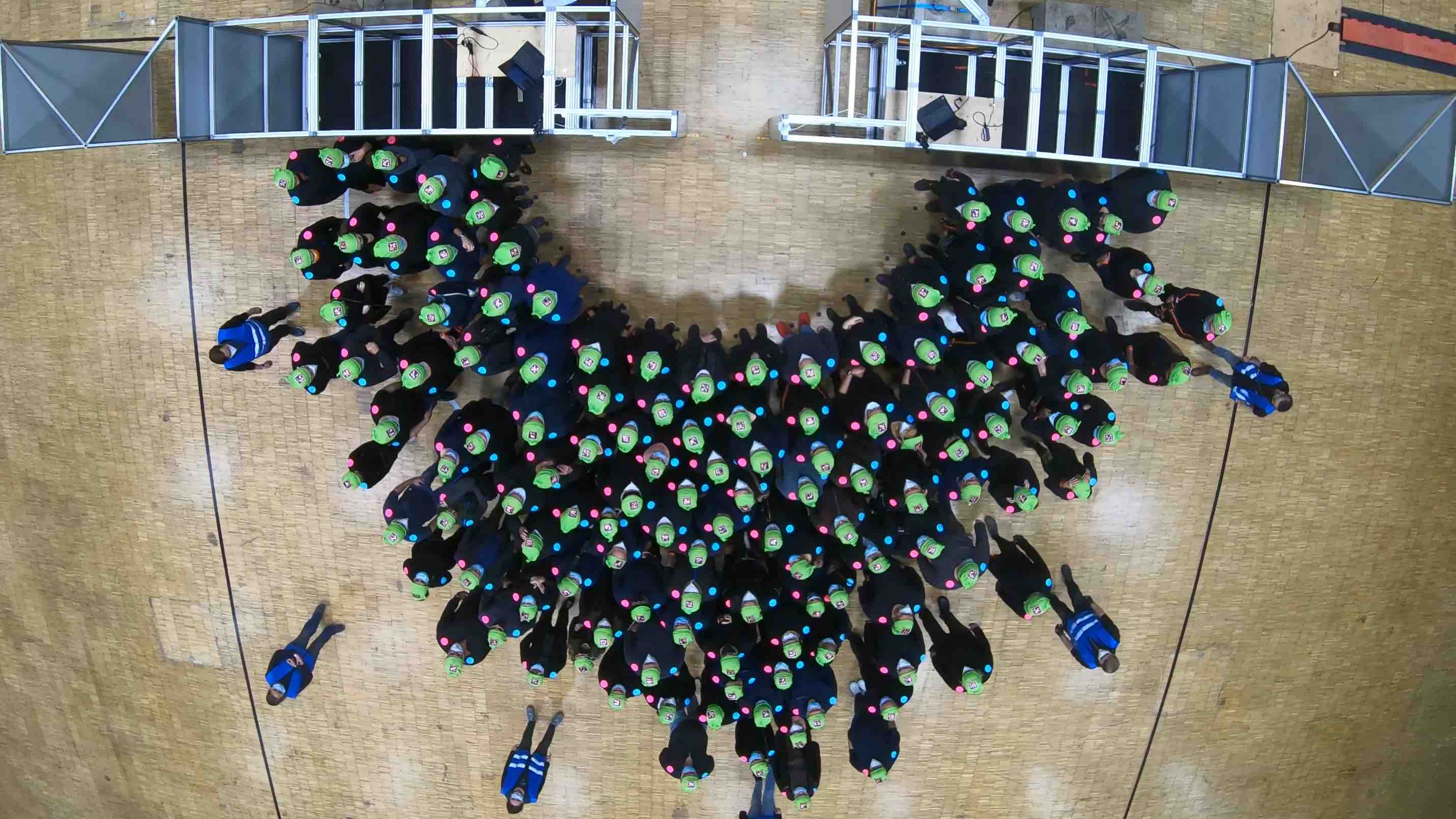}}
 	\subfigure[]{\includegraphics[width=0.48\textwidth]{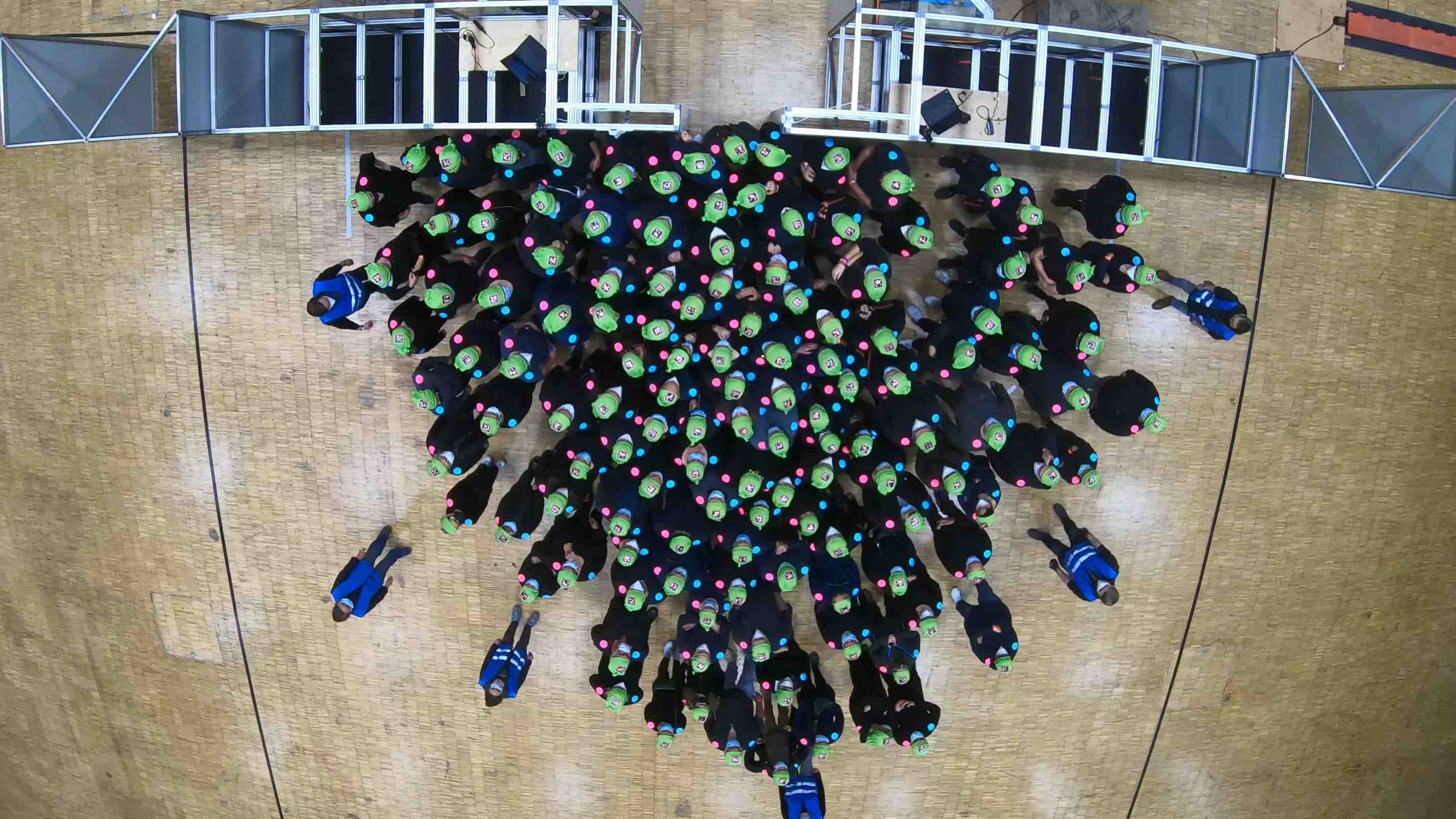}} 	
 	\subfigure[]{\includegraphics[width=0.48\textwidth]{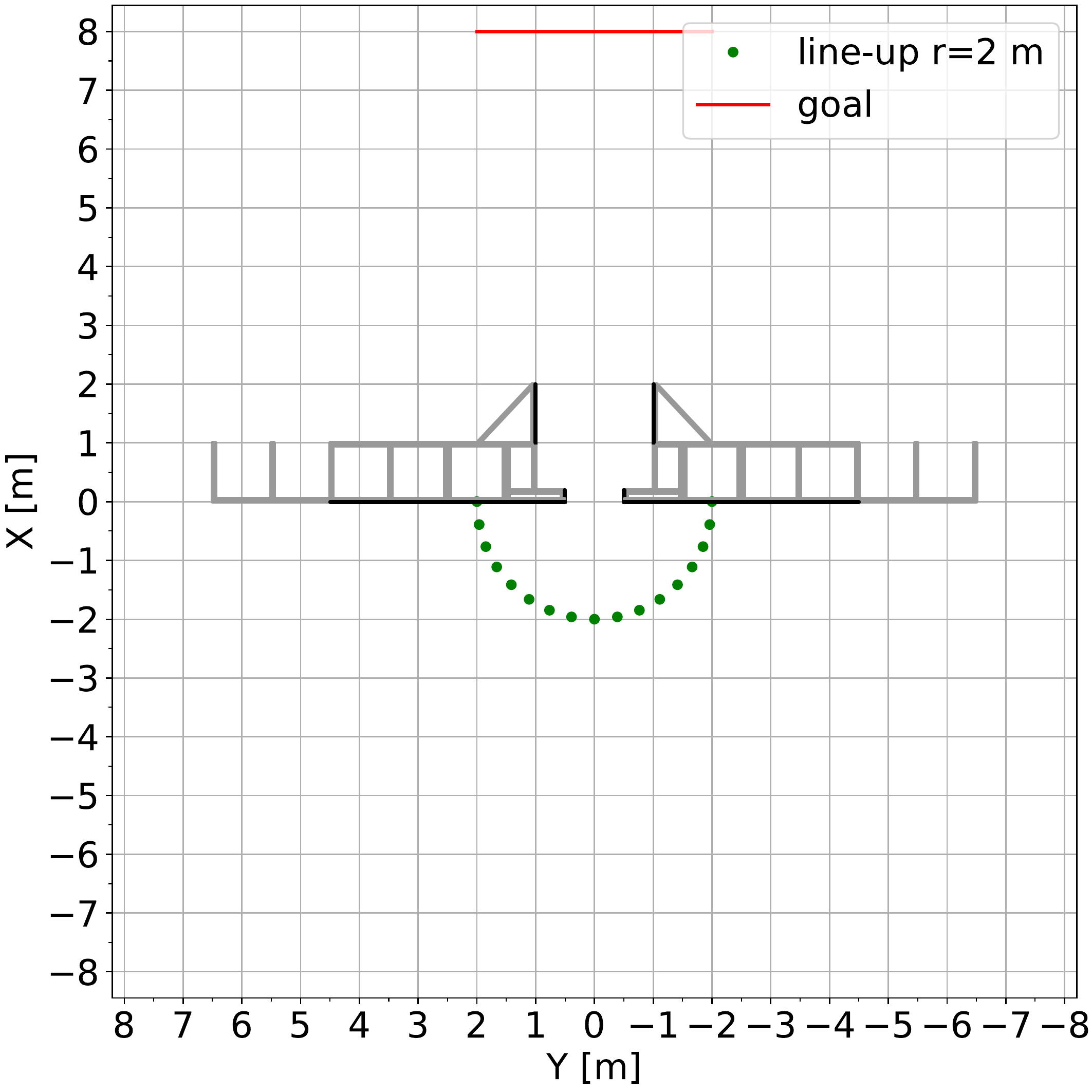}}
 	\subfigure[]{\includegraphics[width=0.48\textwidth]{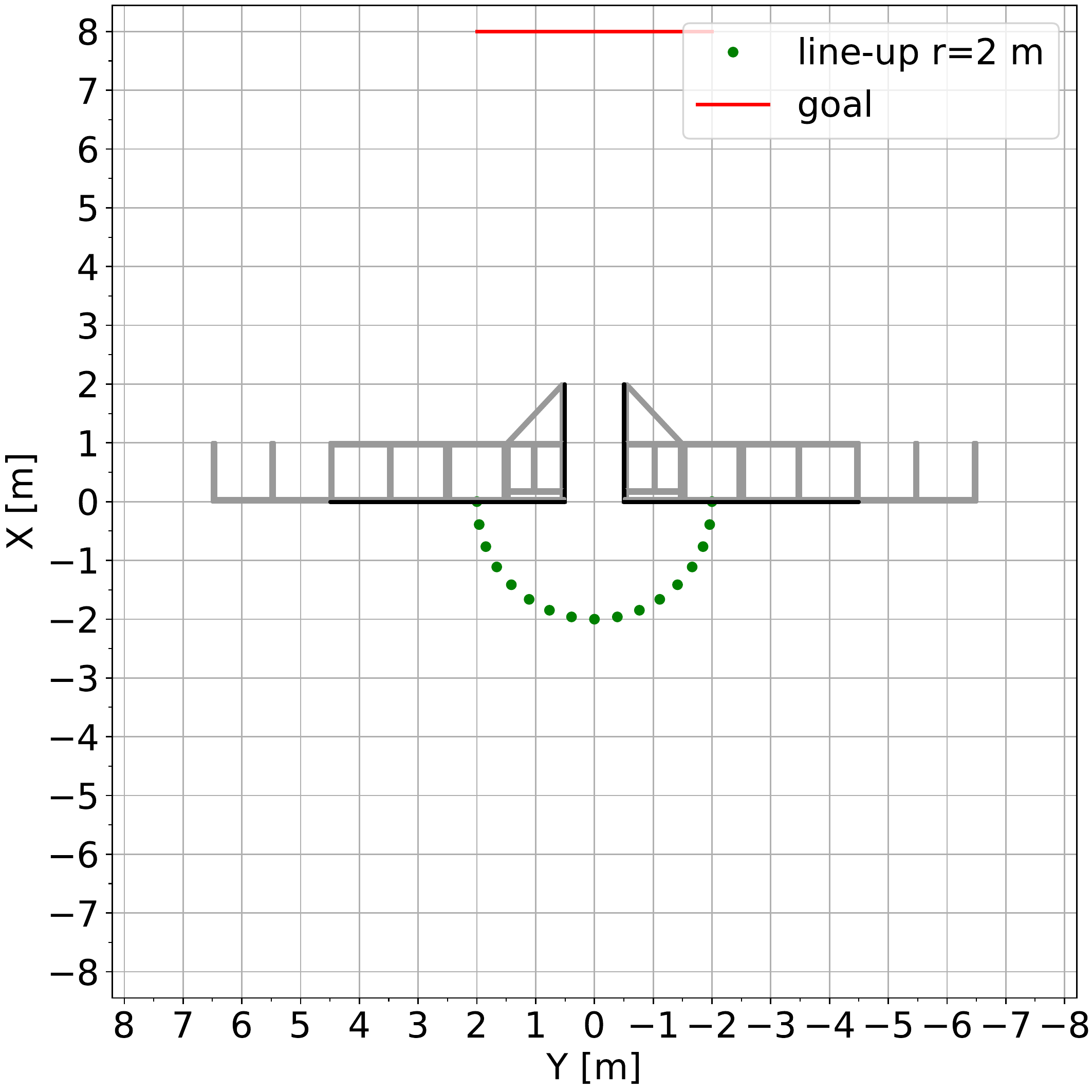}}
 	 \caption{Snapshots of experiments with 0.2\,m bottleneck length and an initial line-up being a) a 2\,m semi-circle and b) standing directly at the bottleneck. Schematics of a setup with a bottleneck length of c) 0.2\,m and d) 2\,m.}
  	\label{fig:setup_siteDd4} 
\end{figure}

\noindent The experiments were performed on day\,4 at experimental site\,D. The announcer and further observers were standing on a scissor lift that was parked behind the bottleneck construction and raised to a height of several meters to have a good overview. All announcements were made with a microphone connected to a portable loudspeaker. To increase the initial density, the participants in the first row were asked to stay in place while everyone else was asked to take one step forward, before each run. The intended initial density was $1\,P/m^2$. Whenever special targeted announcements were necessary to achieve the study objective, they were made via slips of paper or by the investigators directly addressing individuals. In `normal' condition runs participants were instructed as follows: ``You are in a crowd where people walk through a door at a normal pace. You yourself move purposefully, but without haste.'' In `hurry' conditions the instructions were ``You are in a crowd where people are in a hurry to pass a door. You yourself are also moving briskly'', and in `full commitment' conditions ``You are in a crowd where everyone wants to pass through a door as quickly as possible and pushes their way through. You yourself do everything you can to get to the front and through quickly as well.''

\noindent Figure \ref{fig:setup_siteDd4} shows snapshots of two example experiments and sketches of the bottleneck construction. The outer dimensions of the experimental area were 20\,m\,by\,20\,m. The bottleneck construction consisted of an 4\,m\,x\,2\,m\,x\,1\,m aluminium frame with gray plastic panels, weighing 250 kg per side. They were each visually extended to be 6\,m long by adding trade fair walls. Each side was secured against slipping with anti-slip mats and 750 kg concrete blocks which were bolted to the bottleneck construction. Participants were to maintain their motivation until they crossed a finish line 8\,m behind the bottleneck. The way to the finish line was marked with barrier tape. Beyond the finish line, participants could return to the line-up area by turning to either sides. Where questionnaires were to be completed, this was done after the respective runs in an area to the left of the experimental area. \\

\noindent To take safety precautions, a practised crowd manager was present at the experiments equipped with an air pressure horn. The horn was activated whenever a participant indicated discomfort during an experiment by calling out `stop' aloud or if the crowd manager himself identified a situation as critical or potentially harmful. At the beginning of the day, all participants were trained in what was to be done in the case of the horn being activated. The procedure included immediately stopping in the current position without further movement. Designated helpers that were close to the crowd at all times started tapping people at the shoulder once the crowd had come to a full stop. On the signal of 'shoulder tapping' participants were allowed to turn around and move to the far back of the experimental site. The procedure was continued until every person was tapped at the shoulder. Apart from the test run where the horn was activated on purpose, two runs (4D250, 4D280) were aborted and resolved in the way described above. \\

\noindent Cameras were mounted to record the experiment and are listed in Table\,\ref{tab:cameras_siteDd4}. 
Experimental runs in which 3D motion capturing and pressure sensor data were recorded are listed in Table\,\ref{tab:mocap}. Pressure sensor data (cf.\,\sref{subsec:Pressure}) and the mood of the participants (cf.\,\sref{subsec:moodbutton}) were recorded for all runs. Trajectories were generated as described in Section\,\ref{subsec:CameraAndTrajectories}. The coordinate origin was located on the axis of the front of the bottleneck walls in the middle between the two bottleneck sides. The data of the Bottleneck Experiments are provided online \cite{ForschungszentrumJulich2022g}. The scientific content of some of the bottleneck experiments in this series is part of the CrowdDNA project. 

\section{Sensors}
Different combinations of sensory systems were used in the different experiments. This included camera recordings, electrodermal activity sensors, heart rate variability sensors, pressure sensors (at the wall as well as on participants), 3D-motion capturing systems and mood buttons. The design and use of the individual sensors as well as their synchronization in time will be described in the following sections. 
\subsection{Time Synchronization Between Sensors}
\label{subsec:TimeSync}

Accurate time synchronization is required for reliable connection of data from multiple sensor sources. Depending on the sensors' technical settings, different technical solutions need to be adopted to enable synchronization. In an ideal setup, all sensors operate with the same frequency, are connected to the same metronome to capture the exact same instance in time and have the same time code. However, in reality different sensors operate on different frame rates, are not pairable with a metronome or might show a drift in time.
In order to keep the deviations as small as possible, a global time was introduced and distributed by Tentacle timecode generators \cite{TentacleSyncGmbH} and as many sensors as possible were attached to submodule metronomes.

\paragraph{Camera Synchronization with Global Time} All Marshall cameras (\sref{subsec:CameraAndTrajectories}) were operated with a multi-camera video review system Simplylive Ref\&Box8 Mini and the Rosendahl NanoSync \cite{RosendahlStudiotechnikGmbH} metronome. Sony RX0M2 cameras were operated with the connected control box CCD\_WD1 and handled remotely via direct web access. All cameras were synchronized with the Tentacle timecode by filming the Tentacle global time (approx. every 45 minutes) and shifting time data to coincide with the global time in the aftermath. If the global time could not be read in the video (GP4\_1, GP4\_2), synchronization was performed based on movement patterns of individual participants. Due to the regular recording of the timecode and subsequent adjustment, the drift in time was negligible.

\paragraph{3D Motion Capturing System Synchronization with Global Time} The 3D Motion Capturing System (\sref{subsec:MoCap}) was attached to the Tentacle timecode generator for the course of the experiments. The MVN software syncs the time between the Tentacle timecode generator and the Xsens hardware once every second. The timecode is stored in the data set for every sample point.

\paragraph{HRV/EDA and Pressure Sensor Synchronization with Global Time} The computers controlling the other sensors (wall pressure sensor  (\sref{subsec:Pressure}), body pressure sensor, EDA (\sref{subsec:Electrodermal-Activity}), HRV (\sref{subsec:HeartRateVariability})) were filmed each morning and evening together with the Tentacle timecode generator. The time difference $\Delta t$ between the tentacle time and the internal time of the corresponding computers was determined (hh:mm:ss:ff) and data was shifted to coincide with the global time. No drift over time was detected.

\subsection{Camera and Trajectories}
\label{subsec:CameraAndTrajectories}
Camera recordings can be used for experiments in many ways. On the one hand, they make it possible to get a qualitative view of behaviour and to reconstruct the actual execution of the day and any deviations from the plan that may have occurred, as well as to reconstruct announcements and their intonation via the audio track. On the other hand, cameras can be used specifically to obtain measurement results such as extracting walking paths (trajectories) or documenting facial expressions of participants in response to the experiments. \\

\begin{figure}[h!]
	\centering
 	\subfigure[]{\includegraphics[height=0.3\textheight]{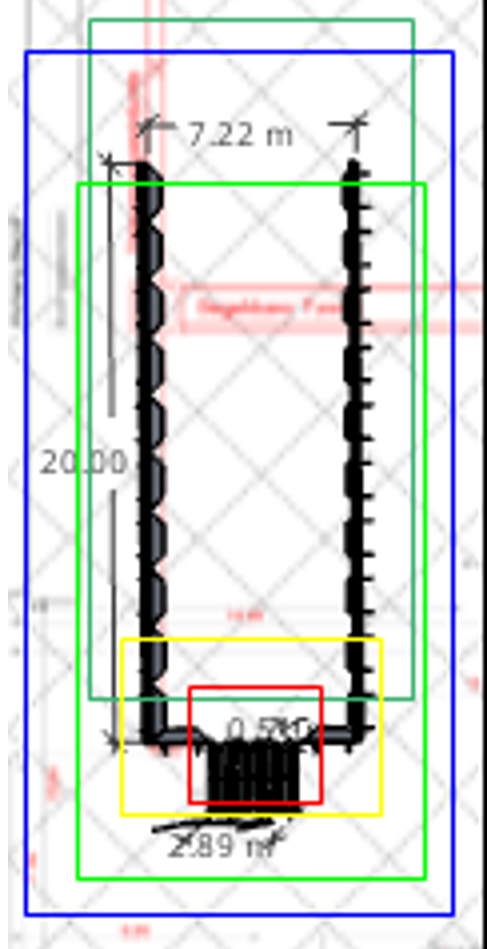}}
 	\subfigure[]{\includegraphics[height=0.3\textheight]{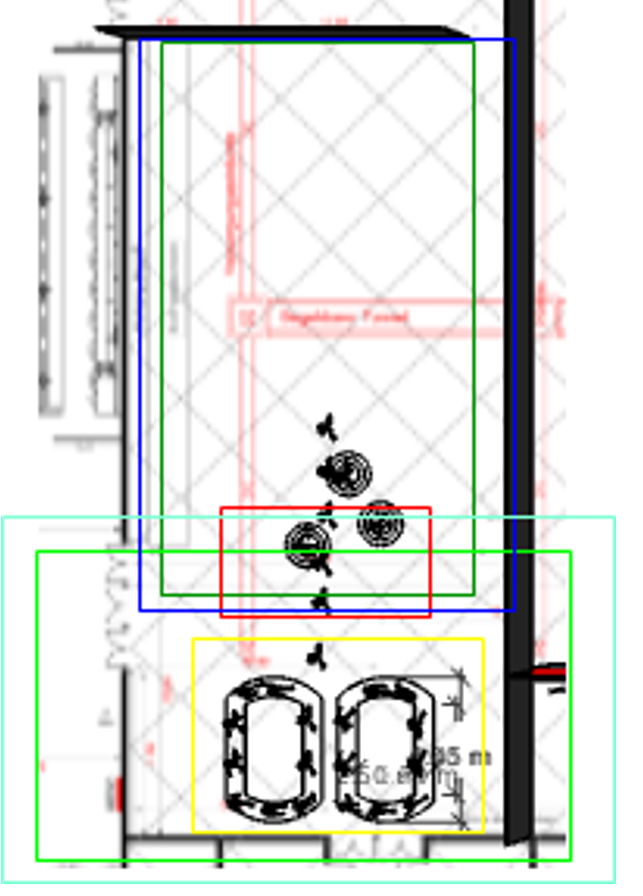}}
 	\subfigure[]{\includegraphics[height=0.3\textheight]{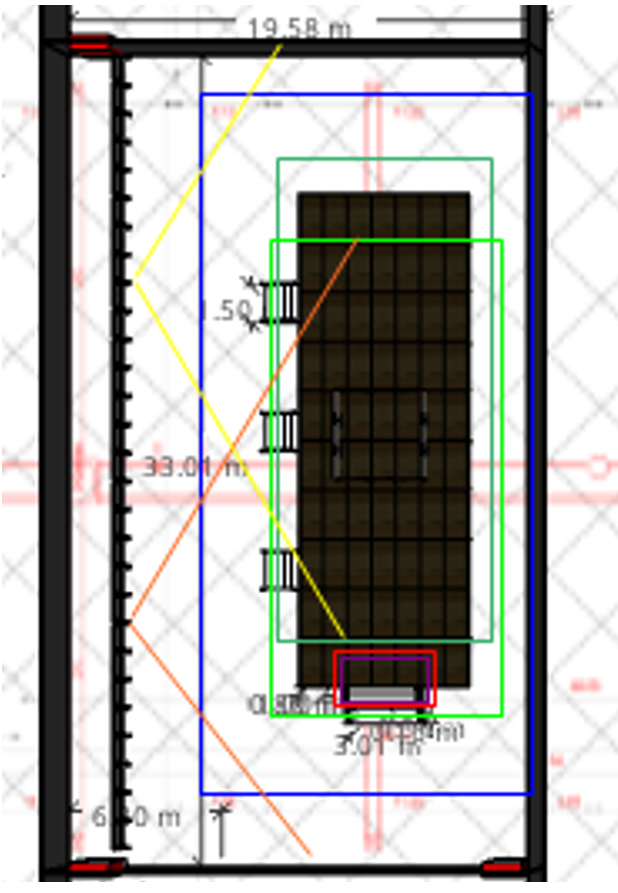}}
 	\subfigure[]{\includegraphics[width=0.4\textwidth]{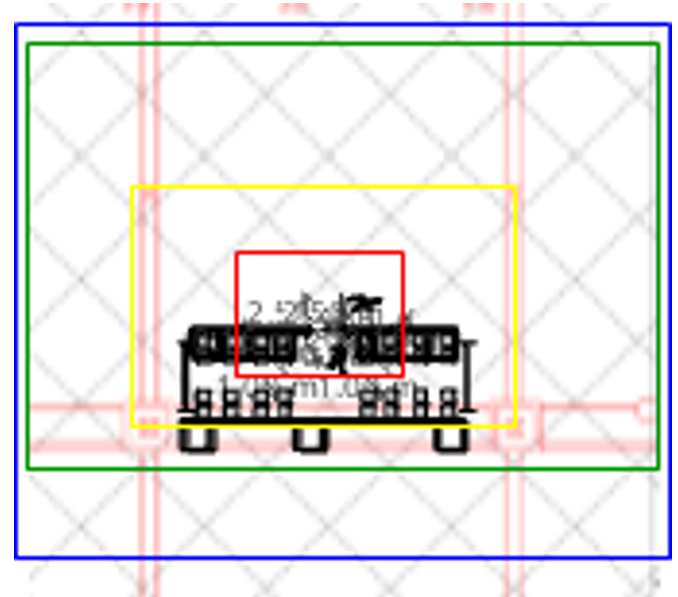}}
 	 \caption{Schematic showing approximate field of view at head height for mounted cameras in each experimental site.\\ a) experimental site\,C on day\,1-3; dark green: SL\_cam4, light green: SL\_cam5, red: SL\_cam6, blue: GP4\_1, yellow: RX0\_cam2;\\ b) experimental site\,C on day\,4; dark green: SL\_cam5, light green: SL\_cam4, red: SL\_cam6, dark blue: GP4\_1, light blue: GP4\_2, yellow: RX0\_cam2;\\ c) experimental site\,B on day\,1-3; dark green: SL\_cam1, light green: SL\_cam2, red: SL\_cam3, purple: PJ740, blue GP7\_2, yellow: X3000\_1, orange: X3000\_2;\\ d) experimental site\,D on day\,1-4; dark green: SL\_cam7, red: SL\_cam8, dark blue: GP7\_1, yellow: RX0\_cam1}
  	\label{fig:fov} 
\end{figure}

\noindent In total, 21 cameras were mounted in order to perform the above tasks. Cameras intended for extracting trajectories and serving documentation purposes were mounted under the ceiling ($\approx$ 8.65\,m) facing straight downwards. Camera views were overlapping and mounted in such a way that the occlusion of people was minimized. All cameras used for trajectory extraction were backed up since image loss would have been fatal for the experiments. The approximate fields of view at head height for cameras mounted in the main experimental sites are shown in Figure\,\ref{fig:fov} and listed in Tables\,\ref{tab:cameras_siteBd1-3}-\ref{tab:cameras_siteDd4}. The camera types and settings are listed in Table\,\ref{tab:cameras_all}.\\

\begin{table}[h!]
\centering
\caption{Camera models of the 21 cameras mounted in the experimental sites, with short name of cameras used in the text, their type, the amount of cameras used and settings the cameras were operated with.}
\label{tab:cameras_all}
\resizebox{\textwidth}{!}{%
\begin{tabular}{llll}\toprule
\textbf{camera name} & \textbf{camera type} & \textbf{no.} & \textbf{camera settings} \\ \midrule
\begin{tabular}[c]{@{}l@{}}SL\_cam3, SL\_cam6, \\ SL\_cam8\end{tabular} & \begin{tabular}[c]{@{}l@{}}Marshall CV365-CGB camera \\ with lens VS-M2812-2\end{tabular} & 3 & \begin{tabular}[c]{@{}l@{}}HD (1920\,x\,1080p), 50 fps, \\shutter: 1/300, brightness: 20\end{tabular}\\[.8\normalbaselineskip]
\begin{tabular}[c]{@{}l@{}}SL\_cam1, SL\_cam2, \\ SL\_cam4, SL\_cam5, \\ SL\_cam7\end{tabular} & \begin{tabular}[c]{@{}l@{}}Marshall CV365-CGB camera\\  with lens VS-M226-A\end{tabular} & 5 & \begin{tabular}[c]{@{}l@{}}HD (1920\,x\,1080p), 50 fps, \\shutter: 1/300, brightness: 20\end{tabular} \\[.2\normalbaselineskip]
\begin{tabular}[c]{@{}l@{}}RX0\_cam1, \\ RX0\_cam2\end{tabular} & Sony RX0M2 & 2 & \begin{tabular}[c]{@{}l@{}}4K (3840\,x\,2160p), 25fps,\\ RX0\_cam1: shutter: 1/320,  ISO: 320,\\ RX0\_cam2: shutter: 1/400, ISO:1000\end{tabular} \\[.2\normalbaselineskip]
\begin{tabular}[c]{@{}l@{}}GP7\_1,  GP7\_2, \\ GP7\_3,  GP7\_4,  \\ GP7\_5\end{tabular} & GroPro 7 & 5 & \begin{tabular}[c]{@{}l@{}}4K (4000\,x\,3000p, wide), 25fps,\\shutter: - , ISO: -  (Protune: off)\end{tabular}\\[1.5\normalbaselineskip]
GP4\_1,  GP4\_2 & GroPro 4 & 2 & \begin{tabular}[c]{@{}l@{}}4K (4000\,x\,3000p, wide), 25fps,\\shutter: - , ISO: -  (Protune: off)\end{tabular} \\[.3\normalbaselineskip]
X3000\_1, X3000\_2 & Sony X3000 & 2 & \begin{tabular}[c]{@{}l@{}}4K (3840\,x\,2160p), 25fps, \\shutter: - , ISO: -\end{tabular}\\[.3\normalbaselineskip]
Sony PJ & Sony PJ740 & 1 & \begin{tabular}[c]{@{}l@{}}HD (1920\,x\,1080p), 25fps, \\shutter: - , ISO: -\end{tabular}\\[.3\normalbaselineskip]
Sony PXW & Sony PXW Z150 & 1 & \begin{tabular}[c]{@{}l@{}}4K (3840\,x\,2160p), 25fps\\shutter: - , ISO: -\end{tabular}\\ \bottomrule
\end{tabular}%
}
\end{table}

\begin{table}[]
    \centering
    \caption{Cameras mounted in experimental site\,D during the Train Platform Experiments (\sref{subsec:TrainPlatformExperiments}) and their purpose.}
    \label{tab:cameras_siteBd1-3}
    \begin{tabular}{lccc}
    \toprule
	\rule[-1ex]{0pt}{2.5ex} \textbf{camera type} & \textbf{purpose} &\textbf{trajectories exported} & \textbf{sound} \\ 
	\midrule 
	\rule[-1ex]{0pt}{2.5ex} SL\_cam1 & trajectories & x & - \\ 
	\rule[-1ex]{0pt}{2.5ex} SL\_cam2 & trajectories & x & - \\ 
	\rule[-1ex]{0pt}{2.5ex} SL\_cam3 & code reading & x & - \\ 
	\rule[-1ex]{0pt}{2.5ex} GP7\_2 & overview (top-side view) & - & x \\ 
	\rule[-1ex]{0pt}{2.5ex} Sony PJ & code reading backup & - & - \\ 
	\rule[-1ex]{0pt}{2.5ex} X3000\_1 & record facial expressions & - & - \\ 
	\rule[-1ex]{0pt}{2.5ex} X3000\_2 & record facial expressions & - & - \\
	\bottomrule 
    \end{tabular}
\end{table}

\begin{table}[]
    \centering   
    \caption{Cameras mounted in experimental site\,C during the Crowd Management Experiments (\sref{subsec:CrowdManagementExperiments}) and their purpose.}
    \label{tab:cameras_siteCd1-3}
    \begin{tabular}{lccc}
    \toprule
	\rule[-1ex]{0pt}{2.5ex} \textbf{camera type} & \textbf{purpose} &\textbf{trajectories exported} & \textbf{sound} \\ 
	\midrule
	\rule[-1ex]{0pt}{2.5ex} SL cam4 & trajectories & x & - \\ 
	\rule[-1ex]{0pt}{2.5ex} SL cam5 & trajectories & x & - \\ 
	\rule[-1ex]{0pt}{2.5ex} SL cam6 & code reading & x & - \\ 
	\rule[-1ex]{0pt}{2.5ex} GP4\_1 & overview, backup & - & x \\ 
	\rule[-1ex]{0pt}{2.5ex} RX0\_cam2 & code reading backup & - & - \\ 
	\bottomrule
    \end{tabular} 
\end{table}
\begin{table}[]
    \centering    
    \caption{Cameras mounted in experimental site\,C during the Single-File Experiments (\sref{subsec:OvalExperiments}) and their purpose.}
    \label{tab:cameras_siteCd4_oval}
    \begin{tabular}{lccc}
    \toprule
	\rule[-1ex]{0pt}{2.5ex} \textbf{camera type} & \textbf{purpose} &\textbf{trajectories exported} & \textbf{sound} \\ 
	\midrule
	\rule[-1ex]{0pt}{2.5ex} SL\_cam4 & trajectories & x & - \\ 
	\rule[-1ex]{0pt}{2.5ex} SL\_cam6 & code reading & - & - \\ 
	\rule[-1ex]{0pt}{2.5ex} GP4\_2 & overview & - & x \\ 
	\rule[-1ex]{0pt}{2.5ex} RX0\_cam2 & code reading backup & x & - \\ 
	\bottomrule
    \end{tabular} 
\end{table}

\begin{table}[]
    \centering
    \caption{Cameras mounted in experimental site\,C during the Personal Space Experiments (\sref{subsec:PersonalSpaceExperiments}) and their purpose.}
    \label{tab:cameras_siteCd4_pspace}
    \begin{tabular}{lccc}
    \toprule
	\rule[-1ex]{0pt}{2.5ex} \textbf{camera type} & \textbf{purpose} &\textbf{trajectories exported} & \textbf{sound} \\ 
	\midrule
	\rule[-1ex]{0pt}{2.5ex} SL\_cam5 & trajectories & - & - \\ 
	\rule[-1ex]{0pt}{2.5ex} SL\_cam6 & code reading & - & - \\ 
	\rule[-1ex]{0pt}{2.5ex} GP4\_1 & overview & - & x \\ 
	\bottomrule
    \end{tabular} 
\end{table}

\begin{table}[]
    \centering
    \caption{Cameras mounted in experimental site\,D during the Boarding and Alighting Experiments (\sref{subsec:BoardingAndAlightingExperiments}) and their purpose.}
    \label{tab:cameras_siteDd1-3}
    \begin{tabular}{lccc}
    \toprule
	\rule[-1ex]{0pt}{2.5ex} \textbf{camera type} & \textbf{purpose} &\textbf{trajectories exported} & \textbf{sound} \\ 
	\midrule
	\rule[-1ex]{0pt}{2.5ex} SL\_cam7 & trajectories & x & - \\ 
	\rule[-1ex]{0pt}{2.5ex} SL\_cam8 & trajectories & x & - \\ 
	\rule[-1ex]{0pt}{2.5ex} GP7\_1 & overview & - & x \\ 
	\rule[-1ex]{0pt}{2.5ex} RX0\_cam1 & code reading backup & - & - \\ 
	\bottomrule
    \end{tabular}
\end{table}

\begin{table}[]
    \centering
    \caption{Cameras mounted in experimental site\,D during the Tiny House Experiments (\sref{subsec:TinyHouseExperiments}) and their purpose.}
    \label{tab:cameras_siteTinyhouse}
    \begin{tabular}{lccc}
    \toprule
	\rule[-1ex]{0pt}{2.5ex} \textbf{camera type} & \textbf{purpose} &\textbf{trajectories exported} & \textbf{sound} \\ 
	\midrule
	\rule[-1ex]{0pt}{2.5ex} GP7\_3 & overview & - & x \\ 
	\bottomrule
    \end{tabular}
\end{table}

\begin{table}[]
    \centering
    \caption{Cameras mounted in experimental site\,D during the Bottleneck Experiments (\sref{subsec:BottleneckExperiments}) and their purpose.}
    \label{tab:cameras_siteDd4}
    \begin{tabular}{lccc}
    \toprule
	\rule[-1ex]{0pt}{2.5ex} \textbf{camera type} & \textbf{purpose} &\textbf{trajectories exported} & \textbf{sound} \\ 
	\midrule
	\rule[-1ex]{0pt}{2.5ex} SL\_cam7 & trajectories & - & - \\ 
	\rule[-1ex]{0pt}{2.5ex} SL\_cam8 & trajectories & - & - \\ 
	\rule[-1ex]{0pt}{2.5ex} GP7\_1 & overview & x & x \\ 
	\rule[-1ex]{0pt}{2.5ex} RX0\_cam1 & code reading backup & x & - \\ 
	\rule[-1ex]{0pt}{2.5ex} GP7\_4 & record facial expressions & - & x \\ 
	\rule[-1ex]{0pt}{2.5ex} GP7\_5 &  record facial expressions  & - & x \\ 
	\rule[-1ex]{0pt}{2.5ex} Sony PXW &  faces bottleneck exit  & - & x \\ 
	\bottomrule
    \end{tabular} 
\end{table}

\paragraph{Trajectories}
The trajectory extraction was performed with the pedestrian tracking software PeTrack \cite{Boltes2022, Boltes2013} for cameras indicated in Sections\,\ref{subsec:TrainPlatformExperiments}\,-\,\ref{subsec:BottleneckExperiments}.
Cameras operated with the Simplylive system produced frame drops, double and black frames. The reason could not be reproduced unequivocally. These artifacts were detected and treated before continuing with trajectory extraction. 
Black frames were detected by applying a binary filter on the grey-scale frames of the video and checking if all pixels were black. 
Duplicated frames were detected by computing the difference between each frame with the previous one in greyscale. On these differences, DBSCAN \cite{Ester1996, Schubert2017} was used to detect clusters with camera-based parameters. Each frame which did not belong to a cluster was considered to be a duplication of the previous one. 
Afterward, the videos were reencoded with ffmpeg skipping these erroneous frames.
Before exporting the trajectories from PeTrack, the results of the tracking were used to further improve the output data to interpolate the movement between dropped frames. For this,  the displacement of each pedestrian in a frame was computed using the Lucas-Kanade method \cite{Lukas1981, Lucas1984}. Computing the ratio between these displacements and the average displacement of the previous frame gave the number of missing frames.\\
To implement the mapping from pixel to real world coordinates two types of calibration \cite{Boltes2010a} had to be performed. Intrinsic calibration was performed to take into account the distortion of the lenses and internal hardware combinations. Extrinsic calibration was performed to create a transformation map between the camera and real world coordinate system and was performed every morning with a ranging pole and attached levelling unit. The resulting mean re-projection error  for all calibration points for all days and cameras was 1.1\,cm with a standard deviation of 0.6\,cm and a maximum error of 2.2\,cm. However, values differ greatly depending on the camera. The values for the individual cameras are shown in the appendix in Table\,\ref{tab:reprojectionerror}.\\
For cameras used for code reading, recognition in the software PeTrack was performed with the code marker method using Aruco Code dictionary dict\_6X6\_1000 \cite{Garrido-Jurado2016}. All other cameras' recognition was performed using the multicolor marker method within PeTrack. After the automatic extraction of trajectories, all runs were manually corrected. To handle the perspective distortion of the cameras for a correct head position in space, the individual heights of each person were accounted for and if a code could not be read a default height of 1.75\,m was applied. The different camera views of each experimental area were combined into one single dataset by linear interpolation from the trajectory of one camera view to the trajectory of the other camera view in the overlap region.

\subsection{Electrodermal Activity}
\label{subsec:Electrodermal-Activity}
Ambulatory sensors (EDA Move4) from the Movisens company \cite{MovisensEdaMove4UserManual} were used for measuring electrodermal activity. A total of 28\,sensors were used, which were activated every morning and their data saved every evening. The EDA Move recorded electrodermal activity using the exosomatic method at a constant voltage of 0.5\,V. The measurement range is 2\,-\,100 micro Siemens and the sampling frequency is 32\,Hz. 

\begin{table}[hb!]
\centering
\caption{Experiment runs in which EDA sensors were used. Experimental configurations of runs are shown in the appendix.}
\label{tab:eda}
{\small
\begin{tabular}{clll}
        \toprule
		\textbf{day} &
		  \multicolumn{1}{c}{\textbf{\begin{tabular}[c]{@{}c@{}}Tiny Box\\ Experiment\end{tabular}}} &
		  \multicolumn{1}{c}{\textbf{\begin{tabular}[c]{@{}c@{}}Oval\\ Experiment\end{tabular}}} &
		  \multicolumn{1}{c}{\textbf{\begin{tabular}[c]{@{}c@{}}Personal Space\\ Experiment\end{tabular}}} \\ \midrule
		\textbf{1} &
		  \begin{tabular}[c]{@{}l@{}}1 -- 85\end{tabular} &
		  \begin{tabular}[c]{@{}l@{}}--\end{tabular} &
		  -- \\ 
		\textbf{2} &
		  \begin{tabular}[c]{@{}l@{}}86 --170\end{tabular} &
		  \begin{tabular}[c]{@{}l@{}}--\end{tabular} &
		  -- \\ 
		\textbf{3} &
		  \begin{tabular}[c]{@{}l@{}}171 -- 204\end{tabular} &
		  \begin{tabular}[c]{@{}l@{}}--\end{tabular} &
		  -- \\
		\multicolumn{1}{c}{\textbf{4}} &
		  -- &
		  \begin{tabular}[c]{@{}l@{}}single\_file\_gender\_2\_6, \\single\_file\_gender\_2\_11,\\single\_file\_gender\_random\_2\_6,\\ single\_file\_gender\_random\_2\_11\\  \end{tabular}&
		  \begin{tabular}[c]{@{}l@{}}4C1010, 4C1020,\\ 4C1030, 4C1040, \\4C2010, 4C2020, \\4C2030, 4C2040\end{tabular}
		   \\ \bottomrule
\end{tabular}%
}
\end{table}

\noindent The sensor was attached to the non-dominant hand of the subjects using a wristband. There were two cables attached to the wristband. These cables connected the two measurement electrodes to the sensor. The electrodes were structural non-woven electrodes with special gel/solid gel and a diameter of 55\,mm, which was cut to size if necessary. The electrodes were glued to the palm of the hand below the little finger so that the gel surfaces of the electrodes did not overlap. If the electrodes did not hold well, they were fixed with leukotape.
The EDA sensors were always attached by the experimenters and worn for a maximum of one hour. Between different subjects, the sensors were not read. The separation of the data was done in the aftermath by cutting up the individual experiment blocks. The sensor number and the subject number for the day were noted and thus the data of the sensor and the remaining data of the subjects could be linked.
EDA data was recorded in runs listed in Table\,\ref{tab:eda}. 

\subsection{Heart Rate Variability}
\label{subsec:HeartRateVariability}
Movisens ambulatory sensors (ECG Move4) \cite{MovisensEcgMove4UserManual} were used for heart rate measurements. A total of 28\,sensors were used, which were activated every morning and their data saved every evening. The ECG Move records the heart rate with a resolution of 12\,bit. The input range is 560\,mV (CM), $\pm$\,mV (DM) and 3\,db bandwidth from 1.6\,-\,33\,Hz. The sampling rate is 1024\,Hz. In addition to the ECG sensor, the ECG Move contains a number of other sensors. These include a 3D-acceleration sensor, which records with 64\,Hz and has a measuring range of $\pm$\,16\,g, and a rotation rate sensor with a measuring range of $\pm$\,2000\,dps and a resolution of 70\,mdps, with an output rate of 64\,Hz. It also has a pressure sensor with a range of 300\,-\,1100\,hPa at a resolution of 0.03\,hPa, a sampling rate of 8\,Hz and a temperature sensor that measures ambient temperature at a frequency of 1\,Hz. 

\begin{table}[h!]
\centering
	\caption{Experiment runs in which HRV sensors were used. Experimental configurations of runs are shown in the appendix.}
	\label{tab:hrv}
	{\small
	\begin{tabular}{cllll}
	    \toprule
		\textbf{day} &
		  \multicolumn{1}{c}{\textbf{\begin{tabular}[c]{@{}c@{}}Tiny Box\\ Experiment\end{tabular}}} &
		  \multicolumn{1}{c}{\textbf{\begin{tabular}[c]{@{}c@{}}Crowd Management \\ Experiment\end{tabular}}} &
		  \multicolumn{1}{c}{\textbf{\begin{tabular}[c]{@{}c@{}}Oval\\ Experiment\end{tabular}}} &
		  \multicolumn{1}{c}{\textbf{\begin{tabular}[c]{@{}c@{}}Personal Space\\ Experiment\end{tabular}}} \\ \midrule
		\textbf{1} &
		  \begin{tabular}[c]{@{}l@{}}1 -- 85\end{tabular} &
		  \begin{tabular}[c]{@{}l@{}}--\end{tabular} &
		  -- &  
		  --\\ 
		\textbf{2} &
		  \begin{tabular}[c]{@{}l@{}}86 --170\end{tabular} &
		  \begin{tabular}[c]{@{}l@{}}--\end{tabular} &
		  -- & 
		  --\\ 
		\textbf{3} & 
		  \begin{tabular}[c]{@{}l@{}}171 -- 204\end{tabular} &
		  \begin{tabular}[c]{@{}l@{}}3C070, 3C080, \\3C081, 3C090, \\3C100, 3C101, \\3C110, 3C120, \\3C121\end{tabular} &
		    \begin{tabular}[c]{@{}l@{}}--\end{tabular} &
		    --\\ 
		\multicolumn{1}{c}{\textbf{4}} &
		  -- &
		  -- &
		 \begin{tabular}[c]{@{}l@{}}single\_file\_gender\_2\_6, \\single\_file\_gender\_2\_11,\\single\_file\_gender\\ \hfill\_random\_2\_6,\\ single\_file\_gender\\ \hfill\_random\_2\_11\\  \end{tabular}&
		  \begin{tabular}[c]{@{}l@{}}4C1010, 4C1020,\\ 4C1030, 4C1040, \\4C2010, 4C2020, \\4C2030, 4C2040\end{tabular}
		   \\ \bottomrule
	\end{tabular}%
	}
\end{table} 

\noindent The sensor was placed below the chest using disposable electrodes. The electrodes contained a highly conductive wet gel and a high quality Ag/AgCl sensor. They had a decentred connection to reduce motion artifacts.
The ECG sensors were frequently attached by the subjects themselves. Between different subjects, the sensors were not read. The separation of the data was performed by cutting up the individual experimental blocks. The sensor number and subject number for the day were noted and thus the sensor data and the rest of the subject data could be linked.
Heart rate data was recorded in runs listed in Table\,\ref{tab:hrv}. 

\subsection{Pressure}
\label{subsec:Pressure}
During the bottleneck experiments on day\,4 (Sec.\,\ref{subsec:BottleneckExperiments}), two pressure sensors from Tekscan (Pressure Mapping Sensor 5400N \cite{TekscanPMS5400N}) were employed to estimate normal forces within a crowd.
Each sensor consists of 1768\,measurement cells covering an area of 57.8\,cm $\times$  88.4\,cm.
Before the actual data recording, the sensors must be calibrated.
For this purpose, the sensor was placed horizontally on a table and successively loaded with 10\,kg, 20\,kg, 30\,kg, 40\,kg, 50\,kg, 95\,kg, 110\,kg, or 120\,kg in total.
Corresponding pressure values were measured with a sensitivity of S-40 and used for a multi-point calibration.\\

\begin{figure}[t!]
 	\centering
	\subfigure[]{\includegraphics[width=0.48\textwidth]{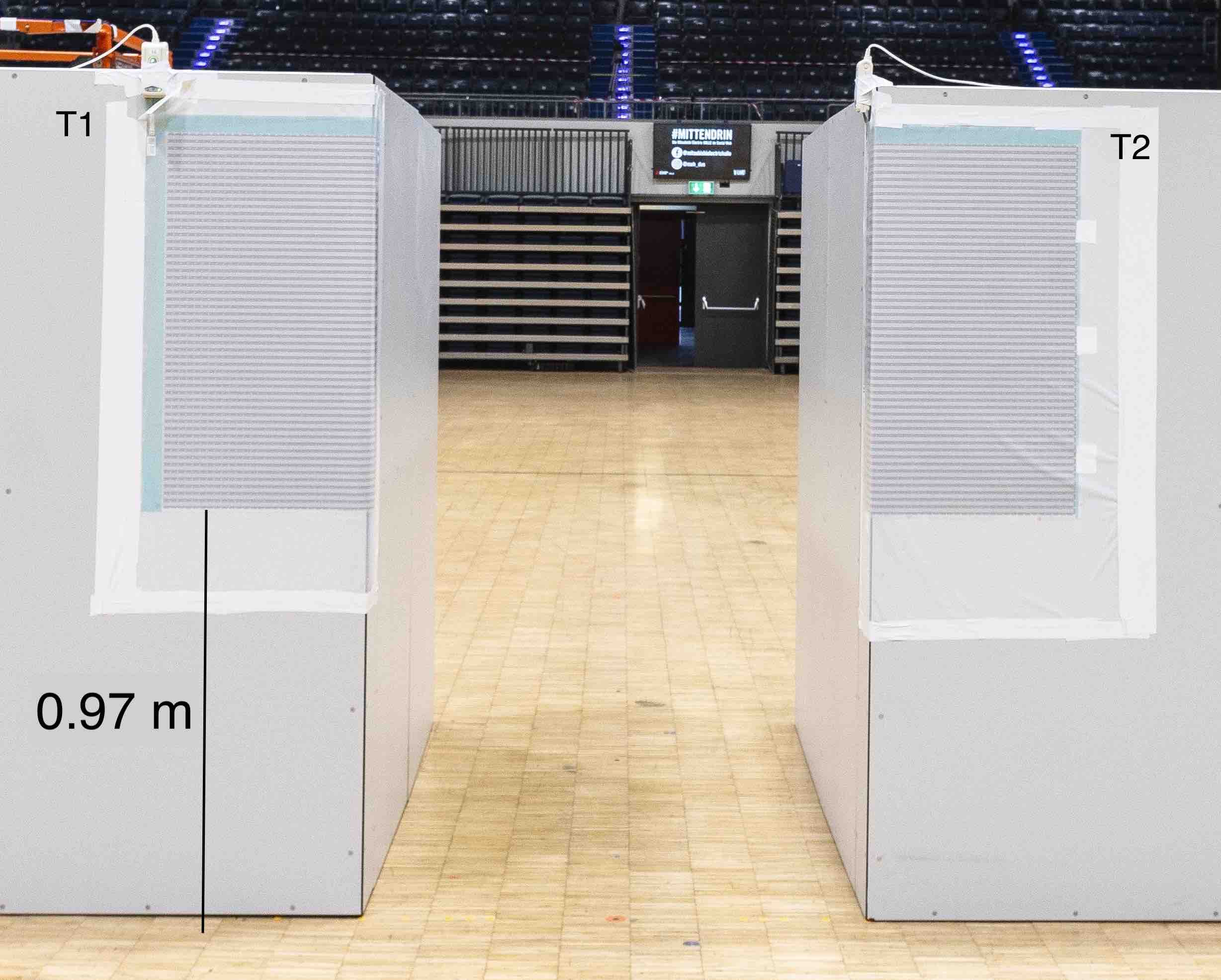}}
	\subfigure[]{\includegraphics[width=0.48\textwidth]{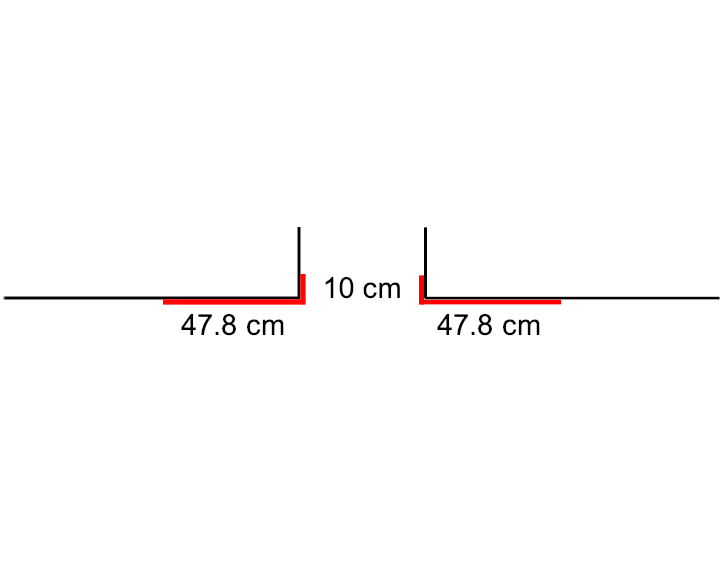}}
	\caption{Attachment of the pressure sensors to the walls of the bottleneck with a) front view and b) top view.}
	\label{fig:pressure_bottleneck}
\end{figure}

\noindent On either side of the bottleneck, a pressure sensor was attached vertically at a height of 0.97\,m for the lower edge (\fref{fig:pressure_bottleneck}\,a).
The short side of the sensor was bent around the corner to place 10\,cm of the measurement area inside the bottleneck and 47.8\,cm in front of it (\fref{fig:pressure_bottleneck}\,b).
Teflon foil was spread over the pressure sensors to reduce shear forces and ensure a secure attachment.
Each sensor was connected to a laptop recording pressure with the I-Scan software using a sampling rate of 60\,fps.

\begin{figure}[h!]
 	\centering
    \includegraphics[width=0.48\textwidth]{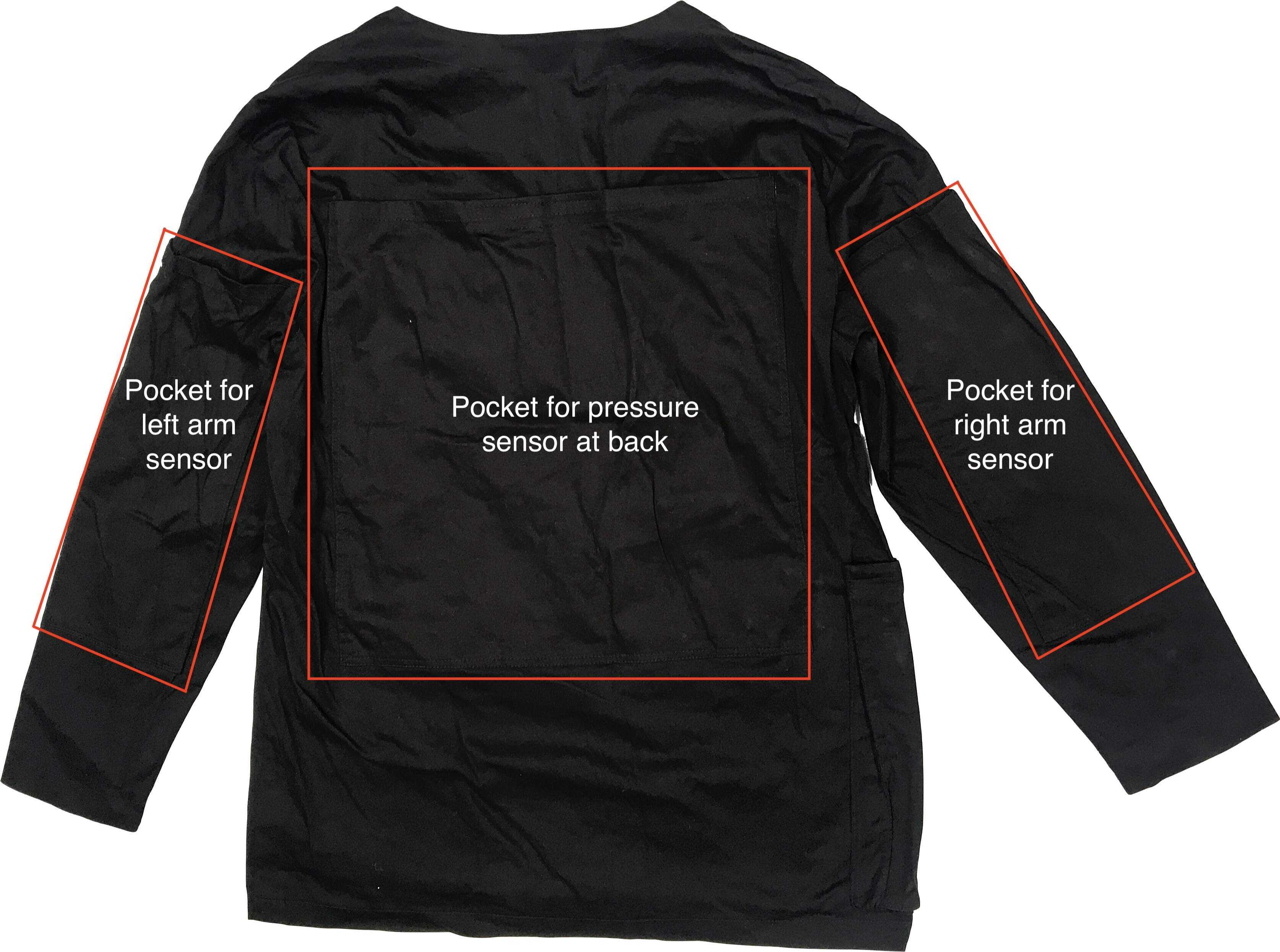}
    \includegraphics[width=0.25\textwidth]{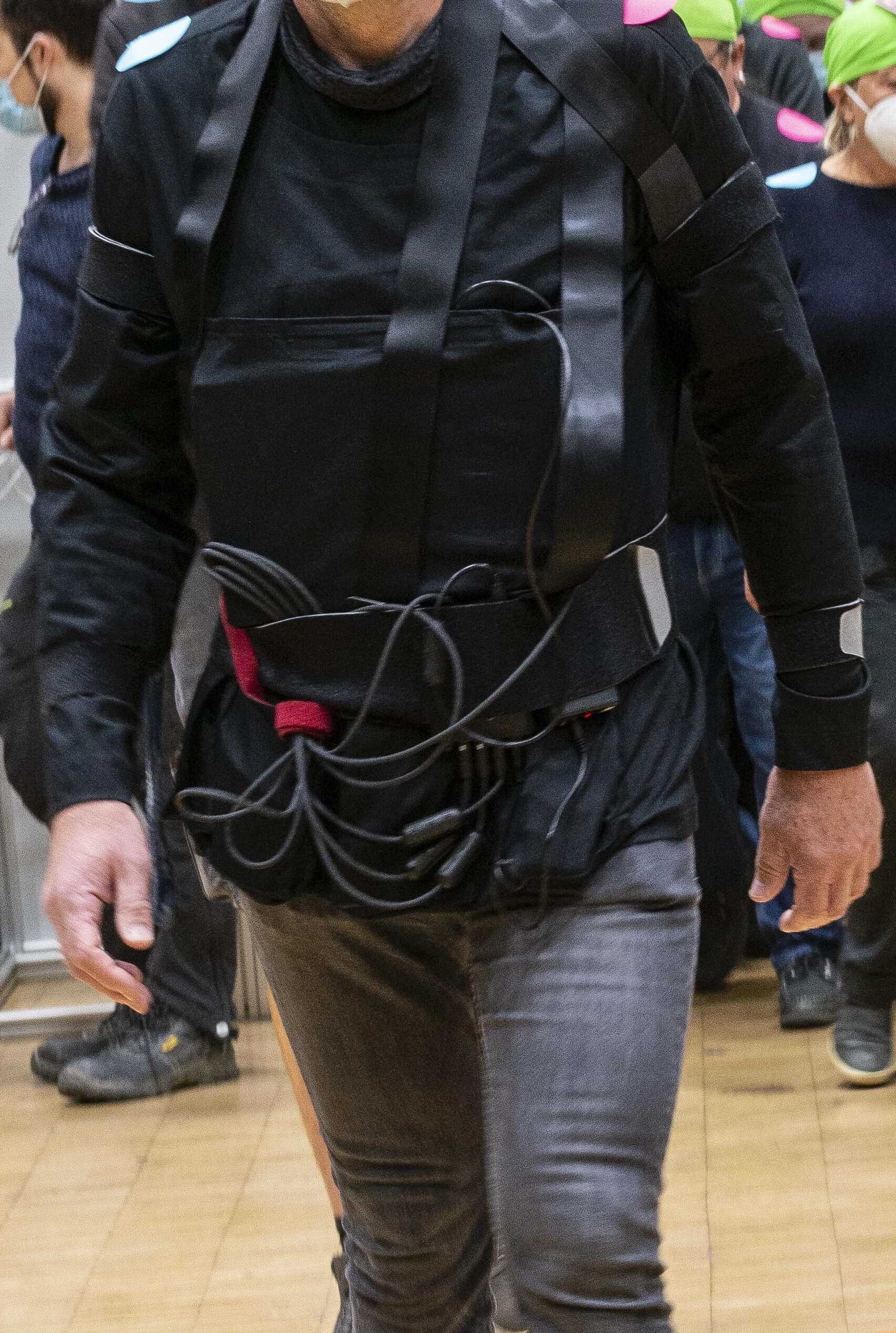}
    \caption{Wearable T-shirts with pockets for the pressure sensors. a) Back view of the T-shirt. Pressure sensors can be stored in the pockets for the left arm, right arm and the back. b) Participant during an experiment. The tablet is stored in the chest pocket.}
	\label{fig:pressure_shirt}
\end{figure}
\noindent Furthermore, two participants were equipped with flexible pressure sensors on their body, each with two upper arm sensors and one sensor for the back.
Xsensor LX210:50.50.05 \cite{XsensorLX210:50.50.05} has 2500 measuring cells providing pressure measurement on an area of 25.4\,cm\,$\times$ 25.4\,cm on the participant's back.
The arm sensor (Xsensor LX210:25.50.05 \cite{XsensorLX210:25.50.05}) covers an area of 12.7\,cm\,$\times$ 25.4\,cm with 1250 measuring cells.
All sensors were calibrated in advance by the manufacturer resulting in a pressure range of 0.14\,Ncm$^{-2}$ -- 10.3\,Ncm$^{-2}$.\\

\noindent For easy wearing, the three pressure sensors were tucked into designated pockets of a specific T-shirt (Fig.\,\ref{fig:pressure_shirt}) and connected to a tablet. 
The tablet, which was carried in the chest pocket throughout the experiments, used the Software Xsensor Pro V8 to capture pressure at a sampling rate of 25\,fps.
In order to receive as much pressure as possible at the central part of the back, the volunteers who wore the shirts were 1.91\,m and 2.04\,m tall. Unfortunately, no pressure data from the Xsensor sensors in the T-shirts were recorded during the experiments.
\subsection{3D-Motion Capturing}
\label{subsec:MoCap}
\noindent We used the 3D motion capturing (MoCap) system MVN Link by Xsens to track the full body motion of a person in the crowd \cite{Schepers2018a}. While optical MoCap Systems need a free line of sight between the tracking points on the body and a set of cameras, the Xsens MoCap system uses inertial measurement units (IMU) as sensors. These IMUs measure the acceleration, the angular rate and the magnetic field strength and a line of sight between the body and a camera is not necessary. Therefore, it is possible to capture the full body motion even in dense crowds.\\

\begin{figure}[h]
	\centering
	\includegraphics[width=0.2\textwidth]{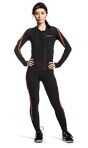}
	\includegraphics[width=0.48\textwidth]{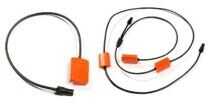}
	\caption{The Xsens MVN Link lycra suit (left) and the associated IMU sensors (right) (reproduced Fig. 2 and 13 from \cite{Xsens2022}).}
	\label{fig:xsens1}
\end{figure}	

\noindent Each MVN Link suit (\fref{fig:xsens1}) is equipped with 17\,IMU sensors on predefined independently moving body segments. The measurement can be triggered manually and the recorded data is stored locally on a body-pack in the suit. Thus, the measurement is self-contained and the data can be downloaded afterwards.\\
After doing a calibration procedure and taking detailed body dimension measurements, the MVN Analyze software then calculates the full body motion based on a biomechanical model from the measured data set. The processed data includes the orientation, position, velocity, acceleration, angular velocity and angular acceleration of each body segment as well as the angles of joints and the location of the centre of mass. The data can be exported either as xml file or as biomechanical c3d file. \\
%

\noindent Because the IMU based motion capturing is self-sufficient and based on relative measurements only, the absolute positioning in space suffers from a drift which can accumulate over time. The head trajectories extracted from camera recordings, however, have a small positioning error. Therefore, we used a hybrid tracking algorithm \cite{Boltes2021} to combine both data sets. In particular, that means that the position of the biomechanical model was shifted and rotated to match the head position and orientation of the camera trajectories.\\

\noindent On all days, we equipped 20\,people with an Xsens MVN Link Motion Capturing system. On experiment days 1-3, these persons were part of the red group (Fig.\,\ref{fig:timeschedule}), and on day\,4 they took part on experiment site\,D, namely the bottleneck experiment. 3D motion capturing data were therefore recorded in runs listed in Table\,\ref{tab:mocap}.
\begin{table}[h!]
\centering	
\caption{Experiment runs in which participants took part, that were equipped with MoCap suits. Experimental configurations of runs are shown in the appendix.}
	\label{tab:mocap}
	{\small
	\begin{tabular}{cllll}
		\toprule
		\textbf{day} &
		  \multicolumn{1}{c}{\textbf{\begin{tabular}[c]{@{}c@{}}Train Platform\\ Experiment\end{tabular}}} &
		  \multicolumn{1}{c}{\textbf{\begin{tabular}[c]{@{}c@{}}Crowd Management \\ Experiment\end{tabular}}} &
		  \multicolumn{1}{c}{\textbf{\begin{tabular}[c]{@{}c@{}}Boarding and Alighting\\ Experiment\end{tabular}}} &
		  \multicolumn{1}{c}{\textbf{\begin{tabular}[c]{@{}c@{}}Bottleneck\\ Experiment\end{tabular}}} \\ \midrule
		\textbf{1} &
		  \begin{tabular}[c]{@{}l@{}}1B110, 1B120, \\ 1B130\end{tabular} &
		  \begin{tabular}[c]{@{}l@{}}1C030, 1C040, \\ SOLO\_REF11, \\ 1C100, 1C110\end{tabular} &
		   D1\_3\_* , D1\_6\_*   &
		   --\\ 
		\textbf{2} &
		  \begin{tabular}[c]{@{}l@{}}2B110, 2B120, \\ 2B130, 2B131\end{tabular} &
		  \begin{tabular}[c]{@{}l@{}}2C040, 2C050, \\ SOLO\_REF21, \\ 2C110, 2C120, \\ 2C130\end{tabular} &
		   D2\_3\_* , D2\_6\_* &
		   --\\
		\textbf{3} &
		  \begin{tabular}[c]{@{}l@{}}3B110, 3B120, \\ 3B130, 3B131\end{tabular} &
		  \begin{tabular}[c]{@{}l@{}}3C030, 3C040, \\ SOLO\_REF31, \\ 3C090, 3C100, \\ 3C101\end{tabular} &
		  D3\_3\_* , D3\_6\_* &
		  --\\ 
		\multicolumn{1}{c}{\textbf{4}} &
		  -- &
		  -- &
		  -- &
		  4D000 -- 4D340
		   \\ \bottomrule
	\end{tabular}%
    }
\end{table}

%
\subsection{Mood-Buttons}
\label{subsec:moodbutton}
\begin{figure}[h!]
	\centering
	\includegraphics[width=0.8\textwidth]{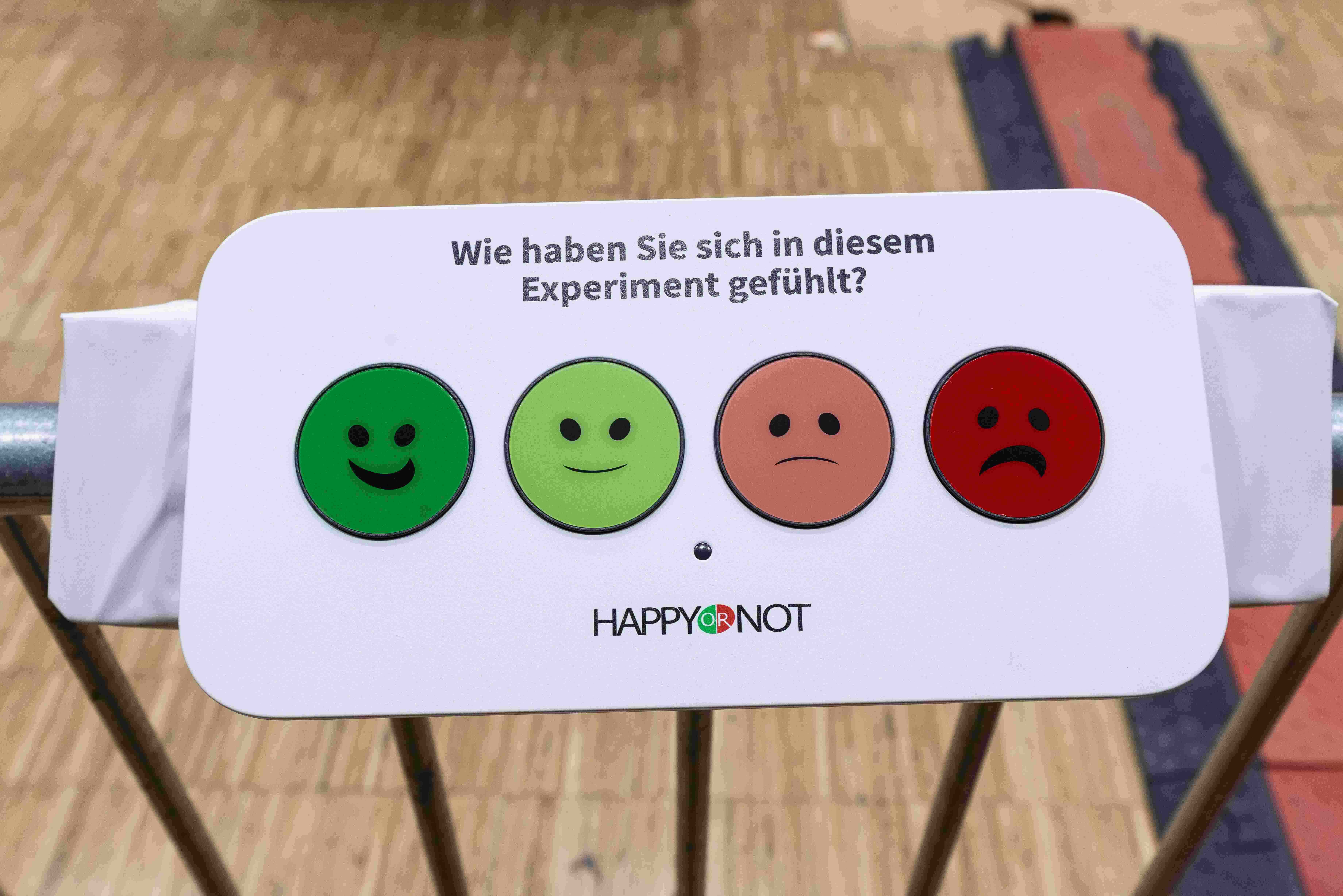}
	\caption{Terminal showing Mood Buttons placed in some experiments to compare the well-being of the participants between the experimental runs. The question on top translates as ``How did you feel in the last run?" from German. }
	\label{fig:moodbutton}
\end{figure}

In order to be able to classify the mood of the test subjects over the course of the day in the individual experiments, we installed simple mood button terminals (Happy-or-not \cite{HappyOrNot}). The terminals consist of four smiley-faced buttons with a sign saying ``how did you feel in the last run?" (Fig.\ref{fig:moodbutton}) that participants were invited to press after every run. The system saved a time stamp for each pressed button. 

\noindent The terminals were attached to man-rails and placed so that participants passed them after each run. Care was taken to ensure that the grids were positioned in such a way that the walking path was affected as little as possible and no backlog was created. Participants were actively asked to press a button after each run. 
In the Train Platform Experiments, the terminal was placed in the corridor leading participants from the area where they filled out questionnaires back to the waiting area in front of the experiment. In the Crowd Management Experiments, the terminal was positioned 15\,m after passing the entry gate on the way back to the line-up area. In the Boarding and Alighting Experiments, the terminal was placed next to the waiting area (behind the train car for runs marked as `reversedirection'). In the Bottleneck Experiments, one terminal was placed at each side of the bottleneck. Participants passed the terminal on their way back to the line-up area, regardless of whether they turned right or left after passing the finish line. No mood buttons were placed in the Tiny House, Oval or Personal Space Experiments.

\section{Summary and Discussion}
This paper presents pedestrian experiments conducted as part of the CroMa project aimed at increasing the robustness and efficiency of transport infrastructure.
Even though the planning and execution of large-scale experiments requires far-reaching planning and organizational steps that go far beyond the scientific content, experiments under laboratory conditions offer the opportunity to control factors and can therefore be worth the effort involved.

\noindent This publication provides an overview of the individual experiments carried out as well as descriptions of sensor techniques applied, as the contents and goals of the experiments were planned and evaluated by different disciplines and had to be coordinated and combined with each other. Furthermore, it presents the context in which the individual experimental runs and experimental sites were intertwined. The results of the scientific analyses will be published in subsequent content papers.  \\

\noindent Even though the experiments took place during a global pandemic, the questionnaire results as well as the evaluation of well-being during the experiments (mean value of mood buttons over all days) show that the overall concept of communication, hygiene and safety measures as well as the slow acclimatization to density (queuing, measurement course, waiting area, icebreaker) led to the participants feeling confident. As a result, they felt good and the thought of a potential infection seemed to have no meaningful influence on their actions. This is consistent with the impressions of the organizers regarding the mood of the subjects during the experiments. \\

\noindent For each of the conducted experiments, the goal of the study is described along with which parameters were varied, how participants were approached and which dimensions the experimental areas and geometries had. The description is supplemented by impressions of the experiments given through sketches and snapshots. \\
\noindent In the chapters about the sensors, the technical specifications are listed. Furthermore, it is documented how the sensors were synchronized with each other, how many of the sensors were used, with which settings they were operated and which basic processing steps were carried out if necessary. For each sensor there is an overview of the runs in which the sensors were used. 
For each experiment, a link to the data archive is given under which the respective complete data will be made freely available after publication of the respective content paper. \\

\begin{acknowledgements}
We thank all staff members of Forschungszentrum Jülich, University of Wuppertal, Ruhr-Universität Bochum and D.LIVE for their valuable help during the experimental days, without whom it would not have been possible to fulfil all tasks and ensure a smooth process on all days.\\

\noindent \textbf{Ethical Review} The application of ethical approval for the experiments ``Crowd Management'', ``Single-File'', ``Personal-Space'', ``Train Platform'' and ``Boarding and Alighting'' were submitted by A.\,Sieben to the German Psychological Society (DGPs, the Society) and approved in December 2019 (file reference \mbox{SiebenAnna2019-10-22VA}). The ``Bottleneck'' experiment was submitted to the ethical review committee of the University of Wuppertal (German: Bergische Universität Wuppertal) by A.\,Seyfried and was approved in January 2020 (file reference \mbox{MS/BBL 191213 Seyfried}).\\

\noindent \textbf{Funding} The experiments were financially supported by the German Federal Ministry of Education and Research (BMBF) within the project  CroMa (Crowd Management in Verkehrsinfrastrukturen / Crowd Management in transport infrastructures) under grant number 13N14530 to 13N14533 and by the European Union’s Horizon 2020 research and innovation program within the project CrowdDNA under grant agreement number 899739.
\end{acknowledgements}
\begin{contributions}
Juliane Adrian: technical responsibility 3D-motion capturing system, scientific content planning bottleneck experiment, software combination of trajectories, data curation (3D-motion capturing system, manual correction trajectories, combination of trajectories), writing (original draft chapter 4.6)\,/ 
Maik Boltes: supervision, technical planning camera, data curation camera, responsible for daily briefing and adherence of safety measures, scientific content planning bottleneck experiment, writing (original draft chapter 1, review and editing)\,/
Ann Katrin Boomers: coordinating and technical realization of experiments, technical planning camera, data curation (camera, processing to trajectories, manual correction trajectories, combination of trajectories), scientific content planning bottleneck experiment, writing (original draft chapter 2 - 4.2, 4.7, 5)\,/
Mira Beermann: technical responsibility EDA/HRV sensors, data curation EDA/HRV, scientific content planning tiny box experiment, boarding-alighting experiment and single-file experiment, writing (original draft chapter 4.3 + 4.4)\,/
Mohcine Chraibi: scientific content planning single-file\,/
Sina Feldmann: technical responsibility pressure sensors, data curation (pressure sensors, manual correction trajectories), scientific content planning bottleneck, writing (original draft chapter 4.5)\,/
Frank Fiedrich: scientific content planning visual inspection of density\,/
Niklas Frings: technical support data collection visual inspection of density\,/
Arne Graf: technical support construction and testing camera system\,/
Alica Kandler: technical responsibility camera and time synchronization, data curation (camera, processing to trajectories), writing (original draft chapter 4.1 + 4.2)\,/ 
Deniz Kilic: software recruiting process\,/ 
Krisztina Konya: scientific content planning train platform experiment, data curation (manual correction trajectories), writing (review and editing)\,/
Mira Küpper: scientific content planning train platform experiment and boarding-alighting experiment, data curation (manual correction trajectories), writing (review and editing)\,/
Andreas Lotter: scientific content planning visual inspection of density\,/
Helena Lügering: scientific content planning bottleneck experiment, data curation (manual correction trajectories)\,/
Francesca Müller: scientific content planning visual inspection of density\,/
Sarah Paetzke: scientific content planning single-file experiment, data curation (manual correction trajectories)\,/
Anna-Katharina Raytarowski: testing pressure sensor system\,/
Olga Sablik: scientific content planning crowd management experiment, data curation (manual correction trajectories), writing (review and editing)\,/
Tobias Schrödter: technical support construction\,/
Armin Seyfried: supervision, scientific content planning all experiments, writing (review and editing)\,/
Anna Sieben: supervision, scientific content planning all experiments, writing (review and editing)\,/
Ezel Üsten: scientific content planning crowd management experiment (interruption), data curation (HRV sensors, manual correction trajectories)\,/
\end{contributions}


\newpage
\appendix
\pretocmd{\section}{%
  \pagenumbering{arabic}%
  \renewcommand*{\thepage}{\thesection\arabic{page}}%
}{}{}

\begin{landscape}
\section{Appendices}
\label{Appendices}

\subsection{Test Person Sample: Statistics per Day}
\label{app:testperson_stats}
\begin{figure}[h!]
	\centering
 	\includegraphics[width=1.1\textwidth]{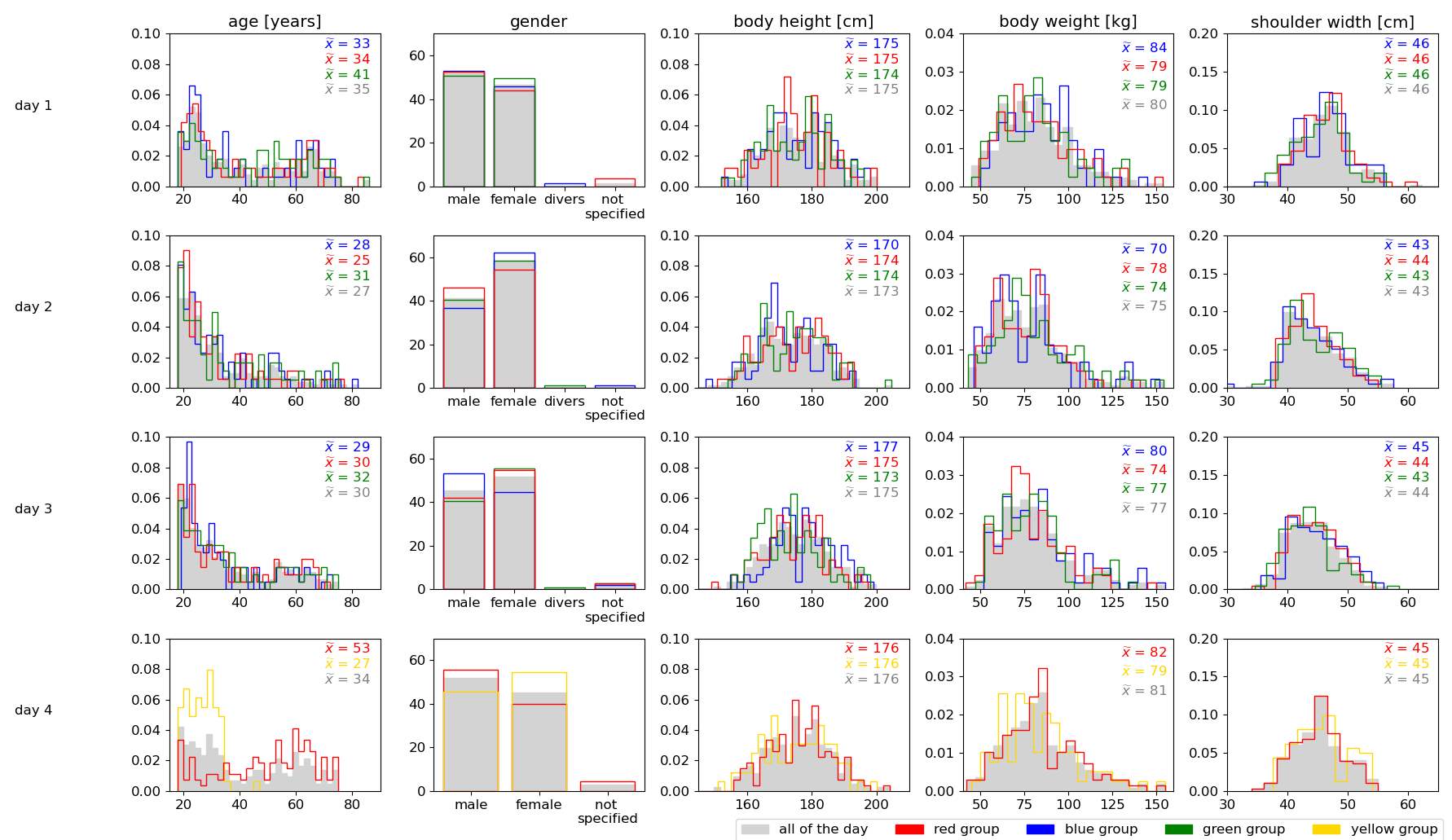}
 	 \caption{Panel of different histograms showing demographic data of the participants for each day. Row one refers to data of participants on day 1, row 2 to day 2, row 3 to day 3 and row 4 to day 4. Column 1 shows age, column 2 gender, column 3 body height, column 4 body weight and column 5 shoulder width. Data shown in grey includes data from all participants of the respective day, data in green data of participants belonging to the green experimental group and data in red, blue and yellow to data of participants belonging to experimental group of the respective color accordingly. Respective Medians are shown in the same color as the data.}
  	\label{fig:testperson_sample_appendix} 
\end{figure}
\end{landscape}


\begin{table} 
\subsection{Covid-Related Questionnaire Data}
\label{app:covidStatistics}
\small
\centering
\caption{Descriptive statistics of Covid-related questionnaire data.}
\label{tab:covidStatistics}
\begin{tabular}{l c c c c c }
\toprule
 & N & Min. & Max. & \begin{tabular}[c]{@{}c@{}}Mean value\\M\end{tabular} & \begin{tabular}[c]{@{}c@{}}Standard \\deviation\end{tabular} \\ \midrule
%
%
\begin{tabular}[l]{@{}l@{}}I found it...\\ ...uncomfortable to be in a crowd \\ \,\,\,during the experiments.\end{tabular} & 992 & 1 & 7 & 2,61 & 1,748 \\
\begin{tabular}[l]{@{}l@{}}...uncomfortable because of crowding.\end{tabular} & 937 & 1 & 7 & 3,37 & 2,148 \\
\begin{tabular}[l]{@{}l@{}}...uncomfortable because of Covid \\\,\,\,infection concerns.\end{tabular} & 940 & 1 & 7 & 2,45 & 1,879 \\
\begin{tabular}[l]{@{}l@{}}...uncomfortable because of infection \\ \,\,\,concerns with other illness.\end{tabular} & 940 & 1 & 7 & 2,19 & 1,693 \\
\begin{tabular}[l]{@{}l@{}}...uncomfortable because of unclear \\ \,\,\,instructions.\end{tabular} & 938 & 1 & 7 & 2,34 & 1,679 \\
\begin{tabular}[l]{@{}l@{}}...uncomfortable because of physical \\ \,\,\,exertion.\end{tabular} & 946 & 1 & 7 & 3,49 & 2,068 \\
\begin{tabular}[l]{@{}l@{}}...uncomfortable because of stress \\ \,\,\,caused by the experimenter.\end{tabular} & 940 & 1 & 7 & 1,80 & 1,378 \\
\begin{tabular}[l]{@{}l@{}}...uncomfortable because of being around \\ \,\,\,many people.\end{tabular} & 941 & 1 & 7 & 2,92 & 1,975 \\
\begin{tabular}[l]{@{}l@{}}I have often thought about Covid.\end{tabular} & 999 & 1 & 7 & 2,20 & 1,690 \\
\begin{tabular}[l]{@{}l@{}}I have often worried about catching Covid.\end{tabular} & 997 & 1 & 7 & 1,98 & 1,559 \\
\begin{tabular}[l]{@{}l@{}}I would have behaved differently before \\ \,\,\,the pandemic.\end{tabular} & 996 & 1 & 7 & 2,69 & 1,959 \\
\begin{tabular}[l]{@{}l@{}}I have also been in crowds elsewhere \\ \,\,\,since the pandemic started.\end{tabular} & 994 & 1 & 7 & 4,02 & 2,361\\
\bottomrule
\end{tabular}
\end{table}
\normalsize

\begin{landscape}
\subsection{Experimental Configurations Train Platform Experiments}
\label{app:Trainplatform}
\tiny
\begin{longtable}{llllllllllll}
\caption{Table showing experimental runs performed within the scope of the Train Platform Experiments (Sec.\,\ref{subsec:TrainPlatformExperiments}). The names of the experiments are given together with the colour of the group that participated in the respective runs as well as varied parameters.}\\
		\multicolumn{1}{l|}{} &  & \begin{tabular}[c]{@{}l@{}}run name\\ (short)\end{tabular} & \begin{tabular}[c]{@{}l@{}}run name\\ (descriptive)\end{tabular} & \begin{tabular}[c]{@{}l@{}}no. of \\ participants\end{tabular} & \begin{tabular}[c]{@{}l@{}}setup 1:\\ building constructions\\ \\ {[}none (-); wall;\\ house; marked areas{]}\end{tabular} & \begin{tabular}[c]{@{}l@{}}setup 2: \\ stairs present\\ at beginning \\ of run \\ {[}yes; no{]}\end{tabular} & \begin{tabular}[c]{@{}l@{}}inflow \\ participants\end{tabular} & \begin{tabular}[c]{@{}l@{}}waiting time \\ on platform\\ {[}mm:ss{]}\end{tabular} & \begin{tabular}[c]{@{}l@{}}induced anonymity\\ {[}none (-); \\ anonym;\\ group identity{]}\end{tabular} & \begin{tabular}[c]{@{}l@{}}special announcements\\ {[}none (-);\\ no speaking; \\ speaking allowed{]}\end{tabular} & \begin{tabular}[c]{@{}l@{}}questionnaire\\ {[}yes; no{]}\end{tabular} \\ \hline
		\endfirsthead
		\multicolumn{12}{c}%
		{{\bfseries Table \thetable\ continued from previous page}} \\
		\multicolumn{1}{l|}{} &  & \begin{tabular}[c]{@{}l@{}}run name\\ (short)\end{tabular} & \begin{tabular}[c]{@{}l@{}}run name\\ (descriptive)\end{tabular} & \begin{tabular}[c]{@{}l@{}}no. of \\ participants\end{tabular} & \begin{tabular}[c]{@{}l@{}}setup 1:\\ building constructions\\ \\ {[}none (-); wall;\\ house; marked areas{]}\end{tabular} & \begin{tabular}[c]{@{}l@{}}setup 2: \\ stairs present\\ at beginning \\ of run \\ {[}yes; no{]}\end{tabular} & \begin{tabular}[c]{@{}l@{}}inflow \\ participants\end{tabular} & \begin{tabular}[c]{@{}l@{}}waiting time \\ on platform\\ {[}mm:ss{]}\end{tabular} & \begin{tabular}[c]{@{}l@{}}induced anonymity\\ {[}none (-); \\ anonym;\\ group identity{]}\end{tabular} & \begin{tabular}[c]{@{}l@{}}special announcements\\ {[}none (-);\\ no speaking; \\ speaking allowed{]}\end{tabular} & \begin{tabular}[c]{@{}l@{}}questionnaire\\ {[}yes; no{]}\end{tabular} \\ \hline
		\endhead
		\hline
		\endfoot
		\endlastfoot
		 & \cellcolor[HTML]{FF0000} & 1B010 & ano\_no-speak\_small & 10 & - & yes & all at once & 05:00 & anonym & no speaking & yes \\
		 & \cellcolor[HTML]{FF0000} & 1B020 & ano\_speak\_small & 10 & - & yes & all at once & 05:00 & anonym & speaking allowed & yes \\
		 & \cellcolor[HTML]{FF0000} & 1B030 & ano\_no-speak\_large & 50 & - & yes & all at once & 05:00 & anonym & no speaking & yes \\
		 & \cellcolor[HTML]{FF0000} & 1B040 & ano\_speak\_large & 50 & - & yes & all at once & 05:00 & anonym & speaking allowed & yes \\
		 & \cellcolor[HTML]{92D050} & 1B050 & blank\_Aa & 40 & - & no & every 2-3 sec & 02:00 & - & - & - \\
		 & \cellcolor[HTML]{92D050} & 1B060 & blank\_Ab & 40 & - & no & every 2-3 sec & 04:00 & - & - & - \\
		 & \cellcolor[HTML]{92D050} & 1B070 & blank\_Ba & 85 & - & no & every 2-3 sec & 02:00 & - & - & yes \\
		 & \cellcolor[HTML]{5B9BD5} & 1B080 & wall\_Aa & 40 & wall & no & every 2-3 sec & 02:00 & - & - & yes \\
		 & \cellcolor[HTML]{5B9BD5} & 1B090 & wall\_Ab & 35 & wall & no & every 2-3 sec & 04:00 & - & - & yes \\
		 & \cellcolor[HTML]{5B9BD5} & 1B100 & wall\_Ba & 83 & wall & no & every 2-3 sec & 02:00 & - & - & - \\
		 & \cellcolor[HTML]{FF0000} & 1B110 & house\_Aa & 40 & house & no & every 2-3 sec & 02:00 & - & - & yes \\
		 & \cellcolor[HTML]{FF0000} & 1B120 & house\_Ab & 38 & house & no & every 2-3 sec & 04:00 & - & - & yes \\
		 & \cellcolor[HTML]{FF0000} & 1B130 & house\_Ba & 82 & house & no & every 2-3 sec & 02:00 & - & - & - \\
		 & \cellcolor[HTML]{92D050} & 1B140 & SI\_no-speak\_small & 10 & - & yes & all at once & 05:00 & group identity & no speaking & yes \\
		 & \cellcolor[HTML]{92D050} & 1B150 & SI\_speak\_small & 10 & - & yes & all at once & 05:00 & group identity & speaking allowed & yes \\
		 & \cellcolor[HTML]{92D050} & 1B160 & SI\_no-speak\_large & 50 & - & yes & all at once & 05:00 & group identity & no speaking & yes \\
		 & \cellcolor[HTML]{92D050} & 1B170 & SI\_speak\_large & 50 & - & yes & all at once & 05:00 & group identity & speaking allowed & yes \\
		 & \cellcolor[HTML]{5B9BD5} & 1B180 & blank\_Aa\_inflow10 & 40 & - & no & groups of 10 & 02:00 & - & - & - \\
		 & \cellcolor[HTML]{5B9BD5} & 1B190 & blank\_Ab\_inflow10 & 40 & - & no & groups of 10 & 04:00 & - & - & - \\
		\multirow{-20}{*}{day 1} & \cellcolor[HTML]{5B9BD5} & 1B200 & blank\_Ba\_inflow10 & 82 & - & no & groups of 10 & 02:00 & - & - & yes \\ \hline
		 & \cellcolor[HTML]{FF0000} & 2B010 & ano\_no-speak\_small & 10 & - & yes & all at once & 05:00 & anonym & no speaking & yes \\
		 & \cellcolor[HTML]{FF0000} & 2B020 & ano\_speak\_small & 10 & - & yes & all at once & 05:00 & anonym & speaking allowed & yes \\
		 & \cellcolor[HTML]{FF0000} & 2B030 & ano\_no-speak\_large & 50 & - & yes & all at once & 05:00 & anonym & no speaking & yes \\
		 & \cellcolor[HTML]{FF0000} & 2B040 & ano\_speak\_large & 50 & - & yes & all at once & 05:00 & anonym & speaking allowed & yes \\
		 & \cellcolor[HTML]{92D050} & 2B050 & blank\_Aa & 40 & - & no & every 2-3 sec & 02:00 & - & - & - \\
		 & \cellcolor[HTML]{92D050} & 2B060 & blank\_Ab & 40 & - & no & every 2-3 sec & 04:00 & - & - & - \\
		 & \cellcolor[HTML]{92D050} & 2B070 & blank\_Ba & 85 & - & no & every 2-3 sec & 02:00 & - & - & yes \\
		 & \cellcolor[HTML]{92D050} & 2B071 & blank\_C & 145 & - & no & every 2-3 sec & 02:00 & - & - & - \\
		 & \cellcolor[HTML]{5B9BD5} & 2B080 & wall\_Aa & 40 & wall & no & every 2-3 sec & 02:00 & - & - & yes \\
		 & \cellcolor[HTML]{5B9BD5} & 2B090 & wall\_Ab & 40 & wall & no & every 2-3 sec & 04:00 & - & - & yes \\
		 & \cellcolor[HTML]{5B9BD5} & 2B100 & wall\_Ba & 87 & wall & no & every 2-3 sec & 02:00 & - & - & - \\
		 & \cellcolor[HTML]{FF0000} & 2B110 & house\_Aa & 40 & house & no & every 2-3 sec & 02:00 & - & - & yes \\
		 & \cellcolor[HTML]{FF0000} & 2B120 & house\_Ab & 40 & house & no & every 2-3 sec & 04:00 & - & - & yes \\
		 & \cellcolor[HTML]{FF0000} & 2B130 & house\_Ba & 87 & house & no & every 2-3 sec & 02:00 & - & - & - \\
		 & \cellcolor[HTML]{FF0000} & 2B131 & house\_C & 147 & house & no & every 2-3 sec & 02:00 & - & - & - \\
		 & \cellcolor[HTML]{92D050} & 2B140 & SI\_no-speak\_small & 10 & - & yes & all at once & 05:00 & group identity & no speaking & yes \\
		 & \cellcolor[HTML]{92D050} & 2B150 & SI\_speak\_small & 10 & - & yes & all at once & 05:00 & group identity & speaking allowed & yes \\
		 & \cellcolor[HTML]{92D050} & 2B160 & SI\_no-speak\_large & 50 & - & yes & all at once & 05:00 & group identity & no speaking & yes \\
		 & \cellcolor[HTML]{92D050} & 2B170 & SI\_speak\_large & 50 & - & yes & all at once & 05:00 & group identity & speaking allowed & yes \\
		 & \cellcolor[HTML]{5B9BD5} & 2B180 & blank\_Aa\_inflow10 & 50 & - & no & groups of 10 & 02:00 & - & - & - \\
		 & \cellcolor[HTML]{5B9BD5} & 2B190 & blank\_Ab\_inflow10 & 40 & - & no & groups of 10 & 04:00 & - & - & - \\
		\multirow{-22}{*}{day 2} & \cellcolor[HTML]{5B9BD5} & 2B200 & blank\_Ba\_inflow10 & 89 & - & no & groups of 10 & 02:00 & - & - & yes \\ \hline
		 \pagebreak & \cellcolor[HTML]{FF0000} & 3B010 & ano\_no-speak\_small & 10 & - & yes & all at once & 05:00 & anonym & no speaking & yes \\
		 & \cellcolor[HTML]{FF0000} & 3B020 & ano\_speak\_small & 10 & - & yes & all at once & 05:00 & anonym & speaking allowed & yes \\
		 & \cellcolor[HTML]{FF0000} & 3B030 & ano\_no-speak\_large & 50 & - & yes & all at once & 05:00 & anonym & no speaking & yes \\
		 & \cellcolor[HTML]{FF0000} & 3B040 & ano\_speak\_large & 50 & - & yes & all at once & 05:00 & anonym & speaking allowed & yes \\
		 & \cellcolor[HTML]{92D050} & 3B050 & blank\_Aa & 40 & - & no & every 2-3 sec & 02:00 & - & - & - \\
		 & \cellcolor[HTML]{92D050} & 3B060 & blank\_Ab & 40 & - & no & every 2-3 sec & 04:00 & - & - & - \\
		 & \cellcolor[HTML]{92D050} & 3B070 & blank\_Ba & 102 & - & no & \begin{tabular}[c]{@{}l@{}}groups of 20 \\ every 2-3 sec\end{tabular} & 02:00 & - & - & yes \\
		 & \cellcolor[HTML]{92D050} & 3B071 & blank\_C & 180 & - & no & every 2-3 sec & 02:00 & - & \begin{tabular}[c]{@{}l@{}}'red-group-train' first, \\ 'green-group-train' second\end{tabular} & - \\
		 & \cellcolor[HTML]{5B9BD5} & 3B080 & wall\_Aa & 40 & wall & no & every 2-3 sec & 02:00 & - & - & - \\
		 & \cellcolor[HTML]{5B9BD5} & 3B090 & wall\_Ab & 40 & wall & no & every 2-3 sec & 04:00 & - & - & - \\
		 & \cellcolor[HTML]{5B9BD5} & 3B100 & wall\_Ba & 104 & wall & no & \begin{tabular}[c]{@{}l@{}}groups of 20\\ every 2-3 sec\end{tabular} & 02:00 & - & - & yes \\
		 & \cellcolor[HTML]{FF0000} & 3B110 & house\_Aa & 40 & house & no & every 2-3 sec & 02:00 & - & - & - \\
		 & \cellcolor[HTML]{FF0000} & 3B120 & house\_Ab & 40 & house & no & every 2-3 sec & 04:00 & - & - & - \\
		 & \cellcolor[HTML]{FF0000} & 3B130 & house\_Ba & 101 & house & no & every 2-3 sec & 02:00 & - & - & - \\
		 & \cellcolor[HTML]{FF0000} & 3B131 & house\_C & 180 & house & no & \begin{tabular}[c]{@{}l@{}}groups of 20\\ every 2-3 sec\end{tabular} & 02:00 & - & \begin{tabular}[c]{@{}l@{}}'blue-group-train' first, \\ 'red-group-train' second\end{tabular} & yes \\
		 & \cellcolor[HTML]{92D050} & 3B140 & SI\_no-speak\_small & 10 & - & yes & all at once & 05:00 & group identity & no speaking & yes \\
		 & \cellcolor[HTML]{92D050} & 3B150 & SI\_speak\_small & 10 & - & yes & all at once & 05:00 & group identity & speaking allowed & yes \\
		 & \cellcolor[HTML]{92D050} & 3B160 & SI\_no-speak\_large & 50 & - & yes & all at once & 05:00 & group identity & no speaking & yes \\
		 & \cellcolor[HTML]{92D050} & 3B170 & SI\_speak\_large & 50 & - & yes & all at once & 05:00 & group identity & speaking allowed & yes \\
		 & \cellcolor[HTML]{5B9BD5} & 3B180 & blank\_Aa\_inflow10 & 60 & - & no & groups of 10 & 02:00 & - & - & - \\
		 & \cellcolor[HTML]{5B9BD5} & 3B190 & blank\_Ab\_inflow10 & 40 & - & no & groups of 10 & 04:00 & - & \begin{tabular}[c]{@{}l@{}}Track change,\\ right train first\end{tabular} & - \\
		\multirow{-22}{*}{day 3} & \cellcolor[HTML]{5B9BD5} & 3B200 & blank\_Ba\_inflow10 & 107 & - & no & groups of 10 & 02:00 & - & - & yes \\ \hline
\label{tab:siteBd1-3_runs}\\
\end{longtable}
\end{landscape}
\normalsize


\begin{landscape}
\subsection{Experimental Configurations Crowd Management Experiments}
\label{app:CrowdManagement}
\tiny
\begin{longtable}{l|lllllllllll}
\caption{Table showing experimental runs performed within the scope of the Crowd Management Experiments (Sec.\,\ref{subsec:CrowdManagementExperiments}). The names of the experiments are given together with the colour of the group that participated in the respective runs as well as varied parameters.}\\
		 \multicolumn{1}{l|}{} &  & \begin{tabular}[c]{@{}l@{}}run name \\ (short)\end{tabular} & \begin{tabular}[c]{@{}l@{}}run name \\ (descriptive)\end{tabular} & \begin{tabular}[c]{@{}l@{}}no. of \\ participants\end{tabular} & \begin{tabular}[c]{@{}l@{}}Setup 1:\\ structure grid\\ {[}straight; \\ small bend; \\ 90° bend{]}\end{tabular} & \begin{tabular}[c]{@{}l@{}}no. of open \\ gateways\\ {[}1; 3{]}\end{tabular} & \begin{tabular}[c]{@{}l@{}}Setup 2:\\ signposting\\ {[}none (-); \\ lines;\\ signs{]}\end{tabular} & \begin{tabular}[c]{@{}l@{}}motivation\\ {[}low; high{]}\end{tabular} & \begin{tabular}[c]{@{}l@{}}norm \\ specification\\ {[}none (-); \\ 70\%;85\% ; \\ 95\%; 100\%{]}\end{tabular} & \begin{tabular}[c]{@{}l@{}}special \\ announcement\end{tabular} & \begin{tabular}[c]{@{}l@{}}questionnaire\\ {[}yes; no{]}\end{tabular} \\ \hline
		\endfirsthead
		\multicolumn{12}{c}%
		{{\bfseries Table \thetable\ continued from previous page}} \\
		\multicolumn{1}{l|}{} &  & \begin{tabular}[c]{@{}l@{}}run name \\ (short)\end{tabular} & \begin{tabular}[c]{@{}l@{}}run name \\ (descriptive)\end{tabular} & \begin{tabular}[c]{@{}l@{}}no. of \\ participants\end{tabular} & \begin{tabular}[c]{@{}l@{}}Setup 1:\\ structure grid\\ {[}straight; \\ small bend; \\ 90° bend{]}\end{tabular} & \begin{tabular}[c]{@{}l@{}}no. of open \\ gateways\\ {[}1; 3{]}\end{tabular} & \begin{tabular}[c]{@{}l@{}}Setup 2:\\ signposting\\ {[}none (-); \\ lines;\\ signs{]}\end{tabular} & \begin{tabular}[c]{@{}l@{}}motivation\\ {[}low; high{]}\end{tabular} & \begin{tabular}[c]{@{}l@{}}norm \\ specification\\ {[}none (-); \\ 70\%;85\% ; \\ 95\%; 100\%{]}\end{tabular} & \begin{tabular}[c]{@{}l@{}}special \\ announcement\end{tabular} & \begin{tabular}[c]{@{}l@{}}questionnaire\\ {[}yes; no{]}\end{tabular} \\ \hline
		\endhead
		\hline
		\endfoot
		\endlastfoot
		\multicolumn{1}{c|}{} & \cellcolor[HTML]{34CDF9} & 1C010 & entry3\_ lines\_ nM & tba & straight & 3 & lines & low & - &  & yes \\
		\multicolumn{1}{c|}{} & \cellcolor[HTML]{34CDF9} & 1C020 & entry3\_lines\_hM &  & straight & 3 & lines & high & - &  & yes \\
		\multicolumn{1}{c|}{} & \cellcolor[HTML]{FD6864} & 1C030 & entry3\_signs\_nM &  & straight & 3 & signs & low & - &  & yes \\
		\multicolumn{1}{c|}{} & \cellcolor[HTML]{FD6864} & 1C040 & entry1\_norm95\_nM &  & straight & 1 & signs & low & 95\% &  & yes \\
		\multicolumn{1}{c|}{} & \cellcolor[HTML]{FD6864} & SOLO\_REF11 & SOLO\_REF &  & straight &  &  &  &  &  & no \\
		\multicolumn{1}{c|}{} & \cellcolor[HTML]{67FD9A} & 1C050 & entry3\_blank\_nM &  & straight & 3 & - & low & - &  & yes \\
		\multicolumn{1}{c|}{} & \cellcolor[HTML]{67FD9A} & 1C060 & entry1\_blank\_nM &  & straight & 1 & - & low & - &  & yes \\
		\multicolumn{1}{c|}{} & \cellcolor[HTML]{67FD9A} & 1C070 & entry1\_norm85\_nM &  & straight & 1 & - & low & 85\% &  & yes \\
		\multicolumn{1}{c|}{} & \cellcolor[HTML]{34CDF9} & 1C080 & entry3\_bendSmall\_nM &  & small bend & 3 & - & low & - &  & yes \\
		\multicolumn{1}{c|}{} & \cellcolor[HTML]{34CDF9} & 1C090 & entry3\_bendSmall\_hM &  & small bend & 3 & - & high & - &  & yes \\
		\multicolumn{1}{c|}{} & \cellcolor[HTML]{FD6864} & 1C100 & entry3\_straight\_nM &  & straight & 3 & - & low & - &  & yes \\
		\multicolumn{1}{c|}{} & \cellcolor[HTML]{FD6864} & 1C110 & entry3\_straight\_hM &  & straight & 3 & - & high & - &  & yes \\
		\multicolumn{1}{c|}{} & \cellcolor[HTML]{67FD9A} & 1C120 & entry3\_signs\_nM &  & straight & 3 & signs & low & - &  & yes \\
		\multicolumn{1}{c|}{} & \cellcolor[HTML]{67FD9A} & 1C130 & entry3\_signs\_hM &  & straight & 3 & signs & high & - &  & yes \\
		\multicolumn{1}{c|}{\multirow{-15}{*}{day 1}} & \cellcolor[HTML]{67FD9A} & SOLO\_REF12 & SOLO\_REF &  & straight &  &  &  &  &  & no \\ \hline
		 & \cellcolor[HTML]{34CDF9} & 2C010 & entry3\_blank\_nM &  & straight & 3 & - & low & - &  & yes \\
		 & \cellcolor[HTML]{34CDF9} & 2C020 & entry1\_blank\_nM &  & straight & 1 & - & low & - &  & yes \\
		 & \cellcolor[HTML]{34CDF9} & 2C030 & entry1\_norm85\_nM &  & straight & 1 & - & low & 85\% &  & yes \\
		 & \cellcolor[HTML]{FD6864} & 2C040 & entry3\_bendSmall\_nM &  & small bend & 3 & - & low & - &  & yes \\
		 & \cellcolor[HTML]{FD6864} & 2C050 & entry3\_bendSmall\_hM &  & small bend & 3 & - & high & - &  & yes \\
		 & \cellcolor[HTML]{FD6864} & SOLO\_REF21 & SOLO\_REF &  & small bend &  &  &  &  &  & no \\
		 & \cellcolor[HTML]{67FD9A} & 2C060 & entry3\_blank\_hM &  & straight & 3 & - & high & - &  & yes \\
		 & \cellcolor[HTML]{67FD9A} & 2C070 & entry1\_blank\_hM &  & straight & 1 & - & high & - &  & yes \\
		 & \cellcolor[HTML]{67FD9A} & 2C080 & entry1\_norm85\_hM &  & straight & 1 & - & high & 85\% &  & yes \\
		 & \cellcolor[HTML]{34CDF9} & 2C090 & entry3\_straight\_nM &  & straight & 3 & - & low & - &  & yes \\
		 & \cellcolor[HTML]{34CDF9} & 2C100 & entry3\_straight\_hM &  & straight & 3 & - & high & - &  & yes \\
		 & \cellcolor[HTML]{FD6864} & 2C110 & entry3\_signs\_hM &  & straight & 3 & signs & high & - &  & yes \\
		 & \cellcolor[HTML]{FD6864} & 2C120 & entry1\_blank\_hM &  & straight & 1 & - & high & - &  & yes \\
		 & \cellcolor[HTML]{FD6864} & 2C130 & entry1\_norm70\_nM &  & straight & 1 & - & low & 70\% &  & yes \\
		 & \cellcolor[HTML]{67FD9A} & 2C140 & entry3\_blank\_hM &  & straight & ? & - & high & - &  & yes \\
		 & \cellcolor[HTML]{67FD9A} & 2C150 & entry1\_norm70\_nM &  & straight & ? & - & low & 70\% &  & yes \\
		\multirow{-17}{*}{day 2} & \cellcolor[HTML]{67FD9A} & SOLO\_REF22 & SOLO\_REF &  & straight &  &  &  &  &  & no \\ \hline
		 \pagebreak & \cellcolor[HTML]{34CDF9} & 3C010 & entry3\_bend90\_lines\_nM &  & 90° bend & 3 & lines & low & - &  & yes \\
		 & \cellcolor[HTML]{34CDF9} & 3C020 & entry3\_bend90\_lines\_hM &  & 90° bend & 3 & lines & low & - &  & yes \\
		 & \cellcolor[HTML]{34CDF9} & 3C021 & entry3\_bend90\_lines\_hM &  & 90° bend & 3 & lines & high & - &  & no \\
		 & \cellcolor[HTML]{FD6864} & 3C030 & entry3\_bend90\_nM &  & 90° bend & 3 & - & low & - &  & yes \\
		 & \cellcolor[HTML]{FD6864} & 3C040 & entry3\_bend90\_hM &  & 90° bend & 3 & - & high & - &  & yes \\
		 & \cellcolor[HTML]{FD6864} & SOLO\_REF31 & SOLO\_REF &  & 90° bend &  &  &  &  &  & no \\
		 & \cellcolor[HTML]{67FD9A} & 3C050 & entry3\_bend90\_nM &  & 90° bend & 3 & - & low & - &  & yes \\
		 & \cellcolor[HTML]{67FD9A} & 3C060 & entry3\_bend90\_hM &  & 90° bend & 3 & - & low & - &  & yes \\
		 & \cellcolor[HTML]{67FD9A} & 3C061 & entry3\_bend90\_hM &  & 90° bend & 3 & - & high & - &  & no \\
		 & \cellcolor[HTML]{34CDF9} & 3C070 & entry1\_bend90\_nI\_nM &  & 90° bend & 1 & - & low & - &  & yes \\
		 & \cellcolor[HTML]{34CDF9} & 3C080 & entry1\_bend90\_wI\_nM &  & 90° bend & 1 & - & low & - &  & yes \\
		 & \cellcolor[HTML]{34CDF9} & 3C081 & entry1\_bend90\_slow\_hM &  & 90° bend & 1 & - & high & - &  & no \\
		 & \cellcolor[HTML]{FD6864} & 3C090 & entry1\_bend90\_nI\_hM &  & 90° bend & 1 & - & low & - & nI, inflow side & yes \\
		 & \cellcolor[HTML]{FD6864} & 3C100 & entry1\_bend90\_wI\_hM &  & 90° bend & 1 & - & low & - & wI, side & yes \\
		 & \cellcolor[HTML]{FD6864} & 3C101 & entry1\_bend90\_slow\_nM &  & 90° bend & 1 & - & high & - & slow, inflow side & no \\
		 & \cellcolor[HTML]{67FD9A} & 3C110 & entry1\_bend90\_nI\_nM &  & 90° bend & 1 & - & low & - & nI, inflow & yes \\
		 & \cellcolor[HTML]{67FD9A} & 3C120 & entry1\_bend90\_wI\_hM &  & 90° bend & 1 & - & high & - & wI, straight inflow & yes \\
		 & \cellcolor[HTML]{67FD9A} & 3C121 & entry3\_bend90\_slow\_hM &  & 90° bend & 1 & - & high & - & slow, straight inflow & no \\
		\multirow{-19}{*}{day 3} & \cellcolor[HTML]{67FD9A} & SOLO\_REF\_32 & SOLO\_REF &  & 90° bend &  &  &  &  &  & no \\ \hline
\label{tab:siteCd1-3_runs}\\
\end{longtable}
\end{landscape}
\normalsize

\subsection{Experimental Configurations Oval Experiments}
\label{app:Oval}
\tiny
\begin{longtable}{cccll}
\caption{Table showing experimental runs performed within the scope of the Oval Experiments (Sec.\,\ref{subsec:OvalExperiments}). The names of the experiments are given as well as varied parameters.}\\
\multicolumn{1}{l}{\textbf{Run no.}} & \textbf{\begin{tabular}[c]{@{}c@{}}No. of peds \\ oval 1\\ (left in camera)\end{tabular}} & \textbf{\begin{tabular}[c]{@{}c@{}}No. of peds\\ oval 2\\ (right in camera)\end{tabular}} & \textbf{run name oval 1} & \textbf{run name oval 2} \\ \hline
\endfirsthead
\endhead
\cline{1-4}
\endfoot
\endlastfoot
\multicolumn{5}{l}{female/male} \\ \hline
1 & 40 & 35 & single\_file\_gender\_female\_1\_1 & single\_file\_gender\_male\_2\_1 \\
2 & 36 & 31 & single\_file\_gender\_female\_1\_2 & single\_file\_gender\_male\_2\_2 \\
3 & 32 & 27 & single\_file\_gender\_female\_1\_3 & single\_file\_gender\_male\_2\_3 \\
4 & 24 & 19 & single\_file\_gender\_female\_1\_4 & single\_file\_gender\_male\_2\_4 \\
5 & 20 & 15 & single\_file\_gender\_female\_1\_5 & single\_file\_gender\_male\_2\_5 \\
6 & 20 & 20 & single\_file\_gender\_female\_1\_6 & single\_file\_gender\_male\_2\_6 \\
7 & 16 & 16 & single\_file\_gender\_female\_1\_7 & single\_file\_gender\_male\_2\_7 \\
8 & 8 & 8 & single\_file\_gender\_female\_1\_8 & single\_file\_gender\_male\_2\_8 \\
9 & 4 & 8 & single\_file\_gender\_female\_1\_9 & single\_file\_gender\_male\_2\_9 \\
10 & 0 & 4 &  & single\_file\_gender\_male\_2\_10 \\ \hline
\multicolumn{5}{l}{gender alternating} \\ \hline
1 & 4 & 4 & single\_file\_gender\_1\_1 & single\_file\_gender\_2\_1 \\
2 & 4 & 4 & single\_file\_gender\_1\_2 & single\_file\_gender\_2\_2 \\
3 & 4 & 4 & single\_file\_gender\_1\_3 & single\_file\_gender\_2\_3 \\
4 & 4 & 8 & single\_file\_gender\_1\_4 & single\_file\_gender\_2\_4 \\
5 & 4 & 4 & single\_file\_gender\_1\_5 & single\_file\_gender\_2\_5 \\
6 & 8 & 4 & single\_file\_gender\_1\_6 & single\_file\_gender\_2\_6 \\
7 & 8 & 8 & single\_file\_gender\_1\_7 & single\_file\_gender\_2\_7 \\
8 & 16 & 16 & single\_file\_gender\_1\_8 & single\_file\_gender\_2\_8 \\
9 & 20 & 20 & single\_file\_gender\_1\_9 & single\_file\_gender\_2\_9 \\
10 & 24 & 0 & single\_file\_gender\_1\_10 &  \\
11 & 24 & 16 & single\_file\_gender\_1\_11 & single\_file\_gender\_2\_10 \\
12 & 32 & 8 & single\_file\_gender\_1\_12 & single\_file\_gender\_2\_11 \\
13 & 36 & 0 & single\_file\_gender\_1\_13 &  \\
14 & 40 & 0 & single\_file\_gender\_1\_14 &  \\ \hline
\multicolumn{5}{l}{gender random order} \\ \hline
1 & 4 & 4 & single\_file\_gender\_random\_1\_1 & single\_file\_gender\_random\_2\_1 \\
2 & 4 & 4 & single\_file\_gender\_random\_1\_2 & single\_file\_gender\_random\_2\_2 \\
3 & 4 & 4 & single\_file\_gender\_random\_1\_3 & single\_file\_gender\_random\_2\_3 \\
4 & 4 & 8 & single\_file\_gender\_random\_1\_4 & single\_file\_gender\_random\_2\_4 \\
5 & 4 & 4 & single\_file\_gender\_random\_1\_5 & single\_file\_gender\_random\_2\_5 \\
6 & 8 & 4 & single\_file\_gender\_random\_1\_6 & single\_file\_gender\_random\_2\_6 \\
7 & 8 & 8 & single\_file\_gender\_random\_1\_7 & single\_file\_gender\_random\_2\_7 \\
8 & 16 & 16 & single\_file\_gender\_random\_1\_8 & single\_file\_gender\_random\_2\_8 \\
9 & 20 & 20 & single\_file\_gender\_random\_1\_9 & single\_file\_gender\_random\_2\_9 \\
10 & 24 & 0 & single\_file\_gender\_random\_1\_10 &  \\
11 & 24 & 16 & single\_file\_gender\_random\_1\_11 & single\_file\_gender\_random\_2\_10 \\
12 & 32 & 8 & single\_file\_gender\_random\_1\_12 & single\_file\_gender\_random\_2\_11 \\
13 & 36 & 0 & single\_file\_gender\_random\_1\_13 &  \\
14 & 40 & 0 & single\_file\_gender\_random\_1\_14 &  \\ \cline{1-5}
\label{tab:siteCd4_runs}\\
\end{longtable}
\normalsize

\newpage
\subsection{Experimental Configurations Personal-Space Experiments}
\label{app:PSpace}
\begin{longtable}{lccll}
\caption{Table showing experimental runs performed within the scope of the Personal Space Experiments (Sec.\,\ref{subsec:PersonalSpaceExperiments}). The names of the experiments are given as well as the IDs of the ``standing" participants.}\\
		\multicolumn{1}{c}{run name} & \multicolumn{1}{c}{\begin{tabular}[c]{@{}c@{}}no. of standing \\ people\end{tabular}} & \multicolumn{1}{c}{\begin{tabular}[c]{@{}c@{}}no. of walking\\ people\end{tabular}} & \multicolumn{1}{c}{\begin{tabular}[c]{@{}c@{}}Code ID of \\ standing people\end{tabular}} & \multicolumn{1}{c}{} \\ \hline
		\endfirsthead
		\endhead
		\hline
		\endfoot
		\endlastfoot 	\hline 
		4C1010 & 7 & 10 & 770, 710, 775, 785, 884, 897, 957 &  \\
		4C1020 & 7 & 10 & 731, 735, 784, 793, 859, 959, 960 &  \\
		4C1030 & 7 & 10 & 670, 713, 714, 718, 837, 945, 963 &  \\
		4C1040 & 7 & 10 & 645, 657, 669, 901, 902, 903, 953 &  \\
		4C2010 & 7 & 10 & 673, 711, 738, 783, 801, 857, 942 &  \\
		4C2020 & 7 & 10 & 666, 766, 789, 835, 843, 880, 946 &  \\
		4C2030 & 7 & 10 & 649, 708, 761, 771, 773, 889, 900 &  \\
		4C2040 & 7 & 10 & 737, 844, 847, 850, 854, 940, 941 &  \\ \hline
\label{tab:siteCd4_pspace}
\end{longtable}

\begin{landscape}
\subsection{Experimental Configurations Boarding And Alighting Experiments}
\label{app:BoardingAlighting}
\tiny
\begin{longtable}{lclccccccccccc}
\caption{Table showing experimental runs performed within the scope of the Boarding and Alighting Experiments (Sec.\,\ref{subsec:BoardingAndAlightingExperiments}). The Names of the experiments are given together with the colour of the group that participated in the respective runs as well as varied parameters.}
\label{tab:my-table}\\
\rowcolor[HTML]{FFFFFF} 
 & \multicolumn{2}{c}{\cellcolor[HTML]{FFFFFF}\textbf{run name}} & \textbf{\begin{tabular}[c]{@{}c@{}}N\_total\\      \\   {[}number of \\ participants{]}\end{tabular}} & \textbf{\begin{tabular}[c]{@{}c@{}}N\_in\\      \\ {[}number of \\ participants \\ boarding{]}\end{tabular}} & \textbf{\begin{tabular}[c]{@{}c@{}}N\_train\\       \\ {[}number of \\ participants\\ staying in train{]}\end{tabular}} & \textbf{\begin{tabular}[c]{@{}c@{}}N\_out\\      \\  {[}number of \\ participants \\ alighting{]}\end{tabular}} & \textbf{\begin{tabular}[c]{@{}c@{}}luggage\_in\\       \\  {[}none, \\ backpack, \\ trolley, \\ buggy, mix{]}\end{tabular}} & \textbf{\begin{tabular}[c]{@{}c@{}}luggage\_out\\      \\ {[}none, \\ backpack, \\ trolley, \\ buggy, mix{]}\end{tabular}} & \textbf{\begin{tabular}[c]{@{}c@{}}norm\\      \\   {[}none, 70, \\ 80, 100{]}\end{tabular}} & \textbf{\begin{tabular}[c]{@{}c@{}}cooperation\\      \\  {[}none, \\ work together{]}\end{tabular}} & \textbf{\begin{tabular}[c]{@{}c@{}}line-up\\      \\ {[}none, \\ throng, \\ corridor, 45°{]}\end{tabular}} & \textbf{\begin{tabular}[c]{@{}c@{}}groups\\      \\ {[}none, pairs, \\ groups of 3, \\ groups of 5{]}\end{tabular}} & \textbf{\begin{tabular}[c]{@{}c@{}}motivation\\      \\ {[}none, hurry, \\ pushing, \\ 50\%/100\% distracted{]}\end{tabular}} \\ \hline
\endfirsthead
\multicolumn{14}{c}%
{{\bfseries Table \thetable\ continued from previous page}} \\
\rowcolor[HTML]{FFFFFF} 
 & \multicolumn{2}{c}{\cellcolor[HTML]{FFFFFF}\textbf{run name}} & \textbf{\begin{tabular}[c]{@{}c@{}}N\_total\\      \\   {[}number of \\ participants{]}\end{tabular}} & \textbf{\begin{tabular}[c]{@{}c@{}}N\_in\\      \\ {[}number of \\ participants \\ boarding{]}\end{tabular}} & \textbf{\begin{tabular}[c]{@{}c@{}}N\_train\\       \\ {[}number of \\ participants\\ staying in train{]}\end{tabular}} & \textbf{\begin{tabular}[c]{@{}c@{}}N\_out\\      \\  {[}number of \\ participants \\ alighting{]}\end{tabular}} & \textbf{\begin{tabular}[c]{@{}c@{}}luggage\_in\\       \\  {[}none, \\ backpack, \\ trolley, \\ buggy, mix{]}\end{tabular}} & \textbf{\begin{tabular}[c]{@{}c@{}}luggage\_out\\      \\ {[}none, \\ backpack, \\ trolley, \\ buggy, mix{]}\end{tabular}} & \textbf{\begin{tabular}[c]{@{}c@{}}norm\\      \\   {[}none, 70, \\ 80, 100{]}\end{tabular}} & \textbf{\begin{tabular}[c]{@{}c@{}}cooperation\\      \\  {[}none, \\ work together{]}\end{tabular}} & \textbf{\begin{tabular}[c]{@{}c@{}}line-up\\      \\ {[}none, \\ throng, \\ corridor, 45°{]}\end{tabular}} & \textbf{\begin{tabular}[c]{@{}c@{}}groups\\      \\ {[}none, pairs, \\ groups of 3, \\ groups of 5{]}\end{tabular}} & \textbf{\begin{tabular}[c]{@{}c@{}}motivation\\      \\ {[}none, hurry, \\ pushing, \\ 50\%/100\% distracted{]}\end{tabular}} \\ \hline
\endhead
\hline
\endfoot
\endlastfoot
 & \cellcolor[HTML]{92D050} & D1\_1\_d\_4\_1 & 30 & 10,10,10 & 0,10,20 & 0,0,10 & -- & -- & -- & -- & -- & -- & -- \\
\multicolumn{1}{c}{} & \cellcolor[HTML]{92D050} & D1\_1\_d\_4\_2 & 50 & 10,10,20 & 10,20,20 & 0,10,10 & -- & -- & -- & -- & -- & -- & -- \\
 & \cellcolor[HTML]{92D050} & D1\_1\_d\_4\_3 & 60 & 10, 20, 10 & 20,20,30 & 10,10,10 & -- & -- & -- & -- & -- & -- & -- \\
 & \cellcolor[HTML]{92D050} & D1\_1\_d\_4\_4 & 50 & 20,10,10 & 20,30,30 & 10,10,20 & -- & -- & -- & -- & -- & -- & -- \\
 & \cellcolor[HTML]{92D050} & D1\_1\_d\_4\_5 & 50 & 10,10,20 & 30,30,20 & 10,20,10 & -- & -- & -- & -- & -- & -- & -- \\
 & \cellcolor[HTML]{92D050} & D1\_1\_d\_4\_6 & 40 & 10,20,10 & 30,20,30 & 20,10,10 & -- & -- & -- & -- & -- & -- & -- \\
 & \cellcolor[HTML]{92D050} & D1\_1\_d\_4\_7 & 40 & 20,10,20 & 20,30,30 & 10,10,20 & -- & -- & -- & -- & -- & -- & -- \\
 & \cellcolor[HTML]{92D050} & D1\_1\_d\_4\_8 & 60 & 10,20,10 & 30,30,30 & 10,20,10 & -- & -- & -- & -- & -- & -- & -- \\
\hline
\end{longtable}
\end{landscape}
\normalsize

\subsection{Experimental Configurations Tiny Box Experiments}
\label{app:TinyHouse}
\tiny
\begin{longtable}[c]{c|llclccl}
\caption{Table showing experimental runs performed within the scope of the Tiny Box Experiments (Sec.\,\ref{subsec:TinyHouseExperiments}). The colour of the group that participants belonged to is given as well as varied parameters.}
\label{tab:tinyHouse}\\
 &
  \multicolumn{2}{c}{run no} &
  \begin{tabular}[c]{@{}c@{}}no. of people \\ in box\end{tabular} &
  \multicolumn{1}{c}{id numbers} &
  \begin{tabular}[c]{@{}c@{}}box no. \\ {[}1,2,3,4{]}\end{tabular} &
  \begin{tabular}[c]{@{}c@{}}special announcement \\      {[}speaking allowed / prohibited{]}\end{tabular} &
  \multicolumn{1}{c}{\begin{tabular}[c]{@{}c@{}}waiting time\\  {[}minutes{]}\end{tabular}} \\ \hline
\endfirsthead
\multicolumn{8}{c}%
{{\bfseries Table \thetable\ continued from previous page}} \\
    &
  \multicolumn{2}{c}{run no} &
  \begin{tabular}[c]{@{}c@{}}no. of people \\ in box\end{tabular} &
  \multicolumn{1}{c}{id numbers} &
  \begin{tabular}[c]{@{}c@{}}box no. \\ {[}1,2,3,4{]}\end{tabular} &
  \begin{tabular}[c]{@{}c@{}}special announcement \\      {[}speaking allowed / prohibited{]}\end{tabular} &
  \multicolumn{1}{c}{\begin{tabular}[c]{@{}c@{}}waiting time\\  {[}minutes{]}\end{tabular}} \\ \hline
\endhead
                         & \cellcolor[HTML]{5B9BD5} & 1   & 2 & 1, 8                   & 1 & speaking allowed      & 5 \\
                         & \cellcolor[HTML]{5B9BD5} & 2   & 2 & 2, 7                   & 2 & speaking   allowed    & 5 \\
                         & \cellcolor[HTML]{5B9BD5} & 3   & 2 & 3, 6                   & 3 & speaking   allowed    & 5 \\
                         & \cellcolor[HTML]{5B9BD5} & 4   & 2 & 4, 5                   & 4 & speaking   allowed    & 5 \\
                         & \cellcolor[HTML]{5B9BD5} & 5   & 1 & 3                      & 1 & speaking   allowed    & 5 \\
                         & \cellcolor[HTML]{5B9BD5} & 6   & 1 & 4                      & 2 & speaking   allowed    & 5 \\
                         & \cellcolor[HTML]{5B9BD5} & 7   & 1 & 5                      & 3 & speaking   allowed    & 5 \\
                         & \cellcolor[HTML]{5B9BD5} & 8   & 1 & 6                      & 4 & speaking   allowed    & 5 \\
\end{longtable}
\normalsize

\newpage
\subsection{Experimental Configurations Bottleneck Experiments.}
\label{app:Bottleneck}
\tiny
\begin{longtable}[c]{llllllll}
\caption{Table showing experimental runs performed within the scope of the Bottleneck Experiments (Sec.\,\ref{subsec:BottleneckExperiments}). The names of the experiments are given as well as varied parameters.\\}
\label{tab:siteDd4_runs}\\
		\multicolumn{1}{c}{\begin{tabular}[c]{@{}c@{}}run name\\ (short)\end{tabular}} & \multicolumn{1}{c}{\begin{tabular}[c]{@{}c@{}}run name\\ (descriptive)\\ {[}width\_length\_motivation\_addition{]}\end{tabular}} & \multicolumn{1}{c}{\begin{tabular}[c]{@{}c@{}}no. of \\ participants\end{tabular}} & \multicolumn{1}{c}{\begin{tabular}[c]{@{}c@{}}width\\ {[}m{]}\end{tabular}} & \multicolumn{1}{c}{\begin{tabular}[c]{@{}c@{}}length\\ {[}m{]}\end{tabular}} & \multicolumn{1}{c}{\begin{tabular}[c]{@{}c@{}}motivation\\ {[}h0: normal;\\ h1: hurry;\\ h2: full commitment;\\ he: normal-hurry{]}\end{tabular}} & \multicolumn{1}{c}{line-up} & \multicolumn{1}{c}{special announcement} \\ \hline
		\endfirsthead
		\endhead
		4D000 & test01 & 136 & 1.2 & 2.0 & h0 & 2\,m semi-circle & explanation h0 \\
		4D001 & test02 & 136 & 1.2 & 2.0 & h1 & 2\,m semi-circle & explanation h1 \\
		4D002 & test03 & 136 & 1.2 & 2.0 & h2 & directly at bottleneck & \begin{tabular}[c]{@{}l@{}}explanation h2,\\ abort signal\end{tabular} \\
		4D010 & w120\_l200\_h0 & 136 & 1.2 & 2.0 & h0 & 2\,m semi-circle & - \\
		4D020 & w120\_l200\_h1 & 136 & 1.2 & 2.0 & h1 & 2\,m semi-circle & - \\
		4D030 & w080\_l200\_h0 & 161 & 0.8 & 2.0 & h0 & 2\,m semi-circle & - \\
		4D040 & w080\_l200\_h1 & 160 & 0.8 & 2.0 & h1 & 2\,m semi-circle & - \\
		4D050 & w160\_l021\_h0 & 125 & 1.6 & 0.2 & h0 & 2\,m semi-circle & - \\
		4D060 & w160\_l021\_h1 & 129 & 1.6 & 0.2 & h1 & 2\,m semi-circle & - \\
		4D070 & w120\_l021\_h0 & 153 & 1.2 & 0.2 & h0 & 2\,m semi-circle & - \\
		4D080 & w120\_l021\_h1 & 129 & 1.2 & 0.2 & h1 & 2\,m semi-circle & - \\
		4D090 & w100\_l021\_h0 & 152 & 1.0 & 0.2 & h0 & 2\,m semi-circle & - \\
		4D100 & w100\_l021\_h1 & 131 & 1.0 & 0.2 & h1 & 2\,m semi-circle & - \\
		4D110 & w080\_l021\_h0 & 132 & 0.8 & 0.2 & h0 & 2\,m semi-circle & mix up: first row queue behind \\
		4D120 & w080\_l021\_h1 & 132 & 0.8 & 0.2 & h1 & 2\,m semi-circle & - \\
		4D130 & w080\_l021\_h2 & 129 & 0.8 & 0.2 & h2 & directly at bottleneck & - \\ \hline
		4D140 & w080\_l021\_h1\_AoO15 & \begin{tabular}[c]{@{}l@{}}2 at a time\\ total: 122\end{tabular} & 0.8 & 0.2 & h1 & \begin{tabular}[c]{@{}l@{}}4\,m semi-circle\\ positions: 0°, 180°\end{tabular} & - \\
		4D150 & w080\_l021\_h1\_AoO13 & \begin{tabular}[c]{@{}l@{}}2 at a time\\ total: 152\end{tabular} & 0.8 & 0.2 & h1 & \begin{tabular}[c]{@{}l@{}}4\,m semi-circle\\ positions: 0°, 90°\end{tabular} & - \\
		4D170 & w080\_l021\_h1\_AoO135 & \begin{tabular}[c]{@{}l@{}}3 at a time\\ total: 75\end{tabular} & 0.8 & 0.2 & h1 & \begin{tabular}[c]{@{}l@{}}4\,m semi-circle\\ positions: 0°, 90°, 180°\end{tabular} & - \\
		4D171 & w080\_l021\_h1\_AoO125 & \begin{tabular}[c]{@{}l@{}}3 at a time\\ total: 57\end{tabular} & 0.8 & 0.2 & h1 & \begin{tabular}[c]{@{}l@{}}4\,m semi-circle\\ positions: 0°, 45°, 180°\end{tabular} & - \\
		4D172 & w080\_l021\_h1\_AoO1245 & \begin{tabular}[c]{@{}l@{}}4 at a time\\ total: 116\end{tabular} & 0.8 & 0.2 & h1 & \begin{tabular}[c]{@{}l@{}}4\,m semi-circle\\ positions: 0°, 45°,\\ 135°, 180°\end{tabular} & - \\
		4D173 & w080\_l021\_h1\_AoO12345 & \begin{tabular}[c]{@{}l@{}}5 at a time\\ total: 115\end{tabular} & 0.8 & 0.2 & h1 & \begin{tabular}[c]{@{}l@{}}4\,m semi-circle\\ positions: 0°, 45°, 90°,\\ 135°, 180°\end{tabular} & - \\ \hline
		4D180 & w070\_l021\_h1\_interrupt & 150 & 0.7 & 0.2 & h1 & 2\,m semi-circle & \begin{tabular}[c]{@{}l@{}}interruption: technical,\\ questionnaire\end{tabular} \\
		4D200 & w070\_l021\_h0 & 168 & 0.7 & 0.2 & h0 & 2\,m semi-circle & - \\
		4D210 & w070\_l021\_h1 & 169 & 0.7 & 0.2 & h1 & 2\,m semi-circle & questionnaire \\
		4D181 & w070\_l021\_h1\_interrupt & 164 & 0.7 & 0.2 & h1 & 2\,m semi-circle & interruption: technical \\
		4D220 & w070\_l021\_h2\_025 & 25 & 0.7 & 0.2 & h2 & directly at bottleneck & - \\
		4D230 & w070\_l021\_h2\_050 & 51 & 0.7 & 0.2 & h2 & directly at bottleneck & - \\
		4D240 & w070\_l021\_h2\_075 & 77 & 0.7 & 0.2 & h2 & directly at bottleneck & - \\
		4D250 & w070\_l021\_h2\_100 & 94 & 0.7 & 0.2 & h2 & directly at bottleneck & abort signal \\
		4D251 & w070\_l021\_h2\_100 & 96 & 0.7 & 0.2 & h2 & directly at bottleneck & - \\
		4D280 & w070\_l021\_h1\_push & 158 & 0.7 & 0.2 & h1 & 2\,m semi-circle & \begin{tabular}[c]{@{}l@{}}20\% active pushing,\\ questionnaire, abort signal\end{tabular} \\
		4D281 & w070\_l021\_h0 & 167 & 0.7 & 0.2 & h0 & - & \begin{tabular}[c]{@{}l@{}}no announced run / \\ without 'go' signal\end{tabular} \\
		4D290 & w070\_l021\_he\_slow & 201 & 0.7 & 0.2 & he & 2\,m semi-circle & 20\% actively slowing down \\
		4D291 & w070\_l021\_he & 168 & 0.7 & 0.2 & he & 2\,m semi-circle & - \\ \hline
		4D300 & w060\_l021\_h0 & 187 & 0.6 & 0.2 & h0 & 2\,m semi-circle &  \\
		4D310 & w060\_l021\_he & 191 & 0.6 & 0.2 & he & 2\,m semi-circle &  \\
		4D320 & w060\_l021\_he\_interrupt & 101 & 0.6 & 0.2 & he & 2\,m semi-circle & \begin{tabular}[c]{@{}l@{}}interruption: information,\\ questionnaire\end{tabular} \\
		4D330 & w060\_l021\_he\_interrupt & 91 & 0.6 & 0.2 & he & 2\,m semi-circle & \begin{tabular}[c]{@{}l@{}}interruption: information,\\ questionnaire\end{tabular} \\
		4D340 & w060\_l021\_h2 & 53 & 0.6 & 0.2 & h2 & directly at bottleneck & \\  \hline
\end{longtable}


\newpage
\begin{table}[]
\subsection{Reprojection Error Camera}
\caption{Table showing the reprojection errors for the calibration points for each camera and day in the persons' default height.}
\centering
\begin{tabular}{|c|c|c|ccc|}
\hline
\multirow{2}{*}{day} & \multirow{2}{*}{area} & \multirow{2}{*}{camera} & \multicolumn{3}{c|}{reprojection error at default height} \\ \cline{4-6} 
 &  &  & \multicolumn{1}{c|}{average / cm} & \multicolumn{1}{c|}{std. dev / cm} & max / cm \\ \hline
1 & B & SL\_cam1 & \multicolumn{1}{c|}{0.99} & \multicolumn{1}{c|}{0.42} & 2.03 \\ \hline
1 & B & SL\_cam2 & \multicolumn{1}{c|}{1.01} & \multicolumn{1}{c|}{0.64} & 3.28 \\ \hline
1 & B & SL\_cam3 & \multicolumn{1}{c|}{0.46} & \multicolumn{1}{c|}{0.16} & 0.67 \\ \hline
1 & C & SL\_cam4 & \multicolumn{1}{c|}{0.98} & \multicolumn{1}{c|}{0.5} & 1.59 \\ \hline
1 & C & SL\_cam5 & \multicolumn{1}{c|}{1.21} & \multicolumn{1}{c|}{0.66} & 2.88 \\ \hline
1 & C & SL\_cam6 & \multicolumn{1}{c|}{0.51} & \multicolumn{1}{c|}{0.25} & 0.92 \\ \hline
1 & D & SL\_cam7 & \multicolumn{1}{c|}{3.15} & \multicolumn{1}{c|}{1.8} & 8.03 \\ \hline
1 & D & SL\_cam8 & \multicolumn{1}{c|}{0.62} & \multicolumn{1}{c|}{0.49} & 1.54 \\ \hline
2 & B & SL\_cam1 & \multicolumn{1}{c|}{0.99} & \multicolumn{1}{c|}{0.53}  & 2.32 \\ \hline
2 & B & SL\_cam2 & \multicolumn{1}{c|}{0.96} & \multicolumn{1}{c|}{0.47} & 2.19 \\ \hline
2 & B & SL\_cam3 & \multicolumn{1}{c|}{0.48} & \multicolumn{1}{c|}{0.19} & 0.68 \\ \hline
2 & C & SL\_cam4 & \multicolumn{1}{c|}{0.98} & \multicolumn{1}{c|}{0.5} & 1.59 \\ \hline
2 & C & SL\_cam5 & \multicolumn{1}{c|}{1.21} & \multicolumn{1}{c|}{0.66} & 2.88 \\ \hline
2 & C & SL\_cam6 & \multicolumn{1}{c|}{0.51} & \multicolumn{1}{c|}{0.25} & 0.92 \\ \hline
2 & D & SL\_cam7 & \multicolumn{1}{c|}{1.51} & \multicolumn{1}{c|}{0.88} & 2.74 \\ \hline
2 & D & SL\_cam8 & \multicolumn{1}{c|}{0.47} & \multicolumn{1}{c|}{0.42}  & 1.18 \\ \hline
3 & B & SL\_cam1 & \multicolumn{1}{c|}{0.99} & \multicolumn{1}{c|}{0.53} & 2.32 \\ \hline
3 & B & SL\_cam2 & \multicolumn{1}{c|}{0.96} & \multicolumn{1}{c|}{0.47} & 2.19 \\ \hline
3 & B & SL\_cam3 & \multicolumn{1}{c|}{0.48} & \multicolumn{1}{c|}{0.19} & 0.68 \\ \hline
3 & C & SL\_cam4 & \multicolumn{1}{c|}{0.98} & \multicolumn{1}{c|}{0.5} & 1.59 \\ \hline
3 & C & SL\_cam5 & \multicolumn{1}{c|}{1.21} & \multicolumn{1}{c|}{0.66}  & 2.88 \\ \hline
3 & C & SL\_cam6 & \multicolumn{1}{c|}{0.55} & \multicolumn{1}{c|}{0.26} & 0.81 \\ \hline
3 & D & SL\_cam7 & \multicolumn{1}{c|}{1.51} & \multicolumn{1}{c|}{0.88} & 2.74 \\ \hline
3 & D & SL\_cam8 & \multicolumn{1}{c|}{0.47} & \multicolumn{1}{c|}{0.42} & 1.18 \\ \hline
4 & C & SL\_cam4 & \multicolumn{1}{c|}{1.21} & \multicolumn{1}{c|}{0.67}  & 2.79 \\ \hline
4 & C & SL\_cam5 & \multicolumn{1}{c|}{1.5} & \multicolumn{1}{c|}{0.73} & 2.85 \\ \hline
4 & C & SL\_cam6 & \multicolumn{1}{c|}{0.6} & \multicolumn{1}{c|}{0.25} & 1.05 \\ \hline
4 & C & RX0\_cam2 & \multicolumn{1}{c|}{1.35} & \multicolumn{1}{c|}{0.5} & 2.33 \\ \hline
4 & D & GP7\_1 & \multicolumn{1}{c|}{3.66} & \multicolumn{1}{c|}{2.1} & 7.47 \\ \hline
4 & D & RX0\_cam1 & \multicolumn{1}{c|}{0.45} & \multicolumn{1}{c|}{0.23} & 0.81 \\ \hline
\end{tabular}

\label{tab:reprojectionerror}
\end{table}


\end{document}